\documentclass[twocolumn,
superscriptaddress,
nofootinbib,
amsmath,amssymb,
aps, physrev,
]{revtex4}

\usepackage{amsthm}
\usepackage{amsmath}
\usepackage{graphicx}
\usepackage{multirow}
\usepackage{tikz}
\usetikzlibrary{arrows.meta}
\usepackage{booktabs}

\begin{document}

\title{Neural network enhanced Bayesian global analysis of relativistic heavy ion collisions}
\author{Jussi Auvinen}
\email{jussi.a.m.auvinen@jyu.fi}
\affiliation{Incubator of Scientific Excellence - Centre for Simulations of Superdense Fluids, University of Wroc\l{}aw, plac Maksa Borna 9, PL-50204 Wroc\l{}aw, Poland}
\affiliation{University of Jyväskylä, Department of Physics, P.O. Box 35, FI-40014 University of Jyväskylä, Finland}
\affiliation{Helsinki Institute of Physics, P.O.Box 64, FI-00014 University of Helsinki, Finland}
\author{Kari J. Eskola}
\affiliation{University of Jyväskylä, Department of Physics, P.O. Box 35, FI-40014 University of Jyväskylä, Finland}
\affiliation{Helsinki Institute of Physics, P.O.Box 64, FI-00014 University of Helsinki, Finland}
\author{Henry Hirvonen}
\affiliation{Department of Mathematics, Vanderbilt University, Nashville, TN 37240, USA}
\affiliation{Department of Physics and Astronomy, Vanderbilt University, Nashville, TN 37240, USA}
\author{Harri Niemi}
\affiliation{University of Jyväskylä, Department of Physics, P.O. Box 35, FI-40014 University of Jyväskylä, Finland}
\affiliation{Helsinki Institute of Physics, P.O.Box 64, FI-00014 University of Helsinki, Finland}

\begin{abstract}

We introduce a novel deep convolutional neural network (NN) -enhanced Bayesian global analysis of bulk observables in highest-energy heavy-ion collisions, using relativistic 2+1 D second-order viscous hydrodynamics with a  dynamical freeze-out, and with perturbative QCD and saturation -based initial conditions from the event-by-event EKRT-model. Our analysis has 13+2 free parameters for the QCD-matter properties + initial state, which are constrained by the experimental data from $\sqrt{s_{NN}}=200$ GeV Au+Au collisions at RHIC and $2.76$ TeV Pb+Pb, $5.02$ TeV Pb+Pb, and $5.44$ TeV Xe+Xe collisions at the LHC.
We replace the computationally demanding hydrodynamical simulations by NNs, which predict bulk observables directly from the initial energy density profiles, event-by-event, and account for the QCD-matter properties. With the NN output, we train the Gaussian process emulators for obtaining centrality-class averaged observables and their uncertainties. The NNs reduce the computing time significantly, enabling us to include also statistics-hungry flow observables like $v_4$ and the normalized symmetric cumulant $NSC(4,2)$ in the analysis. In this paper, we demonstrate the feasibility of the NN based Bayesian global analysis. We find the data favoring a specific shear viscosity $\eta/s$ with a minimum-value plateau at temperatures $150\lesssim T \lesssim 230$ MeV, with $0.12 \lesssim (\eta/s)_{\mathrm{min}} \lesssim 0.18$. The bulk viscous coefficient $\zeta/s$ is non-zero at $200\lesssim T \lesssim 300$ MeV. The Knudsen number at the freeze-out is $0.8-2.3$, while the ratio of the mean free path to the system size at freeze-out is in the range $0.3-1.2$, implying that the freeze-out indeed happens at the expected limit of the applicability of hydrodynamics.

\end{abstract}

\maketitle

\section{Introduction}
\label{S:introduction}

According to Quantum Chromodynamics (QCD), the strongly interacting matter at high temperatures ($T\gtrsim 150$ MeV) and vanishing baryochemical potential takes the form of the quark-gluon plasma (QGP), a fluid of deconfined quarks and gluons \cite{Bazavov:2014pvz, Bazavov:2017dsy, Borsanyi:2013bia, Borsanyi:2010cj}. The properties of hot QCD matter can be currently studied in ultra-relativistic heavy-ion collisions at the CERN Large Hadron Collider (LHC) and at the BNL Relativistic Heavy Ion Collider (RHIC), where such extreme temperatures are reachable.

A cornerstone in modeling QCD matter evolution is dissipative relativistic fluid dynamics, ``hydrodynamics", where the matter properties are embedded in the transport coefficients and QCD equation of state, which includes the cross-over transition from the QGP to a hadron gas. Besides the matter properties, an essential part of the hydrodynamical models are the initial conditions that can be either parametrized, as done e.g.\ in T\raisebox{-.5ex}{R}ENTo \cite{Moreland:2014oya}, or computed from QCD-based initial state models such as color glass condensate based IP-Sat (Impact parameter dependent saturation) \cite{Schenke:2012wb,Gale:2012rq,Schenke:2010nt}, the EKRT (Eskola-Kajantie-Ruuskanen-Tuominen) model \cite{Eskola:1999fc,Eskola:2000xq,Paatelainen:2012at,Paatelainen:2013eea,Niemi:2015qia,Kuha:2024kmq,Hirvonen:2024zne}, and EPOS (Energy conservation + Parallel scattering + factOrization + Saturation) \cite{Pierog:2009zt,Pierog:2013ria,Werner:2013tya, Werner:2023zvo,Werner:2023jps,Werner:2023fne}. The hydrodynamical evolution can be further supplemented by additional modeling for pre-hydrodynamic evolution such as K\o{}MP\o{}ST \cite{Kurkela:2018vqr} and \textsc{Alpaca} \cite{Kurkela:2022qhn}, and for late hadronic evolution that can be modeled using hadron cascade simulations, such as UrQMD \cite{Bass:1998ca,Bleicher:1999xi} or SMASH \cite{SMASH:2016zqf}. Alternatively, one may describe the whole evolution using hydrodynamics, the approach taken in this work, but supplement it with dynamical freeze-out conditions, and chemical freeze-out~\cite{Hirvonen:2022xfv}. Over the last several years hydrodynamics-based approaches have provided a consistent description of the measured bulk (low transverse momentum) observables, and consequently a very strong support for the formation of a hot collectively behaving QGP in these collisions~\cite{Heinz:2024jwu}.

The determination of the QCD matter properties, such as the temperature dependencies of the specific shear  and bulk viscosities, $\eta/s(T)$ and $\zeta/s(T)$, from the LHC and RHIC data is a very active field of research currently. The effective model frameworks mentioned above, which are necessary for describing the full collision event, typically come altogether with a relatively large number of free parameters, ${\cal O}(10-20)$, which must be fitted from the experimental data; the matter properties, together with the unknowns in the initial state dynamics and in the decoupling dynamics, are then expressed in terms of these parameters. Furthermore, many of the fit parameters are correlated in such a way that various different parameter combinations can produce similar effects on the final state observables. For this reason -- to understand these correlations and to have controlled error estimates --  a Bayesian global analysis has become a standard tool in the determination of QCD matter properties and their uncertainties \cite{Novak:2013bqa,Bernhard:2019bmu,Auvinen:2020mpc,JETSCAPE:2020shq,JETSCAPE:2020mzn,JETSCAPE:2024cqe,Nijs:2020ors,Nijs:2020roc,Nijs:2022rme,Nijs:2023yab,Giacalone:2023cet,Mantysaari:2022ffw,Heffernan:2023utr,Heffernan:2023gye,Jahan:2024wpj,Parkkila:2021tqq,Parkkila:2021yha,Virta:2024avu,Gotz:2025wnv}. For a recent review, see Ref.~\cite{Paquet:2023rfd}.

One of the major challenges in these analyses is that typically the final experimental results are averages of millions of collision events, and a meaningful data comparison with the models can require a similar amount of simulated collision events. This is combined with the fact that the parameters for the QCD matter properties, together with those for the decoupling conditions and the initial state,
form a multidimensional parameter space. The Bayesian analysis then requires that we can compute experimentally measured observables fast and easily at an arbitrary point in the parameter space. By a direct model computation this is clearly an impossible task as it would require a new set of at least tens of thousands of model runs, or even significantly more for rare observables, for each new parameter combination. Even with tools like Gaussian process (GP) emulators \cite{RasmussenWilliams,Bernhard:2018hnz,Everett:2021ruv} the task is still computationally very demanding, as it may still require millions of model runs to produce sufficiently comprehensive training data for these emulators.

In this proof-of-principle study, we introduce a novel framework that can speed-up this process potentially by several orders of magnitude. The main new feature in our approach is that instead of computing millions of hydrodynamical simulations, we use the hydrodynamical simulations to train deep convolutional neural networks (NN) that can predict the final experimental observables event-by-event (EbyE) directly from the initial energy density profile. We can then replace the time consuming hydrodynamical simulations by NNs, and use them to generate the training data for the Gaussian process emulators. While the training of the NNs still requires a considerable amount of real hydrodynamical simulations, it is orders of magnitude less than what would be needed for generating data directly to the training of GP emulators using hydrodynamics. The more simulated events are required to obtain meaningful results for a given observable, the more significant is the difference in the computational time. Besides our previous works~\cite{Hirvonen:2023lqy,Hirvonen:2024ycx}, the use of generative neural networks as a fast way to emulate EbyE hydrodynamics simulations has recently been studied in Refs.~\cite{Sun:2024lgo, OmanaKuttan:2025qjj, Stewart:2025vua}. However, to our knowledge, this is the first time that any of these methods have been applied in the context of Bayesian global analysis.

Our main physics goal here is an improved determination of the QCD matter viscosities $\eta/s(T)$ and $\zeta/s(T)$, and an understanding how these properties are correlated with the EKRT initial state calculation and its uncertainties. Our NNs, introduced in \cite{Hirvonen:2023lqy}, are now further developed to accept also the QCD matter parameters as input \cite{Hirvonen:2024ycx}. These new NNs enable a very fast EbyE computation of flow coefficients, average transverse momenta and multiplicities. With the centrality-class averaged results, we then train the GP emulators, which are employed in the Bayesian analysis.

For the EbyE training of the NNs, we employ 2+1 D relativistic shear and bulk viscous hydrodynamics from Ref.~\cite{Hirvonen:2022xfv}, with initial conditions computed from the perturbative QCD + saturation -based EbyE-EKRT (Eskola-Kajantie-Ruuskanen-Tuominen) model \cite{Niemi:2015qia}, and with the decoupling conditions determined dynamically as introduced in Ref.~\cite{Hirvonen:2022xfv}. Instead of coupling our hydrodynamics to a hadron cascade afterburner, we run hydrodynamics in partial chemical equilibrium (PCE) until the final kinetic freeze-out. In doing so, we avoid further assumptions for the various unknown hadronic cross sections and unphysical discontinuities e.g.~in the shear viscosity. Strong and electromagnetic hadron decays, however, we still account for after the freeze-out.

Our Bayesian inference set-up has 15 input parameters in total: 10 for the $\eta/s(T)$ and the $\zeta/s(T)$, 1 for the chemical decoupling temperature, 2 for the dynamical freeze-out conditions, and 2 for the initial state calculation. Exploring such a high-dimensional parameter space with full EbyE 2+1 D hydrodynamic simulations is computationally expensive. Even when utilizing statistical model emulators such as Gaussian processes, full EbyE simulations still need to be run for $\mathcal{O}(1000)$ different parameter combinations, and for each of these the EbyE statistics required by the observables needs to be generated. With the NNs trained in these $\mathcal{O}(1000)$ parameter points, we can, however, easily generate such a large number of events that inclusion of more statistics-hungry flow observables in the analysis, such as here the Fourier coefficient $v_4$ and the normalized symmetric cumulant $NSC(4,2)$,  becomes possible.

This paper is structured as follows: We describe the EbyE-EKRT + viscous hydrodynamics model in Sec.~\ref{S:model} and the neural network framework for predicting model output in Sec.~\ref{S:neuralnetworks}. We go through the statistical analysis procedure in Sec.~\ref{S:statisticalanalysis} and present the analysis results in Sec.~\ref{S:results}. A summary of our findings can be found in Sec.~\ref{S:conclusions}.

\section{Model description}
\label{S:model}

The EbyE-EKRT + viscous hydrodynamics model with dynamical freeze-out utilized in this study has been described in detail in Ref.~\cite{Hirvonen:2022xfv}. Here we summarize the main features and parameters relevant to the current study.

\subsection{EbyE-EKRT initial state}

The initial energy density profiles of QCD matter are obtained from the EbyE-EKRT minijet saturation model \cite{Niemi:2015qia}, which has been developed from Refs.~\cite{Eskola:1999fc,Eskola:2000xq,Paatelainen:2012at,Paatelainen:2013eea}. The model is based on a collinearly factorized next-to-leading order (NLO) perturbative QCD calculation of transverse energy ($E_T$) of minijets that are produced with transverse momenta ($p_T$) above a cut-off scale $p_0$ and rapidities in a mid-rapidity interval $\Delta y=1$ \cite{Niemi:2015qia,Paatelainen:2012at,Eskola:2000ji,Eskola:2000my,Eskola:1988yh}. Saturation of the minijet processes, emerging from QCD non-linearities that become dominant at high parton densities \cite{Gribov:1983ivg,Eskola:1999fc}, is conjectured to suppress the primary parton production towards lower values of $p_T$.  For the transverse density of $E_T$ in an $AA$ collision of an impact parameter {\bf b}, the saturation limit is obtained from a local saturation condition \cite{Paatelainen:2012at},
\begin{equation}
\begin{split}
\frac{dE_T}{d^2\bf{r}}(T_A({\bf r}_1,\sigma_n),T_A({\bf r}_2,\sigma_n), p_0,&\sqrt{s_{NN}},\Delta y,\beta) \\
&=\frac{K_{\rm sat}}{\pi}p_0^3\Delta y,
\label{ETsat}
\end{split}
\end{equation}
where {\bf r} is the transverse coordinate, ${\bf r}_{1,2}={\bf r} \pm {\bf b}/2$, $T_A$'s are the nuclear thickness functions, $\sigma_n$ is the width parameter of the nucleon density distribution, $\sqrt{s_{NN}}$ is the nucleon-nucleon center-of-mass system (cms) collision energy, $\beta$ is a parameter encoding the freedom in the infrared-collinear safe NLO definition of the $E_T$ (see \cite{Paatelainen:2012at,Niemi:2015qia} for the details), and $K_{\rm sat}$ is a parameter controlling the onset of saturation.

The nuclear collision geometry enters the calculation via the nuclear overlap density,
\begin{equation}
\rho_{AA}({\bf b},{\bf r},\sigma_n) = T_A({\bf r}_1,\sigma_n)T_A({\bf r}_2,\sigma_n),
\end{equation}
and also via the $T_A$-dependent nuclear effects in the employed  EPS09s nuclear parton distribution functions \cite{Helenius:2012wd}.
The $T_A$'s here fluctuate due to the EbyE fluctuating configurations of nucleons in the colliding nuclei.  The nucleon positions ${\bf r}_{i1,2}$ around the centers of the nuclei are sampled from the Woods-Saxon densities, with thicknesses $d=0.55,\, 0.55,\, 0.54$~fm and radii $R=6.38,\, 6.7,\, 5.49$~fm for the Au, Pb, Xe nuclei, correspondingly, and for Xe with fixed deformation parameters $\beta_2=0.162$ and $\beta_4=-0.003$ \cite{Moller:2015fba}. The $T_A$'s are obtained as sums of individual nucleon thickness functions, which are assumed to be Gaussians of a width $\sigma_n$,
\begin{equation}
T_A({\bf r}_{1},\sigma_n)
= \sum_{i=1}^A \frac{1}{2\pi\sigma_n^2} \exp\left(-\frac{|{\bf r}_{1}-{\bf r}_{i1}|^2}{2\sigma_n^2}\right),
\end{equation}
and similarly for $T_A({\bf r}_{2},\sigma_n)$. We leave $\sigma_n$ here as a free fit parameter to be determined from the experimental data.

The saturation scale that regulates the parton production at low $p_T$ is obtained as a solution of Eq.~(\ref{ETsat}),
\begin{equation}
p_{\rm sat} = p_{\rm sat}(\rho_{AA}({\bf b},{\bf r}; \sigma_n), \sqrt{s_{NN}},\Delta y,\beta, K_{\rm sat}).
\label{psat}
\end{equation}
As noticed in \cite{Niemi:2015qia}, to a very good precision $p_{\rm sat}$ does not depend individually on $T_A({\bf r}_{1})$ and $T_A({\bf r}_{2})$,  but effectively only on the value of the overlap density $\rho_{AA}$ (as the above list of arguments to $p_{\rm sat}$ is to indicate). Thus, for fixed $A$ and $\sqrt{s_{NN}}$, it has been possible to parametrize the $\rho_{AA}$ dependence of $p_{\rm sat}$ with $K_{\rm sat}$ and $\beta$ dependent coefficients \cite{Niemi:2015qia,Niemi:2015voa,Eskola:2017bup}. To minimize the number of free parameters in our Bayesian inference study, we fix  $\beta=0.8$ as in our earlier works \cite{Paatelainen:2013eea,Niemi:2015qia,Niemi:2015voa,Eskola:2017bup,Hirvonen:2022xfv}, but leave $K_{\rm sat}$  as another free initial-state parameter to be constrained by the data.

Assuming the formation time of the minijet system to be $\tau_{\rm sat}=1/p_{\rm sat}$, the EbyE-EKRT model gives the local energy density at the mid-rapidity unit, for each nucleon configuration, impact parameter and cms-energy, and fixed $\beta$, as \cite{Niemi:2015qia}
\begin{equation}
e({\bf r},\tau_{\rm sat}({\bf r}))
= \frac{K_{\rm sat}}{\pi}[p_{\rm sat}(\rho_{AA}({\bf b},{\bf r}; \sigma_n),K_{\rm sat})]^4.
\end{equation}
where now the dependence on  the fit parameters $\sigma_n$ and $K_{\rm sat}$ is indicated. Concretely, for a fast computation of the EbyE fluctuating energy densities for the present Bayesian inference study, we employ the parametrizations of $p_{\rm sat}$ introduced in Refs.~\cite{Niemi:2015qia,Niemi:2015voa,Eskola:2017bup}.

Following our earlier EbyE-EKRT studies, in the transverse coordinate region where $p_{\rm sat}\ge p_{\rm sat}^{\rm min} = 1$~GeV, we evolve the energy densities from $\tau_{\rm sat}({\bf r})$ to $\tau_0=1/p_{\rm sat}^{\rm min} \approx 0.2$~fm using 0+1 D Bjorken hydrodynamics. In the region where $p_{\rm sat}({\bf r}) < p_{\rm sat}^{\rm min}$, we connect the energy density profile at the time $\tau_0$ smoothly to a binary collision profile ($\propto \rho_{AA}$), see again \cite{Niemi:2015qia} for the details. We then use the obtained full energy density profile $e({\bf r},\tau_0)$, assuming no initial transverse flow, as the initial condition for our 2+1 D viscous hydrodynamics described in the next section.

For triggering the nuclear collisions here, we require that between the colliding two randomly sampled nucleon configurations the transverse distance of at least one pair of nucleons is less than $\sqrt{\sigma_{NN}^{\rm in}/\pi}$, where $\sigma_{NN}^{\rm in}$ is the inelastic nucleon-nucleon cross section. As in \cite{Hirvonen:2022xfv}, we use here the values $\sigma_{NN}^{\rm in}=42,\, 64,\, 70,\, 72$~mb for $\sqrt{s_{NN}}=200$~GeV Au+Au, 2.76~TeV Pb+Pb, 5.02~TeV Pb+Pb, 5.44~TeV Xe+Xe collisions, correspondingly.

\subsection{2+1 D viscous hydrodynamics}

The fluid dynamical equations describe local conservation of energy and momentum, $\partial_{\mu}T^{\mu\nu}=0$. The energy-momentum tensor can be decomposed in the Landau frame as
\begin{equation}
T^{\mu\nu}=eu^{\mu}u^{\nu}-P\Delta^{\mu\nu}+\pi^{\mu\nu},
\end{equation}
where $u^{\mu}$ is the fluid 4-velocity, $e$ is the local energy density, $P$ is the isotropic pressure, and $\pi^{\mu\nu}$ is the shear-stress tensor. Here $P=-\frac{1}{3}\Delta_{\mu\nu}T^{\mu\nu}$, with
$\Delta^{\mu\nu} = g^{\mu\nu} - u^\mu u^\mu$.
The bulk viscous pressure is defined as $\Pi=P-P_0$, where $P_0$ is the equilibrium pressure provided by the equation of state (EoS), $P_0=P_0(e)$, with conserved charges neglected. For the EoS we use the {\em s95p} parametrization \cite{Huovinen:2009yb} with a variable chemical freeze-out temperature $T_{\rm chem}$ \cite{Huovinen:2007xh}.

The evolution of the dissipative quantities is given by second-order transient fluid dynamics, \cite{Israel:1979wp, Denicol:2010xn, Denicol:2012cn}
\begin{equation}
\tau_{\Pi} \frac{d}{d\tau}\Pi+\Pi=-\zeta \theta - \delta_{\Pi\Pi}\Pi\theta + \lambda_{\Pi\pi}\pi^{\mu\nu}\sigma_{\mu\nu},
\end{equation}
\begin{equation}
\begin{split}
\tau_{\pi}\frac{d}{d\tau}\pi^{\langle \mu\nu \rangle}+\pi^{\mu\nu}=
& 2\eta\sigma^{\mu\nu} + 2\tau_{\pi}\pi_{\alpha}^{\langle \mu}\omega^{\nu\rangle \alpha} \\
& -\delta_{\pi\pi}\pi^{\mu\nu}\theta - \tau_{\pi\pi}\pi_{\alpha}^{\langle \mu}\sigma^{\nu\rangle \alpha} \\
& + \varphi_7 \pi_{\alpha}^{\langle \mu}\pi^{\nu\rangle \alpha}+\lambda_{\pi\Pi}\Pi\sigma^{\mu\nu},
\end{split}
\end{equation}
where $\eta$ is the shear viscosity, $\zeta$ is the bulk viscosity, $\tau_{\Pi}=\left[15\left(\frac{1}{3}-c_s^2 \right)^2 (e+P_0) \right]^{-1}$ is the bulk relaxation time with $c_s$ denoting the speed of sound, $\tau_{\pi}=\frac{5\eta}{e+P_0}$ is the shear relaxation time, $\theta=\nabla_{\mu}u^{\mu}$ is the expansion rate, $\sigma^{\mu\nu}=\nabla^{\langle \mu}u^{\nu \rangle}$ is the strain-rate tensor and $\omega^{\mu\nu}=\frac{1}{2}(\nabla^{\mu}u^{\nu}-\nabla^{\nu}u^{\mu})$ is the vorticity tensor. The angular brackets denote the symmetric and traceless projection that is orthogonal to the 4-velocity, and the spatial gradient is defined as $\nabla^\mu = \Delta^{\mu\nu}\partial_\nu$. The second order transport coefficients are obtained from the 14-moment approximation for massless gas \cite{Denicol:2010xn, Denicol:2012cn, Molnar:2013lta} and bulk related coefficients from Ref.~\cite{Denicol:2014vaa}, $\delta_{\Pi\Pi}=\frac{2}{3}\tau_{\Pi}$, $\lambda_{\Pi\pi}=\frac{8}{5}\left(\frac{1}{3}-c_s^2 \right)\tau_{\Pi}$, $\delta_{\pi\pi}=\frac{4}{3}\tau_{\pi}$, $\tau_{\pi\pi}=\frac{10}{7}\tau_{\pi}$, $\varphi_7=\frac{9}{70P_0}$, $\lambda_{\pi\Pi}=\frac{6}{5}\tau_{\pi}$.

The temperature dependence of the specific shear viscosity is parametrized as~\cite{Hirvonen:2022xfv}
\begin{equation}
 (\eta/s)(T)=
 \begin{cases}
  (\eta/s)_{\mathrm{min}} + S_HT\left((\frac{T}{T_H})^{-P_H}-1 \right), \hfill T < T_H\\
  (\eta/s)_{\mathrm{min}}, \hfill T_H \leq  T \leq T_Q \\
  (\eta/s)_{\mathrm{min}} + S_Q(T-T_Q), \hfill T > T_Q
 \end{cases}
\end{equation}
where $s$ is the equilibrium entropy density, and $T_Q = T_H + W_{\rm min}$. The parametrization allows both wide and narrow constant-$\eta/s$ region near the QCD transition temperature, and also a rapid increase in $\eta/s$ towards the lower temperatures in the hadronic matter. The latter is particularly important in connection with the dynamical freeze-out, discussed below, that relates decoupling and the temperature dependence of the hadronic shear viscosity.

The temperature dependence of the specific bulk viscosity is parametrized as~\cite{Hirvonen:2022xfv}
\begin{equation}
 (\zeta/s)(T) =\frac{(\zeta/s)_{\mathrm{max}}}{1+\left(\frac{T-T^{\zeta/s}_{\mathrm{max}}}{w(T)}\right)^2},
 \end{equation}
where
\begin{equation}
 w(T)=\frac{2(\zeta/s)_{\mathrm{width}}}{1+\exp \left( \frac{a_{\zeta/s} (T-T^{\zeta/s}_{\mathrm{max}}) }{(\zeta/s)_{\mathrm{width}}} \right)}.
\end{equation}
The parametrization gives a maximum of $\zeta/s$ at temperature $T^{\zeta/s}_{\mathrm{max}}$, and the width of the peak is controlled by $w(T)$. The peak is allowed to be asymmetric (regulated by $a_{\zeta/s}$) so that the most important part of the bulk viscous effects at temperatures below the QCD transition is given by the chemical freeze-out \cite{Paech:2006st,Dusling:2011fd,Rose:2020lfc}.

Thus, the $\eta/s(T)$ and $\zeta/s(T)$ -related 10 free parameters to be determined from the experimental data are $(\eta/s)_{\rm min}$, $T_H$, $W_{\rm min}$, $P_H$, $S_H$, $S_Q$, and $(\zeta/s)_{\rm max}$, $T^{\zeta/s}_{\mathrm{max}}$, $(\zeta/s)_{\mathrm{width}}$, and $a_{\zeta/s}$.

\subsection{Dynamical freeze-out conditions}

In order to model the system size dependence of freeze-out, while still treating hadronic QCD-matter evolution using fluid dynamics, we employ a dynamical decoupling condition, where the decoupling surface is determined by imposing simultaneously two different conditions\footnote{As an additional constraint, the minimum and maximum temperatures for the decoupling are $T_{\rm min} = 75$ MeV and $T_{\rm max} = 150$ MeV.}. 
We take the microscopic scale to be given by the shear relaxation time
$\tau_{\pi}$ and assume that the fluid decouples when the local Knudsen number exceeds a limit $C_{\rm Kn}$. Thus, at decoupling
\begin{equation}
 {\rm Kn}= \tau_{\pi}\max(\sqrt{|\nabla_\mu e \nabla^\mu e|}/e, \theta)=C_{\rm Kn}.
\end{equation}
In addition, we require that the system decouples when the mean free path, assumed to be proportional to $\tau_{\pi}$, becomes larger than the system size. This gives a non-local system size decoupling condition
\begin{equation}
\frac{\gamma \tau_{\pi}}{R}=C_R,
\end{equation}
where the Lorentz-$\gamma$ factor accounts for $\tau_{\pi}$ being in the fluid rest frame while the system size is determined in the center-of-momentum frame of the nuclear collision, and $R=\sqrt{\frac{A_{\rm Kn}}{\pi}}$, with $A_{\rm Kn}$ denoting the area in
the transverse-coordinate plane where ${\rm Kn} < C_{\rm Kn}$. In the cases where there are multiple separate areas of the fluid, the system size is calculated separately for each region. Both $C_{\rm Kn}$ and $C_R$ are free parameters to be determined from the experimental data.

A clear advantage of treating also the hadronic QCD-matter evolution with fluid dynamics is that it is then possible to avoid artificial discontinuities in the temperature dependence of transport coefficients, and at the same time obtain constraints from the experimental data also for the transport coefficients below the QCD transition temperature.

\section{Predicting hydrodynamic simulation output with neural networks}
\label{S:neuralnetworks}

\begin{table*}[]
    \begin{tikzpicture}[]
    \node [align=center](table1) at (0.0,0.0) {
        \begin{tabular}{||p{2.2cm} | p{2cm} | p{4.1cm}||}         
        \multicolumn{3}{c}{Initial energy density (256x256)}\\
        \hline
        Block & Output size & Layers\\
        \hline
        Convolution & 128x128x64 & 7x7 conv, stride 2  \\ 
        \hline
        Pooling & 64x64x64 & 3x3 max pool, stride 2\\
        \hline
        Dense Block & 64x64x256 & 
        $\begin{bmatrix} 
             \text{1x1 conv}\\  
             \text{3x3 conv}\\  
        \end{bmatrix}$  x 6\\
        \hline
        \multirow{2}{2.2cm}{Transition Layer} & 67x67x128 &  1x1 conv \\\cline{2-3}
        & 32x32x128 & 2x2 average pooling, stride 2\\
        \hline
        Dense Block & 32x32x512 & 
        $\begin{bmatrix} 
             \text{1x1 conv}\\  
             \text{3x3 conv}\\  
        \end{bmatrix}$  x 12 \\
        \hline
        \multirow{2}{2.2cm}{Transition Layer} & 33x33x256 &  1x1 conv \\\cline{2-3}
        & 16x16x256 & 2x2 average pooling, stride 2\\
        \hline
        Dense Block & 16x16x896 & 
        $\begin{bmatrix} 
             \text{1x1 conv}\\  
             \text{3x3 conv}\\  
        \end{bmatrix}$  x 20 \\
        \hline
        \multirow{2}{2.2cm}{Transition Layer} & 16x16x448 &  1x1 conv \\\cline{2-3}
        & 8x8x448 & 2x2 average pooling, stride 2\\
        \hline
        Dense Block & 8x8x1216 & 
        $\begin{bmatrix} 
             \text{1x1 conv}\\  
             \text{3x3 conv}\\  
        \end{bmatrix}$  x 24 \\
        \hline
        Pooling & 1216 &  8x8 global average pooling  \\
        \hline
        \end{tabular}
        };
    \node [align=center](table2) at (10.0,3.1) {
                \begin{tabular}{||p{2.7cm} | p{1.4cm} | p{1.4cm}| p{2.2cm} ||} 
                \multicolumn{4}{c}{Additional inputs}\\
                \hline
                Layer & \multicolumn{3}{c||}{Outputs size}\\
                \hline
                 Input parameters&$\eta/s$~(6)  & $\zeta/s$~(4)  & Decoupling~(3)  \\
                 \hline
                Fully Connected & 128 & 128 & 128  \\
                \hline
                Concatenation & \multicolumn{3}{c||}{384}\\
                \hline
                Fully Connected & \multicolumn{3}{c||}{1024}\\
                \hline       
                \end{tabular}
        };
         \node [align=center](table3) at (10.0,-3.0) {
            \begin{tabular}{||p{3cm} | p{3cm}||} 
            \hline
            Layer & Outputs size\\
            \hline
            Fully Connected & 512\\
            \hline
            Fully Connected & 256\\
            \hline
            Fully Connected & $N_{\rm out}$\\
            \hline
            \end{tabular}
        };

        \node (concat) at (10.0, 0) {\text{Concatenation, output size: 2240}};
    \draw [very thick, -Stealth] (table2) -- (concat);
    \draw [very thick, -Stealth] (table1) -- (concat);
    \draw [very thick, -Stealth] (concat) -- (table3);
\end{tikzpicture}
    \caption{The structure of a neural network with initial energy density and model parameter inputs. The numbers in parentheses next to the input variable describe the dimensions of the input. The output of the initial energy density part of the network is connected to the model parameter input branch via concatenation.}
    \label{tab:nn_structure}
\end{table*}

\begin{figure*}
    \includegraphics[width=0.3\linewidth]{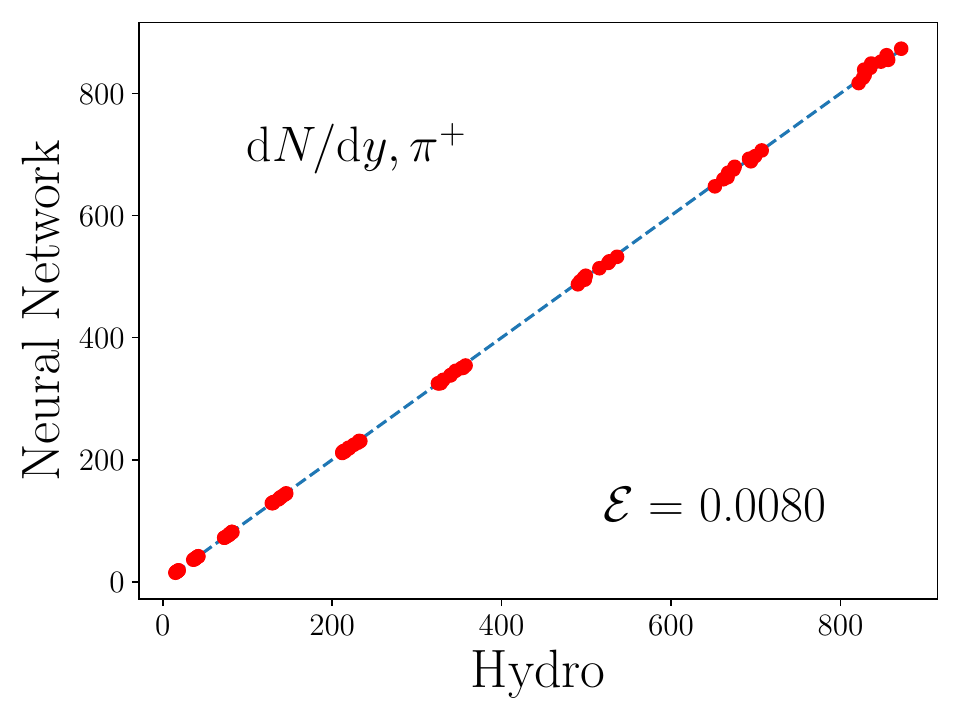}
    \includegraphics[width=0.3\linewidth]{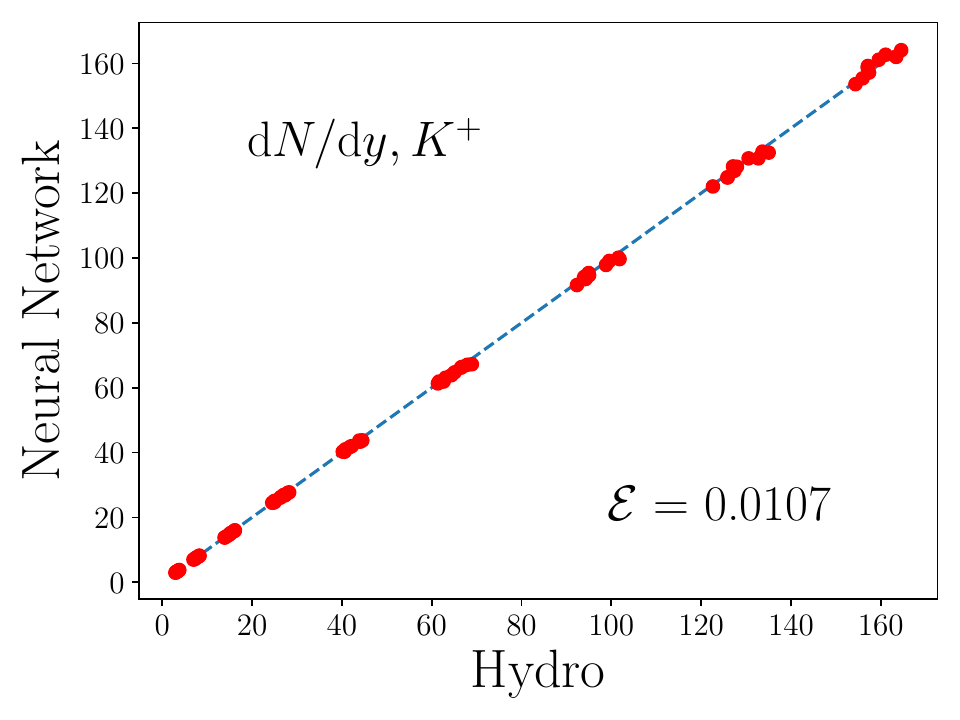}
    \includegraphics[width=0.3\linewidth]{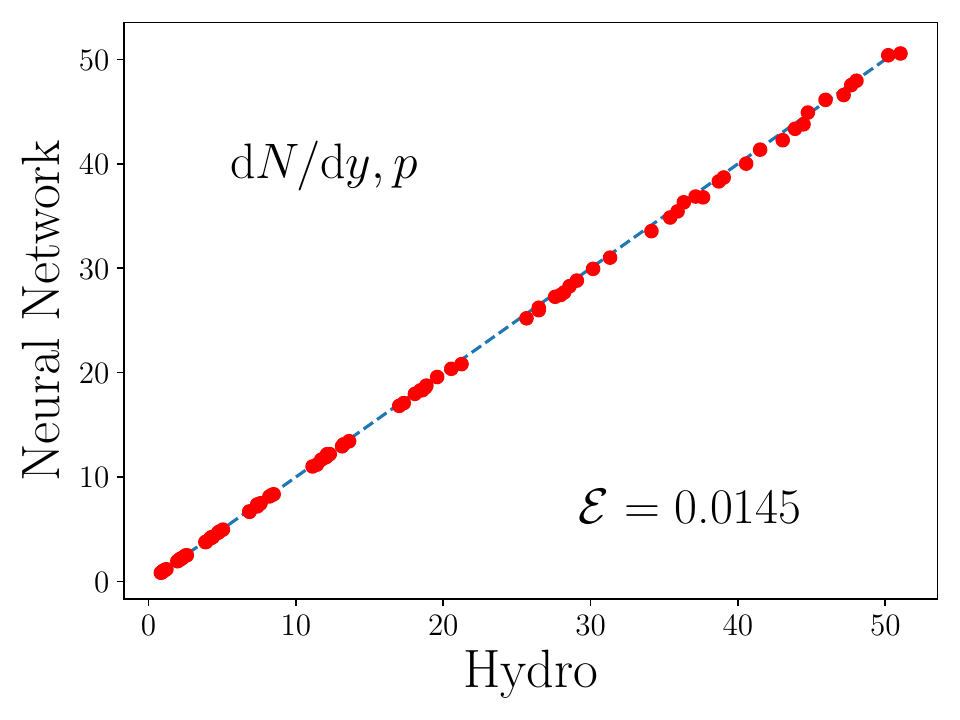} 
    \includegraphics[width=0.3\linewidth]{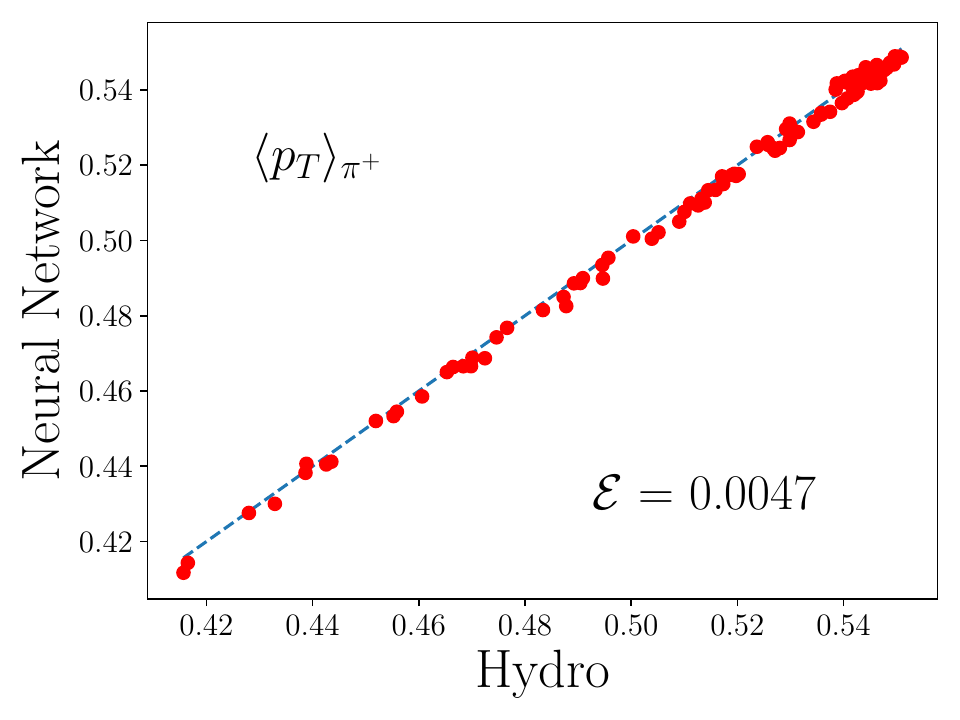}    
    \includegraphics[width=0.3\linewidth]{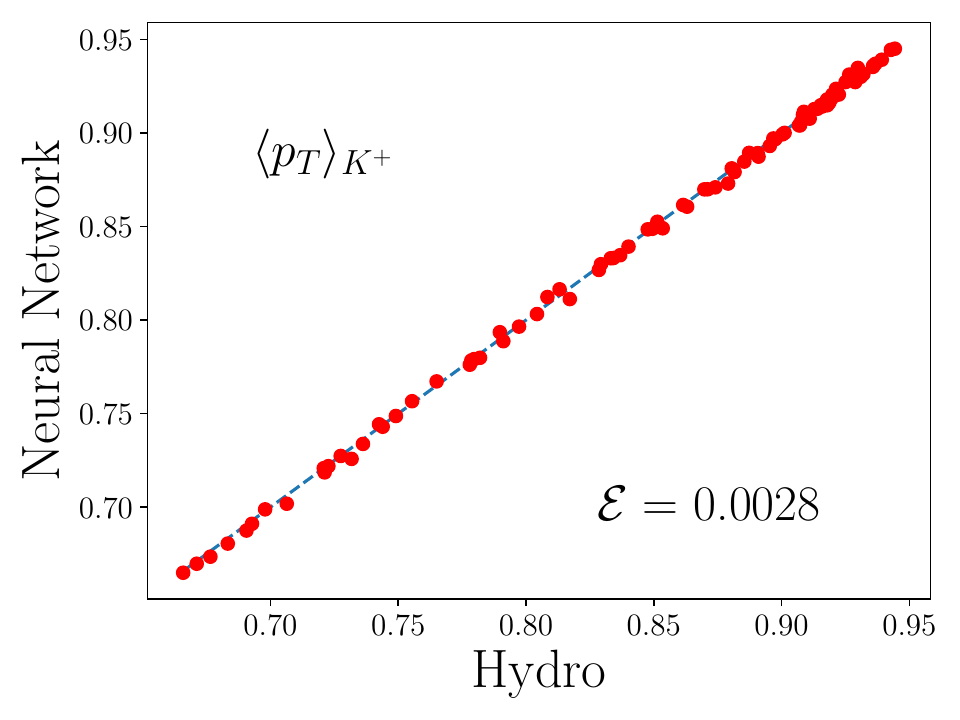}    
    \includegraphics[width=0.3\linewidth]{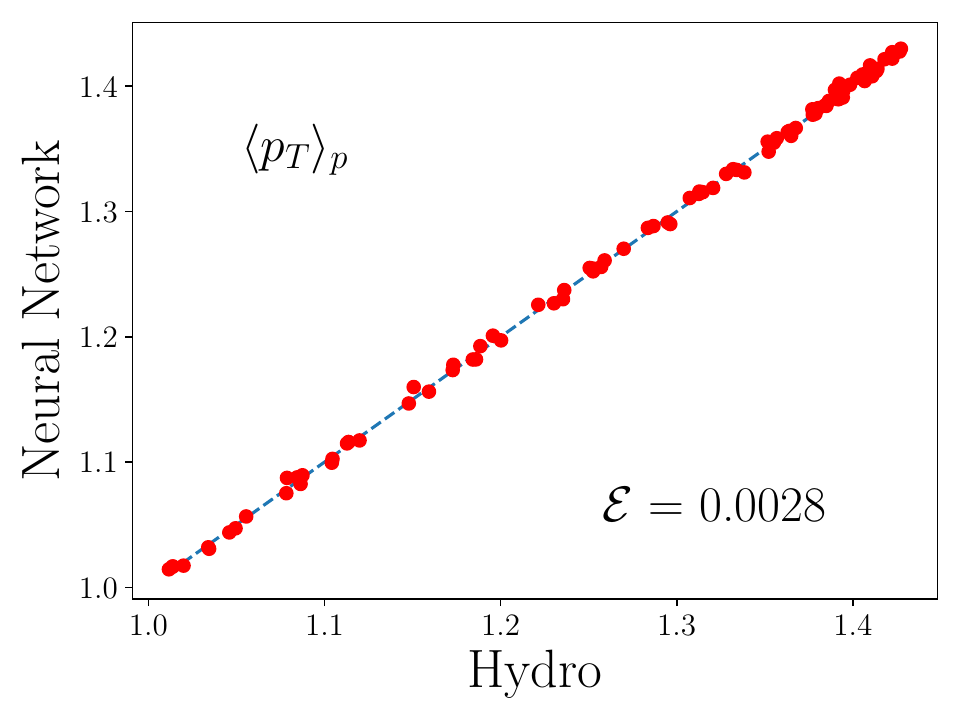}    
    \includegraphics[width=0.3\linewidth]{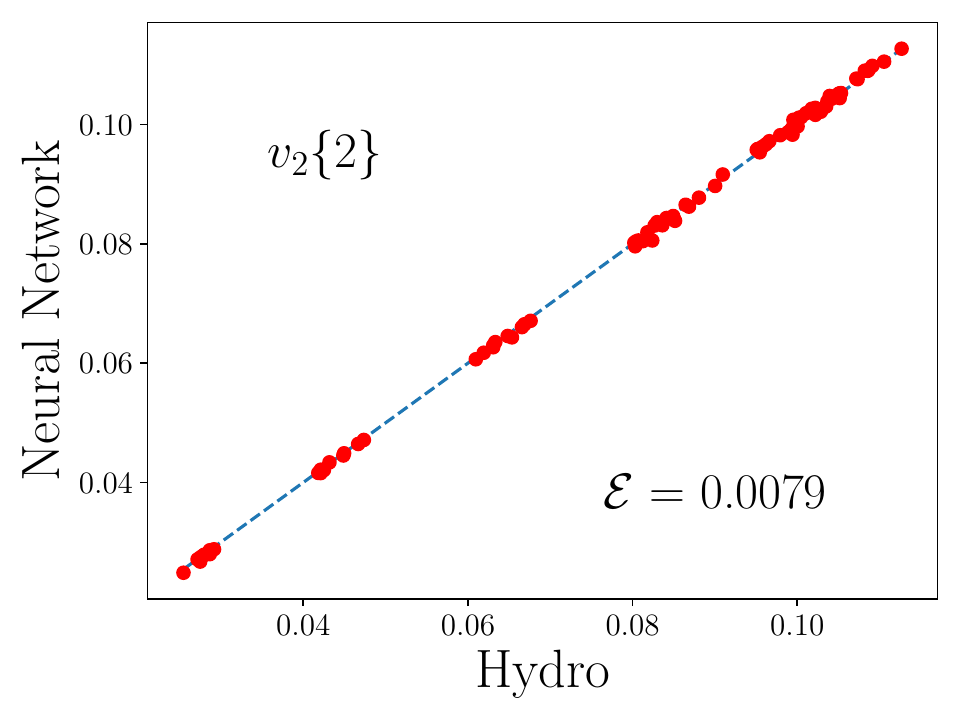}   
    \includegraphics[width=0.3\linewidth]{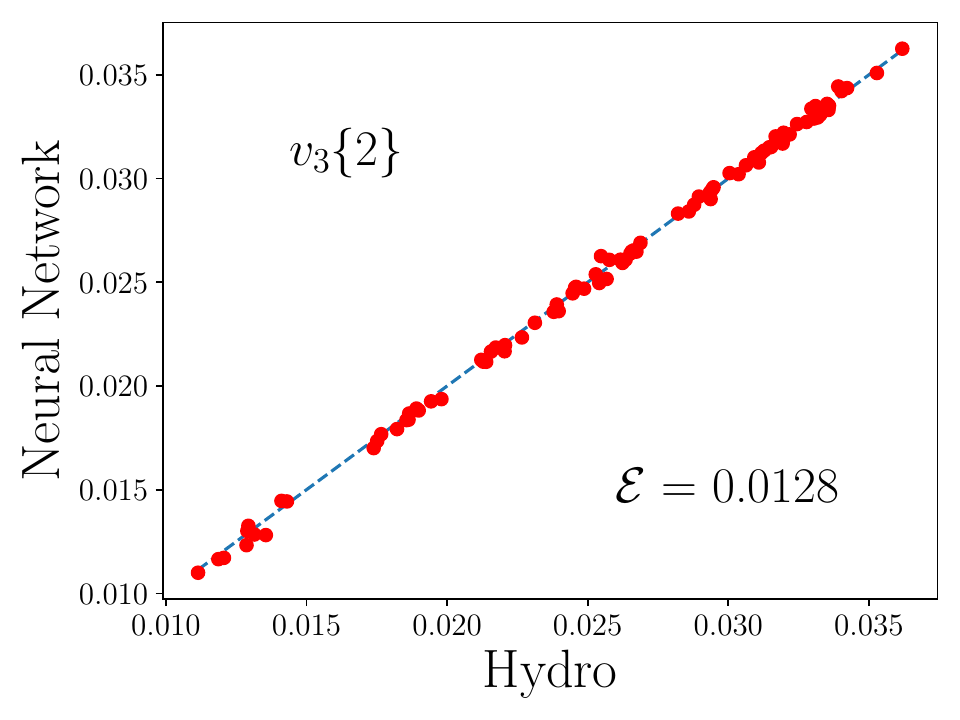}    
    \includegraphics[width=0.3\linewidth]{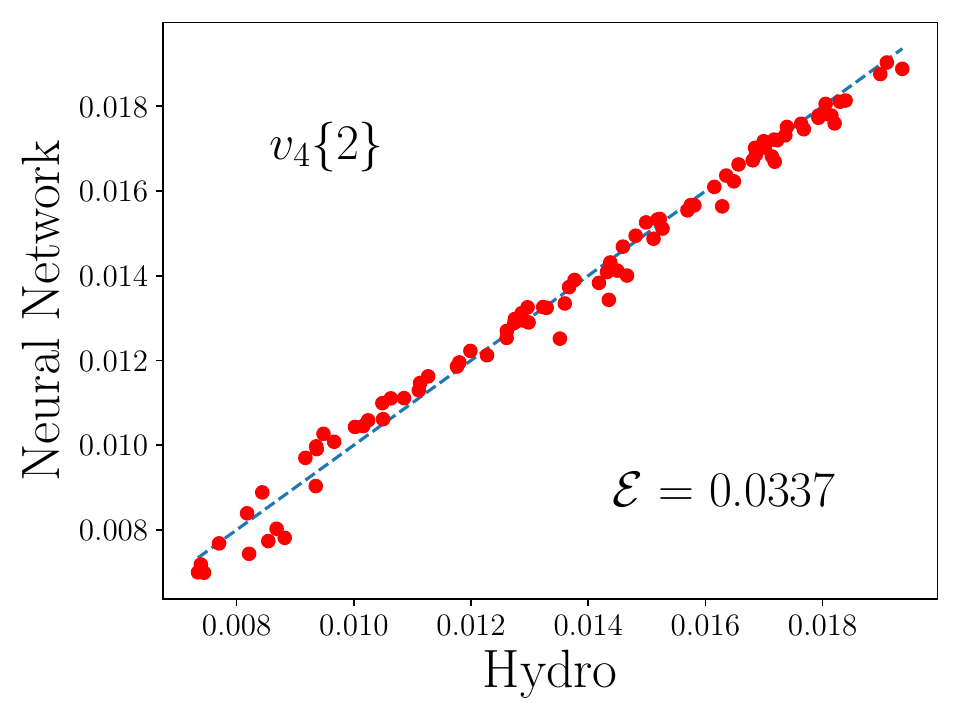}

    \caption{Validation of neural network predictions against full-simulation results for various observables in $\sqrt{s_{NN}}=5.02~\mathrm{TeV}$ Pb+Pb collisions. The predictions are compared with hydrodynamical simulation results for 10 parameter points across the centrality classes listed in Table~III. The straight line is added as a reference to indicate where the two results match exactly. $\mathcal{E}$ is the root-mean-square (RMS) error of the observable, defined in Eq.~\ref{eq:rms_error}.}
    \label{fig:nn_validation}    
\end{figure*}

Some observables, such as higher flow harmonics $v_3,v_4,...$ and their correlations,
require a lot of events to constrain statistical errors, which makes them computationally expensive.
We use neural networks to obtain quick estimates for single-event model output for any given energy density profile and parameter combination. The performance gain is significant; we can run approximately two full simulations in one CPU-hour, while with neural networks we can produce $\sim 100$ events {\em per second} using a GPU (NVIDIA V100).

Following Refs.~\cite{Hirvonen:2024ycx, Hirvonen:2024hkz}, the neural network architecture consists of two main parts. The first part extracts the most relevant information from the initial energy density profile at midrapidity, which is discretized into a \(256 \times 256\) matrix in the transverse \(xy\)-plane with a resolution of 0.07 fm. This segment of the network is built from a series of convolutional layers connected via dense connections. This design, known as the Densely Connected Convolutional Network (DenseNet), was originally developed for computer vision tasks~\cite{DenseNet}. Dense connections preserve information from all preceding convolution operations within a block by concatenating the output channels of each layer with those of all previous layers in that block. Formally, for \(n\) successive convolution layers, the output of one densely connected block is:
\begin{equation}
\begin{split}
    \mathbf{y} = [\mathbf{x}, C_1(\mathbf{x}), &C_2([\mathbf{x}, C_1(\mathbf{x})]), \dots, \\
    &C_n([\mathbf{x}, C_1(\mathbf{x}), C_2([\mathbf{x}, C_1(\mathbf{x})]), \dots ]],
\end{split}
\end{equation}
where \(C_i\) denotes a convolution operation and \(\mathbf{x}\) is the input to the block. Using a convolutional design allows the network to learn local features hierarchically, with smaller-scale patterns building up to recognize larger structures. Furthermore, dense connectivity enhances training stability and mitigates the vanishing gradient problem \cite{DenseNet}.

As in Refs.~\cite{Hirvonen:2023lqy,Hirvonen:2024ycx}, we use a DenseNet-BC variant with a growth rate \(k=32\) and compression parameter $\theta = 0.5$, which reduces the number of feature maps by a factor of 2 after each densely connected block. The DenseNet part contains a total of four densely connected blocks. Before each block, the spatial size is reduced by a factor of two by performing average pooling, except for the first block, where $3\times3$ average pooling is used. Each convolution is followed by batch normalization and a Rectified Linear Unit (ReLU) activation function, \(F(x)=\max(0,x)\). The DenseNet section concludes with a global average pooling layer, after which the condensed information is passed forward.

The second part of the network processes the model parameters (excluding those related to the initial state, as their information is already encoded in the energy density profile). This segment consists of two fully connected layers. The output of the second segment is then concatenated with the output from the DenseNet, followed by two additional dense layers and a final output layer. A ReLU activation is applied after each fully connected layer. The detailed structure of the full neural network is shown in Table~\ref{tab:nn_structure}. 

Separate neural networks are trained to produce different types of $p_T$-integrated event-by-event observables. For this purpose, the observables are divided into the following classes for which the networks are trained:
\begin{itemize}
    \item Particle multiplicities: protons, kaons, pions, charged particles
    \item Mean transverse momenta \(\langle p_T \rangle\): protons, kaons, pions, charged particles
    \item Charged particle flow harmonics \(v_2\), \(v_3\), and \(v_4\), each in a separate class.
\end{itemize}
Each class contains observables with multiple $p_T$ integration ranges matched to the kinematic cuts of various experiments. Furthermore, each observable is normalized such that the typical values are $\mathcal{O}(1)$.

Network inputs are normalized based on the training data. The initial energy density profiles are scaled to have zero mean and unit standard deviation. Other inputs are scaled to the range \([0,1]\).

For training, we run full simulations for 40 events per training point for each investigated collision system. 36 of these events are used for the training, and the remaining 4 are used in validation.

The training points have been sampled using the following iterative procedure:
First, we sample 950 points uniformly from the full input parameter space using Latin hypercube sampling and perform initial training of the neural networks. The training was done using the Adam optimizer \cite{Kingma14} for a period of 120 epochs. After each epoch, the network performance was tested against the validation set, and the network was saved if it outperformed the previously best-performing network. A logarithmic root-mean-squared (RMS) error loss is used for multiplicities, while the standard RMS error is used for other observables.
During the training, the training data is augmented by applying random rotations, flips, and shifts.

Next, we perform the global analysis as described in the following section.
After this initial analysis, we sample an additional 100 training points from the obtained posterior distribution and add these to the training set.
We then repeat the procedure by retraining the neural network with this enhanced training set by an additional 20 epochs to obtain more accurate estimates in the more relevant region of the parameter space.

The neural networks are used solely to increase event statistics for individual parameter points, while interpolation across the parameter space is handled by a Gaussian process emulator, which will be described in the next section.
For each of the (950 + 100) parameter sets, we generate $10^5$ events per collision system utilizing the trained NNs.

To validate the neural network's performance for the event-averaged observables, we selected 10 parameter points from the second training set and performed 5,000 hydrodynamic simulations for 5.02 TeV Pb+Pb collisions at each point. A comparison with neural network predictions is shown in Fig.~\ref{fig:nn_validation}. The accuracy is very good, with an average deviation of \(\sim 1\%\) of the observable value. This error is subleading relative to the error from parameter-space interpolation when the training data for the Gaussian process are sparsely distributed in the parameter space (see appendix~\ref{A:GPtests}).

\begin{figure*}
    \centering
    \includegraphics[width=\linewidth]{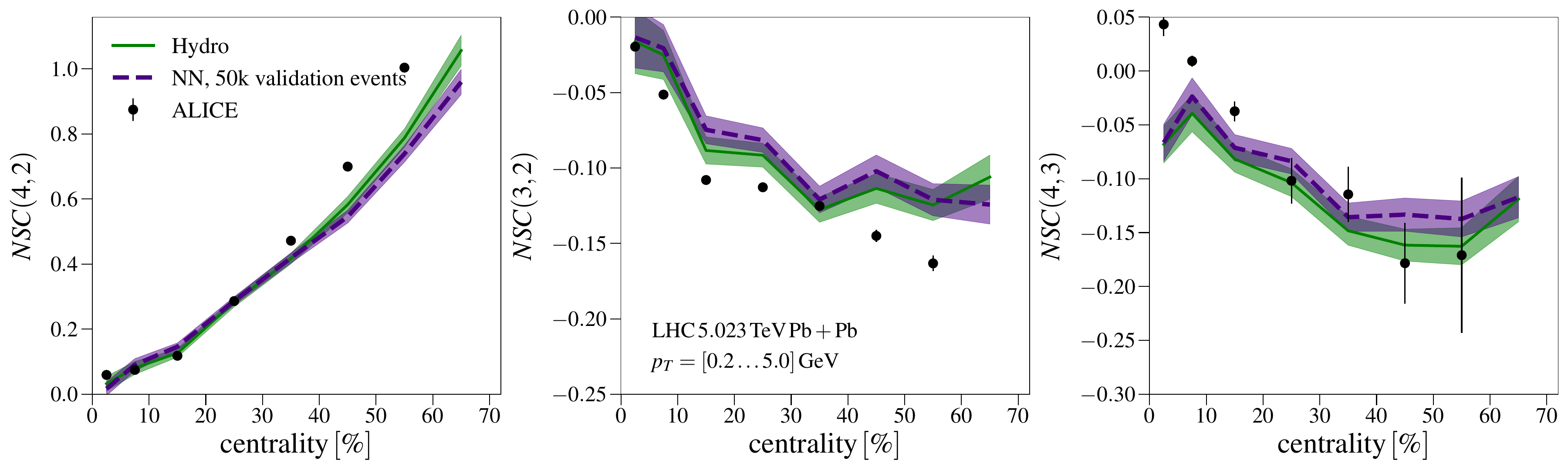}
    \caption{Validation of the neural network prediction for normalized symmetric cumulants against hydrodynamic simulations with 50,000 simulated events. Note that only $NSC(4,2)$ is included in our Bayesian analysis.}
    \label{fig:nsc_validation}
\end{figure*}

Unfortunately, 5,000 collision events are not enough to do the same test for the normalized symmetric cumulants. Therefore, to validate the neural network performance in this case, we chose the training parameter point with the highest value of the likelihood function and ran 50,000 hydrodynamic simulations. The results for this test for $NSC(4,2), NSC(3,2),$ and $NSC(4,3)$ are shown in Fig.~\ref{fig:nsc_validation}. The accuracy of the neural network is of the order 1-10\% for the $NSC(4,2)$ in the 10-70\% centrality range. In the most central collision systems and for other normalized symmetric cumulants, the relative errors can be slightly larger, but these data were not included in the analysis.

The code package used for the neural network training and the event generation is available at \cite{HirvonenNNpackage}.

\section{Statistical analysis}
\label{S:statisticalanalysis}

The goal of a Bayesian global analysis is to obtain the posterior probability distribution of the model parameters. According to Bayes' theorem, the posterior probability $P$ for $x_0$ to be the ``true'' value of parameter $x$, on the condition of the observed data $y_{\text{exp}}$, is proportional to the prior probability $p$ of $x_0$ and the likelihood of observation $y_{\text{exp}}$ if $x=x_0$:
\begin{equation}
P(x=x_0|y_{\text{exp}}) \propto p(x=x_0) \cdot \mathcal{L}(y_{\text{exp}}|x_0).
\label{E:bayes}
\end{equation}
In the case of a uniform prior probability distribution, determining the posterior amounts to determining the likelihood function.

\subsection{Input parameters}

Our EbyE-EKRT + (2+1) D viscous hydrodynamics with dynamical freeze-out model has altogether 15 tunable parameters, which were introduced in Sec.~\ref{S:model}. To focus the sampling in the physically most meaningful areas of the input parameter space, we do the following replacements when constructing our uniform prior:
\begin{itemize}
 \item $S_H \rightarrow \Delta (\eta/s)_{H} \equiv (\eta/s)(T=80\ \mathrm{MeV})-(\eta/s)_{\mathrm{min}}$,
 \item $C_{\mathrm{Kn}} \rightarrow C_{\mathrm{Kn}}/(\eta/s)_{\mathrm{min}}$, 
 \item $C_R \rightarrow C_R/(\eta/s)_{\mathrm{min}}$.
\end{itemize}
The final 15 fit parameters in our Bayesian inference study are summarized in Table~\ref{T:params}, together with their assumed prior ranges. Furthermore, for the model emulation and Markov chain Monte Carlo, the prior ranges have been scaled into a unit hypercube.

\begin{table}[h]
\centering
\begin{tabular}{|c|c|c|}
 \hline
 Parameter & Range\\
 \hline
 \hline
 \multicolumn{2}{|c|}{Initial state}  \\
 \hline
 $K_{\mathrm{sat}}$ & $0.2-1.0$\\
 \hline
 $\sigma_{n}$ & $0.35-0.60$ fm \\
 \hline
 \hline
 \multicolumn{2}{|c|}{QCD matter}  \\
 \hline
 $T_H$ & $120-250$ MeV  \\
 \hline
 $(\eta/s)_{\mathrm{min}}$ & $0.08-0.22$ \\ 
 \hline
 $P_H$ & $2.0-10.0$ \\ 
 \hline
 $\Delta (\eta/s)_{H}$ & $0.1-1.5$ \\ 
 \hline
 $S_Q$ & $0.0-3.0$ GeV$^{-1}$ \\
 \hline
 $W_{\mathrm{min}}$ & $0-300$ MeV \\
 \hline
 $(\zeta/s)_{\mathrm{max}}$ & $0.0-0.15$ \\ 
 \hline
 $T^{\zeta/s}_{\mathrm{max}}$ & $180-350$ MeV \\
 \hline
 $a_{\zeta/s}$ & $-2.0-0.0$ \\ 
 \hline
 $(\zeta/s)_{\mathrm{width}}$ & $10-100$ MeV \\
 \hline
 \hline
 \multicolumn{2}{|c|}{Decoupling}  \\
\hline
 $T_{\mathrm{chem}}$ & $135-165$ MeV \\
 \hline
 $C_{\mathrm{Kn}}/(\eta/s)_{\mathrm{min}}$ & $2.0-15.0$ \\ 
 \hline
 $C_R/(\eta/s)_{\mathrm{min}}$ & $0.5-7.5$ \\ 
 \hline
\end{tabular}
\caption{Free parameters of the model and their prior ranges. }
\label{T:params}
\end{table}

\subsection{Observables}

Simulations have been performed for four collision systems: Au+Au at $\sqrt{s_{NN}}=200$ GeV, Pb+Pb at $\sqrt{s_{NN}}=2.76$ TeV and $5.02$ TeV, and Xe+Xe at  $\sqrt{s_{NN}}=5.44$ TeV. For each collision system, the following centrality classes have been included: 0-5 \%, 5-10 \%, 10-20 \%, 20-30 \%, 30-40 \%, 40-50 \%, 50-60 \%, 60-70 \%, and 70-80 \%.
For each centrality class, the following 15 observables have been determined:
\begin{itemize}
 \item Charged particle multiplicity $dN_{\mathrm{ch}}/d\eta|_{|\eta|<0.5}$
 \item Pion multiplicity $dN(\pi^+)/dy|_{|y|<0.5}$
 \item Kaon multiplicity $dN(K^+)/dy|_{|y|<0.5}$
 \item Proton multiplicity $dN(p)/dy|_{|y|<0.5}$
 \item Charged particle mean transverse momentum $\langle p_T \rangle (\mathrm{ch})$
 \item Pion mean transverse momentum $\langle p_T \rangle (\pi^+)$
 \item Kaon mean transverse momentum $\langle p_T \rangle (K^+)$
 \item Proton mean transverse momentum $\langle p_T \rangle (p)$
 \item Charged particle $v_2\{2\}$ ... $v_4\{2\}$, $p_T$ integrated from 0.2 GeV to 3.0 GeV
 \item Charged particle $v_2\{2\}$ ... $v_4\{2\}$, $p_T$ integrated from 0.2 GeV to 5.0 GeV
 \item Normalized symmetric cumulant $NSC(4,2)=\frac{\langle v_4^2v_2^2 \rangle}{\langle v_4^2 \rangle \langle v_2^2 \rangle}-1$
\end{itemize}
We provide a detailed list of the included experimental data in Table~\ref{T:expdata_summary}.

\begin{table*}[htb]
\centering
\begin{tabular}{|c||c|c||c|c||c|c||c|c||}
\hline
 & \multicolumn{2}{c||}{Au+Au $200$ GeV}
 & \multicolumn{2}{c||}{Pb+Pb $2760$ GeV}
 & \multicolumn{2}{c||}{Pb+Pb $5020$ GeV}
 & \multicolumn{2}{c||}{Xe+Xe $5440$ GeV} \\
\hline
Observable
 & Centrality (\%) & Ref.
 & Centrality (\%) & Ref.
 & Centrality (\%) & Ref.
 & Centrality (\%) & Ref. \\
\hline
$dN_{\mathrm{ch}}/d\eta|_{|\eta|<0.5}$ & $0-5, ..., 50-60$ & \cite{Adler:2004zn} & $0-5, ...,  60-70$ & \cite{Aamodt:2010cz} & $0-5, ..., 60-70$ & \cite{Adam:2015ptt} & $0-5, ..., 70-80$ & \cite{ALICE:2018hza} \\
\hline
$dN(\pi^+)/dy|_{|y|<0.5}$ & $0-5, ..., 70-80$ & \cite{Adler:2003cb} & $0-5, ..., 70-80$ & \cite{Abelev:2013vea} & $0-5, ..., 70-80$  & \cite{ALICE:2019hno} & - & - \\
\hline
$dN(K^+)/dy|_{|y|<0.5}$ & $0-5, ..., 70-80$ & \cite{Adler:2003cb} & $0-5, ..., 70-80$ & \cite{Abelev:2013vea} & $0-5, ..., 70-80$ & \cite{ALICE:2019hno} & - & - \\
\hline
$dN(p)/dy|_{|y|<0.5}$ & $0-5, ..., 70-80$ & \cite{Adler:2003cb} & $0-5, ..., 70-80$ & \cite{Abelev:2013vea} & $0-5, ..., 70-80$ & \cite{ALICE:2019hno} & - & - \\
\hline
$\langle p_T \rangle (\mathrm{ch})$ & - & - & - & - & - & - & $0-5, ..., 70-80$ & \cite{ALICE:2018hza} \\
\hline
$\langle p_T \rangle (\pi^+)$ & $0-5, ..., 70-80$ & \cite{Adler:2003cb} & $0-5, ..., 70-80$ & \cite{Abelev:2013vea} & $0-5, ..., 70-80$ & \cite{ALICE:2019hno} & - & - \\
\hline
$\langle p_T \rangle (K^+)$ & $0-5, ..., 70-80$ & \cite{Adler:2003cb} & $0-5, ..., 70-80$ & \cite{Abelev:2013vea} & $0-5, ..., 70-80$ & \cite{ALICE:2019hno} & - & - \\
\hline
$\langle p_T \rangle (p)$ & $0-5, ..., 70-80$ & \cite{Adler:2003cb} & $0-5, ..., 70-80$ & \cite{Abelev:2013vea} & $0-5, ..., 70-80$ & \cite{ALICE:2019hno} & - & - \\
\hline
$v_2\{2\}$ & $0-5, ..., 70-80$ & \cite{STAR:2017idk} & $0-5, ..., 70-80$ & \cite{ALICE:2018rtz} & $0-5, ..., 70-80$ & \cite{ALICE:2018rtz} & $0-5, ..., 60-70$ & \cite{ALICE:2018lao} \\
\hline
$v_3\{2\}$ & $0-5, ..., 70-80$ & \cite{STAR:2017idk} & $0-5, ..., 60-70$ & \cite{ALICE:2018rtz} & $0-5, ..., 70-80$ & \cite{ALICE:2018rtz} & $0-5, ..., 50-60$ & \cite{ALICE:2018lao} \\
\hline
$v_4\{2\}$ & $0-5, ..., 70-80$ & \cite{STAR:2017idk} & $0-5, ..., 60-70$ & \cite{ALICE:2018rtz} & $0-5, ..., 70-80$ & \cite{ALICE:2018rtz} & $0-5, ..., 50-60$ & \cite{ALICE:2018lao} \\
\hline
$NSC(4,2)$ & $10-20, ..., 40-50$ & \cite{STAR:2018fpo} & $10-20, ..., 40-50$ & \cite{ALICE:2016kpq} & $10-20, ..., 40-50$ & \cite{ALICE:2021adw} & - & - \\
\hline
\end{tabular}
\caption{Summary of experimental data. For Au+Au, the proton target multiplicity is an average of protons and antiprotons and identified particle multiplicities at 10-20 \% are averages of 10-15 \% and 15-20 \% centrality classes; for $\langle p_T \rangle$,  10-20 \% centrality class is not included. For Pb+Pb, in cases where $v_n\{2\}$ data have narrower centrality bins than the model output, an average over bins has been taken.}
\label{T:expdata_summary}
\end{table*}

\subsection{Principal component analysis}

The dimension of the observable space can be reduced considerably via principal component analysis, where the goal is to find the linear combination of observables with the largest variance (so the largest sensitivity to the change in input). For the variance comparison to make sense, the observables are centered so that their mean is 0, and scaled with their sample standard deviation to make them dimensionless and of similar scale with regards to the variance.

Given a matrix $\hat{Y}$ of (centered and scaled) model data, where each row corresponds to one point in input parameter space, with the number of rows corresponding to the number of parameter sets that we will use for the training of the GPs (900 in the first analysis and 980 in the second, see Sec.~IV D below)
and each column corresponds to one of $m$ data points, we find an eigenvalue decomposition of the covariance matrix $\hat{Y}^T\hat{Y}$:
\begin{equation}
\hat{Y}^{T}\hat{Y}=V \Lambda V^{T},
\end{equation}
where $\Lambda$ is the diagonal matrix containing the eigenvalues $\lambda_1,...,\lambda_m$
and $V$ is an orthogonal matrix containing the eigenvectors of the covariance matrix. Since $\lambda_1 \geq \lambda_2 \geq ... \geq \lambda_m$,
the fraction of the total variance explained by the principal components above some index $k$ will be negligible.
This allows us to define a lower-rank approximation of the original transformed data matrix $Z=\hat{Y}V$ as $Z_k=\hat{Y}V_k$,
where $V_k$ contains the first $k$ columns of $V$. The transformation of a vector $\vec{y}$ from the space of observables
to a vector $\vec{z}$ in the reduced-dimension principal component space with $k$ components is then defined as
\begin{equation}
\vec{z} = \vec{y}\,V_k.
\end{equation}

For each collision system we include enough principal components to cover 99\% of the data variance:
\begin{itemize}
\item In Au+Au collisions, 89 data points are reduced to 15 principal components
\item In Pb+Pb collisions at $\sqrt{s_{NN}}=2.76$ TeV, 91 data points are reduced to 12 principal components
\item In Pb+Pb collisions at $\sqrt{s_{NN}}=5.02$ TeV, 93 data points are reduced to 12 principal components
\item In Xe+Xe collisions, 40 data points are reduced to 8 principal components
\end{itemize}

\subsection{Gaussian process emulators}
Given a list of model data $(X,Z_k)$ for a principal component $z(\vec{x}) \in Z_k$ evaluated at parameter points $\vec{x} \in X$,
we obtain the Gaussian process predictive mean and uncertainty for $z$ at a point $\vec{x}_0$ as follows:
\begin{equation}
\label{E:gp_estimate}
 z_{\text{GP}}(\vec{x}_0) = \mathcal{C}_{0,Z}\mathcal{C}_{Z,Z}^{-1}Z,
\end{equation}
\begin{equation}
\label{E:gp_error}
 \sigma_{\text{GP}}^2(\vec{x}_0) = \mathcal{C}_{0,0}-\mathcal{C}_{0,Z}\mathcal{C}_{Z,Z}^{-1}\mathcal{C}_{Z,0}
\end{equation}
where the covariance matrix has the form
\begin{equation}
\mathcal{C} =
  \begin{pmatrix}
  \mathcal{C}_{0,0} & \mathcal{C}_{0,Z}\\
  \mathcal{C}_{Z,0} & \mathcal{C}_{Z,Z}
 \end{pmatrix}
\end{equation}
further defined by the covariance function also known as GP kernel $k$: $\mathcal{C}_{0,0}=k(\vec{x}_0, \vec{x}_0)$, $\mathcal{C}_{Z,Z}=k(X, X)$.
In this work we use the radial basis function with an additional white noise term for the kernel,
\begin{equation}
k(\vec{x}_i, \vec{x}_j)=A\exp\left( -\frac{1}{2}\sum_{k=1}^{15}\frac{(x_{i,k} - x_{j,k})^2}{\ell_k^2} \right) + \delta_{\rm noise}.
\end{equation}
The white noise term $\delta_{\rm noise}$ applies only on the diagonal elements of the covariance matrix, $k(\vec{x}_i, \vec{x}_i)$, and accounts for the statistical and neural network uncertainties in the event averages.
The kernel hyperparameters $A,\vec{\ell},\delta_{\rm noise}$ are initially tuned using 900 points from the neural network training dataset, while the remaining 50 were used for validation.
On the second round, 80 of the additional 100 training points are used for hyperparameter tuning and remaining 20 are added to the validation set.

To compare the emulator predictions against physical observables, the prediction $\vec{z}_{\,\text{GP}}$ in the principal component space is transformed back to the space of observables:
\begin{equation}
\vec{y}_{\,\text{GP}} = \vec{z}_{\,\text{GP}} \, V_k^T.
\end{equation}
One must also remember to invert the scaling and centering of the model data which was performed in preparation for the principal component analysis.

We utilise the {\scshape scikit-learn} \cite{scikit-learn} Python library for both the principal component analysis and the Gaussian process regression. A summary of the emulator quality checks can be found in Appendix~\ref{A:GPtests}.

\subsection{Likelihood function}
For each collision energy, the likelihood function is defined as
\begin{equation}
\label{eq:likelihood}
\mathcal{L}(\vec{z}_{\text{\,exp}}|\vec{x}) = \frac{1}{\sqrt{|2\pi\Sigma|}} e^{-d_M^2/2},
\end{equation}
where $d_M$ is the Mahalanobis distance
\begin{equation}
\label{eq:mahalanobis}
 d_M^2 \equiv (\vec{z}_{\text{GP}}(\vec{x})-\vec{z}_{\text{\,exp}})^T\Sigma^{-1}(\vec{z}_{\text{GP}}(\vec{x})-\vec{z}_{\text{\,exp}}).
\end{equation}
The covariance matrix $\Sigma$ contains three different sources of uncertainty: the GP prediction errors \eqref{E:gp_error}, the experimental uncertainties, 
and an additional theoretical uncertainty $\sigma_{\mathrm{th}}$ reflecting the remaining modeling imperfections. Such a theoretical uncertainty was included also in Refs.~\cite{Bernhard:2019bmu,Virta:2024avu}; we, however, do not treat $\sigma_{\mathrm{th}}$ as a free parameter, but assume it to be $\sim \mathcal{O}(10\%)$ of the measured value for each observable.\footnote{Note that the $\mathcal{O}(10\%)$ theoretical errors are of the same order as the errors in the computed charged multiplicities that arise from the uncertainties of nuclear parton distributions as estimated in Ref.~\cite{Paatelainen:2012at}.}
We investigate three possible error values, $10\%$, $20\%$, and $30\%$, to get a clearer view on its effect on the posterior distribution\footnote{The initial analysis (with 950 parameter combinations sampled from the prior) is done only once, with $\sigma_{\mathrm{th}}=0$, so the GP emulators use the same training set and hyperparameters for all three scenarios.}. Such a study
is relevant since a theoretical model, like EbyE-EKRT here, can be too rigid in some of its predictions. If the model is not matching perfectly with data for all observables, the rigidity easily leads to an artificial bias in an automated global analysis. There is also some degree of uncertainty in the numerical simulations due to the discretization, which is now effectively accounted for by $\sigma_{\mathrm{th}}$.

The total likelihood at input parameter point $\vec{x}$ is then the product of the likelihoods from all collision systems:
\begin{equation}
\mathcal{L}_{\text{total}}=\mathcal{L}_{\text{200}} \times \mathcal{L}_{\text{2760}} \times \mathcal{L}_{\text{5020}} \times \mathcal{L}_{\text{5440}}.
\end{equation}

We use the Markov chain Monte Carlo code {\scshape emcee} \cite{Foreman-Mackey:2012any} to draw samples from the likelihood function. We run an ensemble of 300 walkers for 40000 steps and discard the first 30000 steps to ensure that all samples used in results are drawn after MCMC has converged to the target distribution.

\section{Results}
\label{S:results}

\subsection{Posterior distribution}

We present the summary statistics of the posterior probability distributions in Table~\ref{T:posteriorsummary}, i.e., the median values and 90\% credible intervals for the marginal distributions of each parameter.

\begin{table}[h]
\renewcommand{\arraystretch}{1.3}
\centering
\begin{tabular}{|c|c|c|c|}
\hline
 & \multicolumn{3}{|c|}{Theoretical uncertainty $\sigma_{\rm th}$} \\
\hline
Parameter & 10\% & 20\% & 30\% \\
\hline
$K_{\mathrm{sat}}$ & $0.57_{-0.09}^{+0.10}$ & $0.53_{-0.11}^{+0.12}$ & $0.51_{-0.13}^{+0.16}$ \\[3pt]
\hline
$\sigma_n$ [fm] & $0.38_{-0.02}^{+0.03}$ & $0.38_{-0.03}^{+0.05}$ & $0.40_{-0.04}^{+0.07}$ \\[3pt]
\hline
$T_H$ [MeV] & $181_{-59}^{+61}$ & $178_{-55}^{+61}$ & $178_{-54}^{+63}$ \\[3pt]
\hline
$(\eta/s)_{\mathrm{min}}$ & $0.14_{-0.02}^{+0.01}$ & $0.15_{-0.02}^{+0.02}$ & $0.15_{-0.03}^{+0.03}$ \\[3pt]
\hline
$P_H$ & $8.4_{-3.9}^{+1.5}$ & $8.4_{-3.7}^{+1.5}$ & $8.3_{-3.8}^{+1.5}$ \\[3pt]
\hline
$\Delta (\eta/s)_{H}$ & $0.3_{-0.2}^{+0.3}$ & $0.3_{-0.2}^{+0.6}$ & $0.4_{-0.3}^{+0.8}$ \\[3pt]
\hline
$S_Q$ [GeV$^{-1}$] & $1.2_{-1.1}^{+1.6}$ & $1.3_{-1.2}^{+1.5}$ & $1.4_{-1.2}^{+1.5}$ \\[3pt]
\hline
$W_{\mathrm{min}}$ [MeV] & $225_{-130}^{+68}$ & $183_{-151}^{+104}$ & $165_{-144}^{+120}$ \\[3pt]
\hline
$(\zeta/s)_{\mathrm{max}}$ & $0.09_{-0.05}^{+0.05}$ & $0.09_{-0.06}^{+0.05}$ & $0.08_{-0.06}^{+0.06}$ \\[3pt]
\hline
$T^{\zeta/s}_{\mathrm{max}}$ [MeV] & $215_{-31}^{+66}$ & $223_{-37}^{+59}$ & $230_{-43}^{+76}$ \\[3pt]
\hline
$a_{\zeta/s}$ & $-0.5_{-1.1}^{+0.4}$ & $-0.7_{-1.1}^{+0.7}$ & $-0.8_{-1.0}^{+0.7}$ \\[3pt]
\hline
$(\zeta/s)_{\mathrm{width}}$ [MeV] & $ 57_{-43}^{+38}$ & $ 60_{-43}^{+36}$ & $ 61_{-43}^{+35}$ \\[3pt]
\hline
$T_{\mathrm{chem}}$ [MeV] & $147_{ -4}^{ +3}$ & $148_{ -5}^{ +5}$ & $149_{ -6}^{ +7}$ \\[3pt]
\hline
$C_{\mathrm{Kn}}$ & $1.2_{-0.4}^{+0.6}$ & $1.4_{-0.5}^{+0.7}$ & $1.5_{-0.6}^{+0.8}$ \\[3pt]
\hline
$C_R$ & $0.7_{-0.4}^{+0.3}$ & $0.7_{-0.4}^{+0.3}$ & $0.8_{-0.4}^{+0.4}$ \\[3pt]
\hline
\end{tabular}
\caption{Summary of the median values and 90\% credible intervals from the marginal posterior distributions of the model parameters, for three different values of theoretical uncertainty.}
\label{T:posteriorsummary}
\end{table}

For the initial state parameters, we find the saturation parameter has a median value $K_{\mathrm{sat}} \approx 0.5$ and the full 90\% credible range considering all three theoretical uncertainties is $0.38 < K_{\mathrm{sat}} < 0.67$, while the nucleon width favors the lowest allowed values, $0.35$ fm $ < \sigma_n < 0.47$ fm. These values for $K_{\mathrm{sat}}$ and $\sigma_n$ are consistent with the earlier studies \cite{Niemi:2015qia,Hirvonen:2022xfv}. Interestingly, the $\sigma_n$ obtained here is quite close to the value 0.43 $\pm$ 0.01 fm, extracted from the HERA data for exclusive electroproduction of $J/\Psi$ \cite{ZEUS:2002wfj}.

The chemical freeze-out temperature $T_{\mathrm{chem}}$ is constrained within the range $143-156$ MeV, with the increase in theoretical uncertainty effecting mainly the upper bound.
The current posterior distribution is extending to a slightly lower range than the previously obtained $150-158$ MeV for {\em s95p} EoS from the global analysis presented in Ref.~\cite{Auvinen:2020mpc}.

The Knudsen number for freeze-out is $ 0.8-2.3$, while the ratio of mean free path to system size at freeze-out is in the range $0.3-1.2$ with the median around $C_R\approx 0.7-0.8$. Thus, according to the experimental data,
the freeze-out takes place, as expected, at the very limit of the applicability of hydrodynamics.

For the transport coefficients, rather than investigating each parameter individually, we illustrate their combined effect by plotting the actual temperature dependence of $\eta/s$ and $\zeta/s$ based on samples from the posterior distribution. Even if the individual parameters themselves are not very tightly constrained, they are not independent from each other, and the net effect is that in particular shear viscosity itself is much more tightly constrained than the constraints on the individual parameters would suggest. We also note that the final constraints for the transport properties of QCD matter are very robust against the variations in the assumed theoretical error, which is an encouraging result. The most significant difference is seen in bulk viscosity, with the smallest 10 \% error there is a clear lower limit, but this diminishes with larger theoretical errors. The upper limit for bulk is much less sensitive. On the other hand, the limits for the shear viscosity are practically independent of the assumed error.

\begin{figure}[h]
\includegraphics[width=8.8cm]{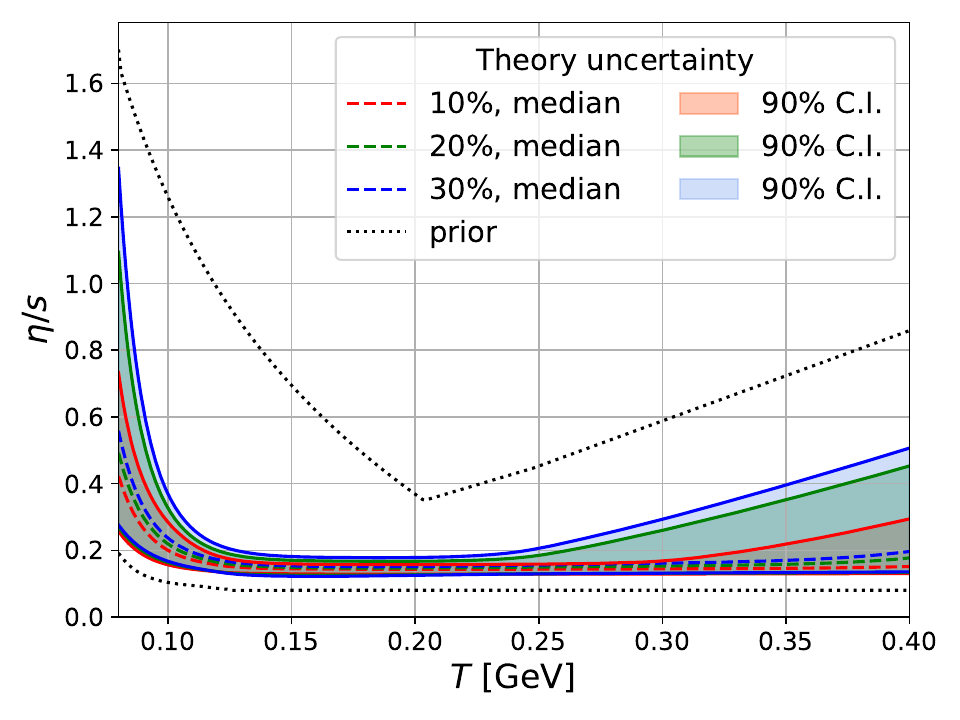}
\caption{The median and 90\% credible interval for the specific shear viscosity $\eta/s$ with respect to temperature $T$.}
\label{F:etasT}
\end{figure}

We see from Fig.~\ref{F:etasT} that $\eta/s$ has a strong temperature dependence below $T\approx 150$ MeV (parameter $P_H$ leaning strongly towards the maximum of the prior range), while the minimum-value plateau is between $150-220$ MeV with $0.12 < (\eta/s)_{\mathrm{min}} < 0.18$. For temperatures higher than 200 MeV, a rising trend is allowed but not favored; for all three cases the median remains close to constant all the way to 350 MeV. This is a clear difference from the previous global analysis with EKRT initial state \cite{Auvinen:2020mpc} where the model with {\em s95p} equation of state had $\eta/s$ increasing with temperature beyond $T>250$ MeV. However, the two analyses are not directly comparable, as the previous analysis was done using averaged initial conditions and without bulk viscosity.

Comparing our result for $\eta/s$ against other recent Bayesian analyses, we obtain quite a good agreement with the JETSCAPE collaboration (Ref.~\cite{JETSCAPE:2020mzn}, figure 6) for the temperature range investigated in that study, $T=150-350$ MeV. While the JETSCAPE parametrization for $\eta/s(T)$ did not explicitly include a constant-value plateau, the slope at high temperatures was allowed to be either positive or negative and no strong preference for the sign of the slope was found; a later Bayesian study with the same parametrisation but wider prior and IP-Glasma initial state also found temperature-independent $\eta/s$ to be consistent with data \cite{Heffernan:2023gye}. Our result also fits within the credible intervals from the various Trajectum analyses presented in Ref.~\cite{Nijs:2022rme}; indeed, our 90\% credible interval for the 10\% theoretical uncertainty is fully consistent with the Trajectum results for $\eta/s$. Meanwhile, the rising trend observed by Virta et al.~in Ref.~\cite{Virta:2024avu} cannot be fully accommodated within the uncertainties of the present analysis; this discrepancy likely stems from the more restricted parametrization of $(\eta/s)(T)$ in Ref.~\cite{Virta:2024avu} which does not allow a constant-value plateau or a decrease of $\eta/s$ with increasing temperature.

\begin{figure}[h]
\includegraphics[width=8.8cm]{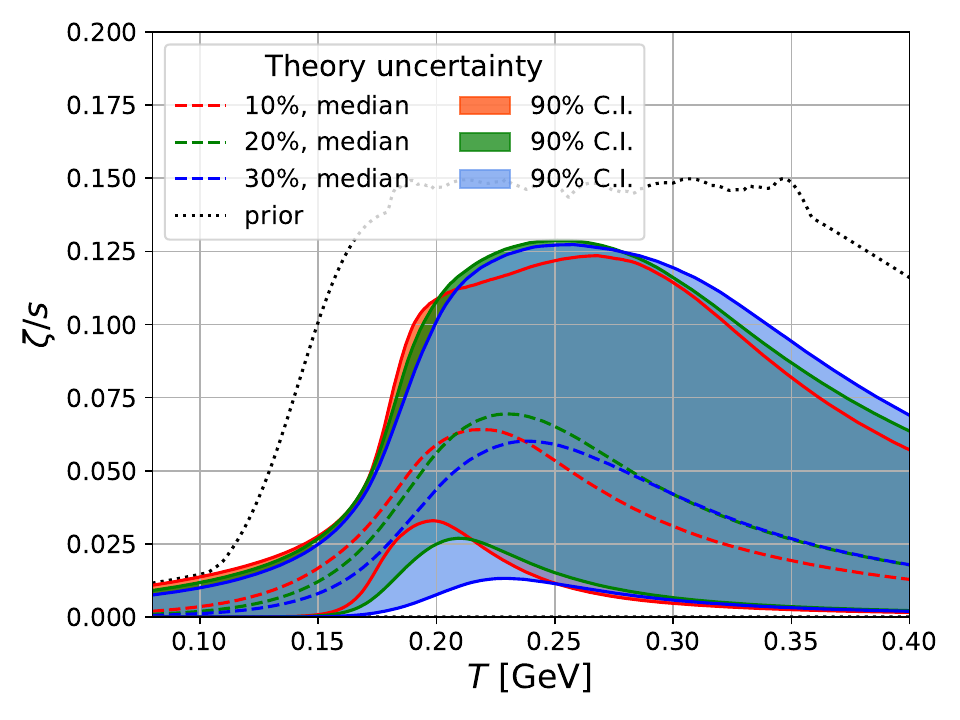}
\caption{The median and 90\% credible interval for the bulk viscous coefficient $\zeta/s$ with respect to temperature $T$. Note that the lower limit of the prior is zero.}
\label{F:zetasT}
\end{figure}

For the bulk viscosity in this study, we obtain much weaker constraints compared to $\eta/s$; see Fig.~\ref{F:zetasT}. We can say with 90\% credibility that $\zeta/s$ is non-zero in the temperature range $T = 200-300$ MeV, but the variation in the exact value remains considerable. This temperature range for nonzero $\zeta/s$ is slightly higher than the range $150-200$ MeV obtained in the analyses of JETSCAPE~\cite{JETSCAPE:2020mzn} and Virta et al.~\cite{Virta:2024avu}. On the other hand, both the IP-Glasma study \cite{Heffernan:2023gye} and the Trajectum analysis with the total hadronic cross section $\sigma_{AA}$ included in the fit data and no additional weighting of observables~\cite{Nijs:2022rme} find a nonzero minimum bound for $\zeta/s$ in a temperature range similar to our result.

We note that bulk viscosity remains small at and below the chemical freeze-out temperature even if the prior range would allow larger values. This is on par with the discussion in Ref.~\cite{Hirvonen:2022xfv}, where the parametrization was constructed so that the main part of the bulk viscous effect in hadron gas is described by chemical freeze-out.

\subsection{Comparison to data}

In the following, for each of the three investigated scenarios, we sample 5000 parameter combinations from the posterior distribution and compare the distribution of emulator predictions for these points against the data.

\begin{figure}
\includegraphics[width=8cm]{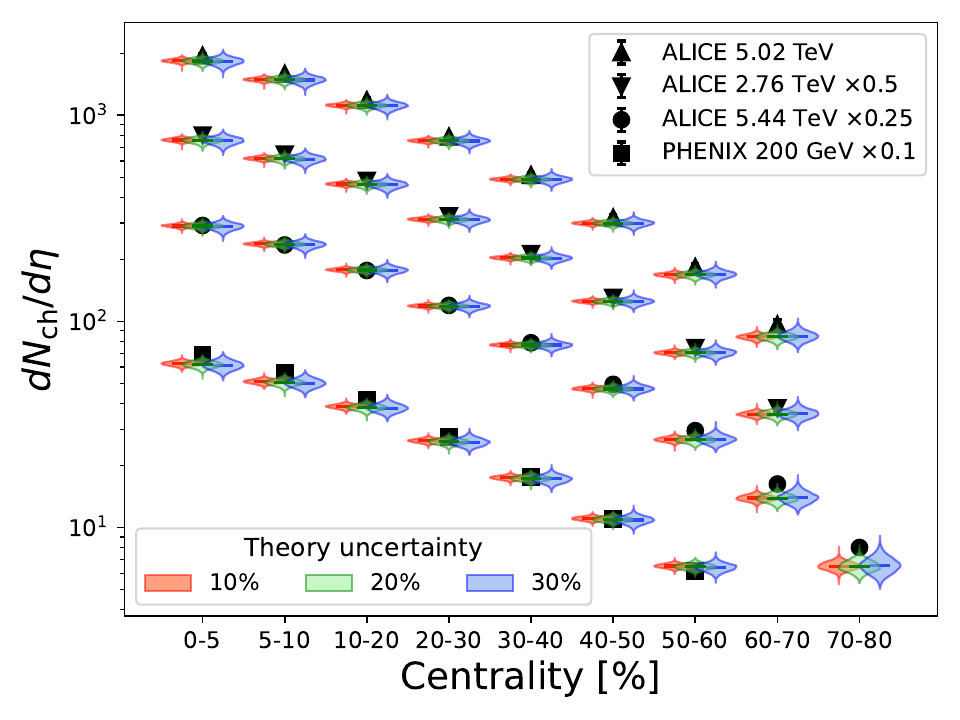}
\includegraphics[width=8cm]{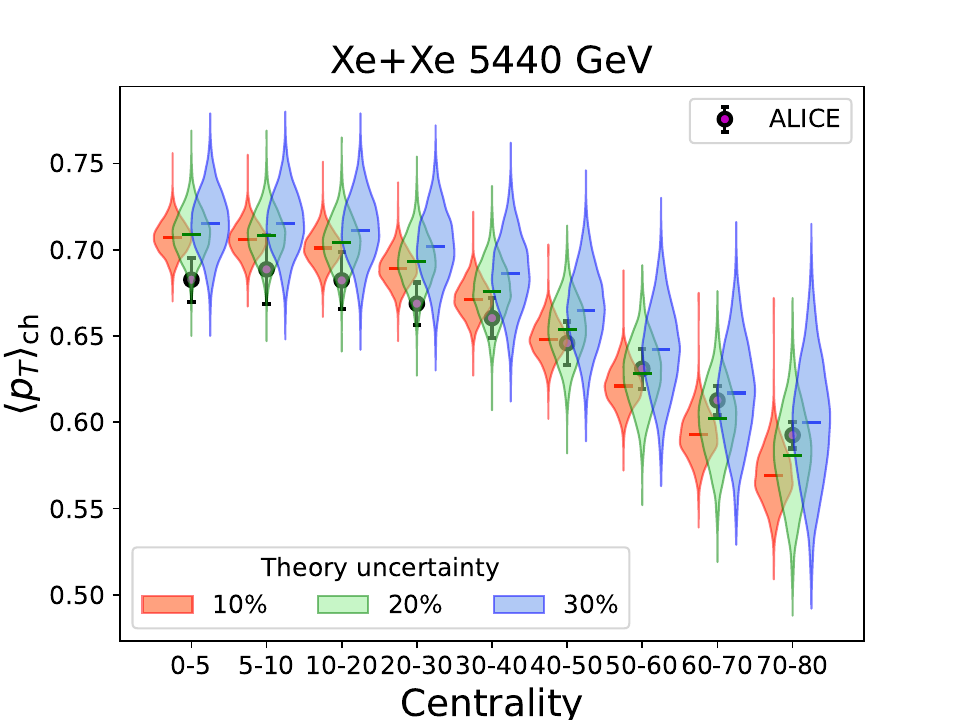}
\caption{Posterior distribution based emulator predictions for the centrality dependence of charged particle multiplicity for all four collision systems (top panel) and average transverse momentum in Xe+Xe collisions at $\sqrt{s_{NN}}=5.44$ TeV (bottom panel). The violin plots reflect the distribution of emulator predictions, with more points concentrated on the broader parts of the violin. The horizontal line within the violin indicates the mean value of predictions. PHENIX data is from Ref.~\cite{Adler:2004zn} and ALICE data from Refs. \cite{Aamodt:2010cz} (2.76 TeV), \cite{Adam:2015ptt} (5.02 TeV), and \cite{ALICE:2018hza} (5.44 TeV).}
\label{F:NchXe}
\end{figure}

The centrality dependence of charged particle multiplicities for the investigated collision systems are presented in Fig.~\ref{F:NchXe}, along with the average transverse momentum in $5.44$ TeV Xe+Xe collisions. The agreement with the multiplicity data is very satisfactory across the whole collision energy range, from $200$ GeV Au+Au up to $5.44$ TeV Xe+Xe collisions, with larger theoretical uncertainty merely increasing the variation in the predictions while the median distance from the data remains very similar. However, in the case of Xe+Xe in particular, we see that the centrality dependence comes off a bit too steep; the multiplicity falls slightly too strongly towards peripheral collisions. This overly strong centrality dependence applies also for the mean transverse momentum of charged particles; in addition, here the uncertainties tend to increase more towards higher values of $\langle p_T \rangle$, and consequently the increase in theoretical uncertainty also increases the median value of $\langle p_T \rangle_{\mathrm{ch}}$ predictions.

Here, we also want to emphasize that none of the model parameters depend on the collision energy $\sqrt{s_{NN}}$ or nuclear mass number $A$, and that the $\sqrt{s_{NN}}$, $A$, and centrality dependencies of multiplicity are mainly  predictions of the EKRT model.

\begin{figure}
\includegraphics[width=8cm]{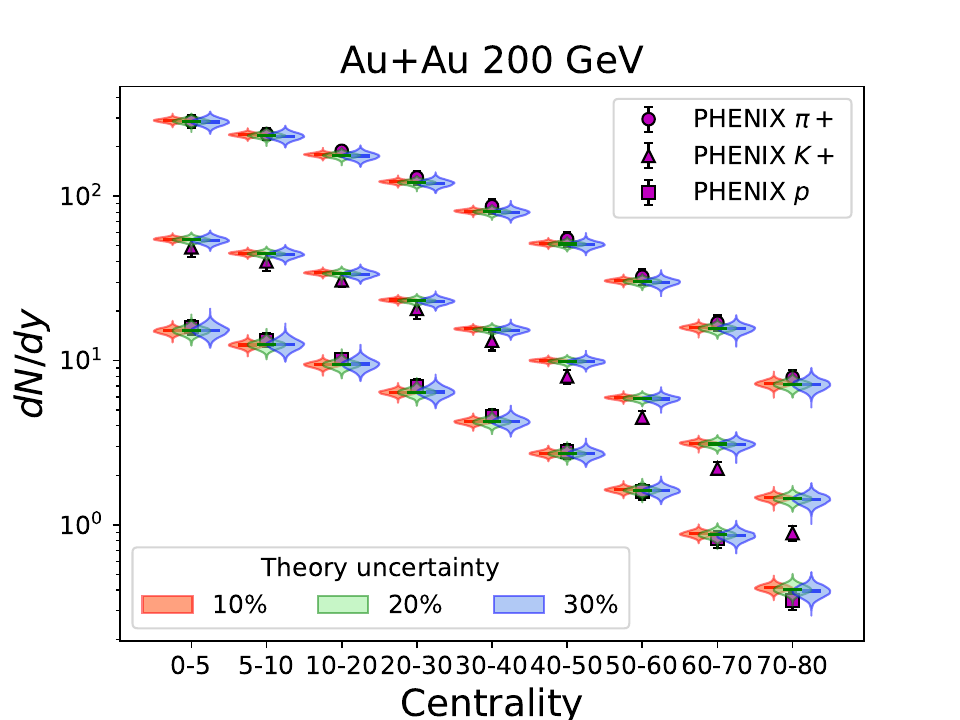}
\includegraphics[width=8cm]{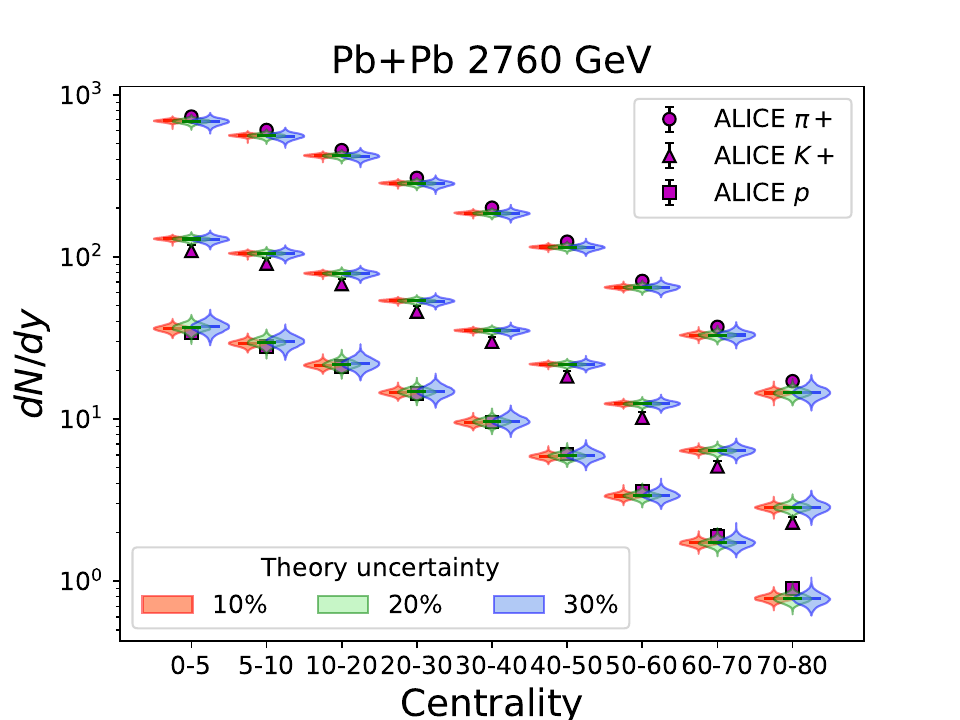}
\includegraphics[width=8cm]{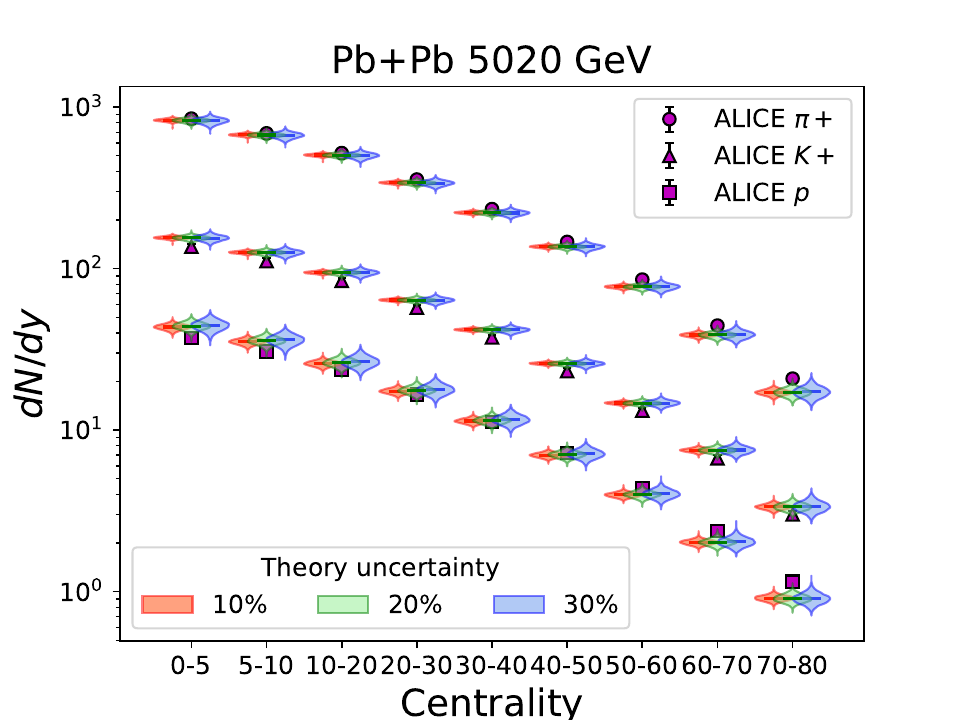}
\caption{Posterior distribution emulator predictions for identified particle multiplicities in 200 GeV Au+Au (top panel), 2.76 TeV Pb+Pb (middle panel), and 5.02 TeV Pb+Pb (bottom panel) collisions. PHENIX data from \cite{Adler:2003cb}. ALICE data from \cite{Abelev:2013vea} (2.76 TeV) and \cite{ALICE:2019hno} (5.02 TeV).}
\label{F:Nid}
\end{figure}

We show the data comparison of emulator predictions for pion, kaon and proton multiplicities in Au+Au and Pb+Pb collisions in Fig.~\ref{F:Nid}. We obtain good descriptions of pion and proton multiplicities in all cases; however, the kaon production is systematically overestimated. As was already seen in the case of charged particle multiplicity, the centrality dependence tends to be too steep at the LHC energies, suggesting perhaps a collision energy dependent nucleon width $\sigma_n$ and/or possibly also that nuclear substructure (``hotspots'') might be required for a better fit.

\begin{figure}
\includegraphics[width=8cm]{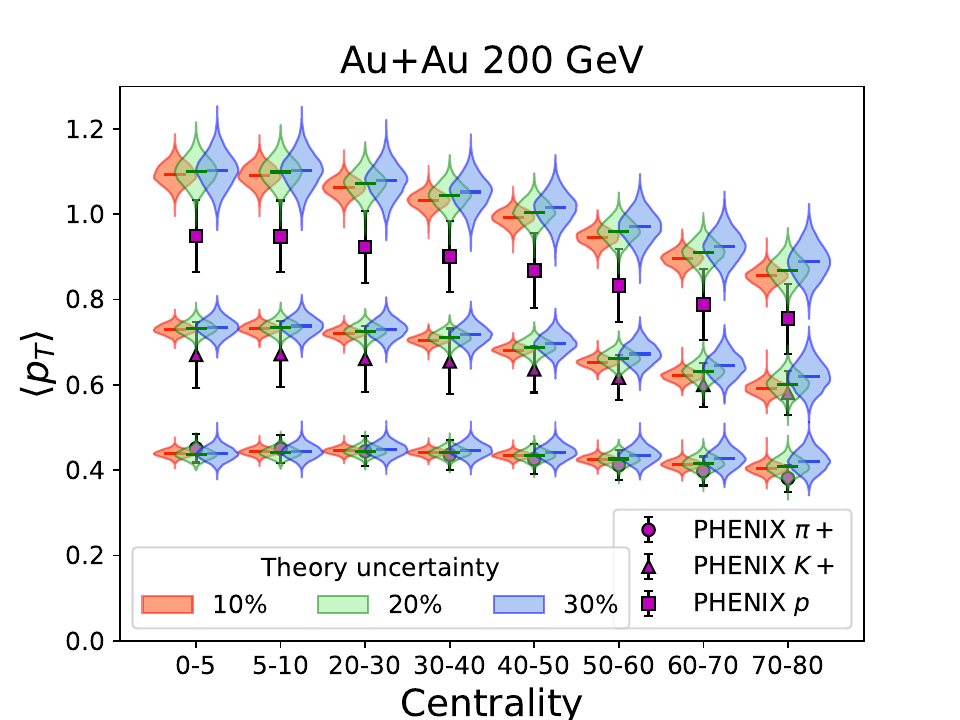}
\includegraphics[width=8cm]{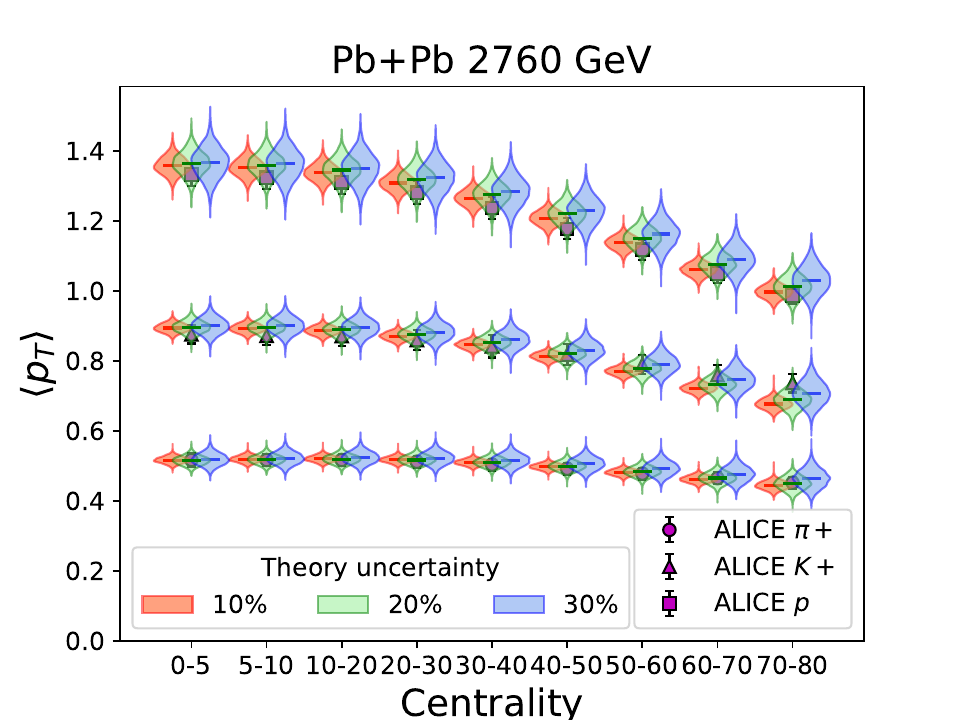}
\includegraphics[width=8cm]{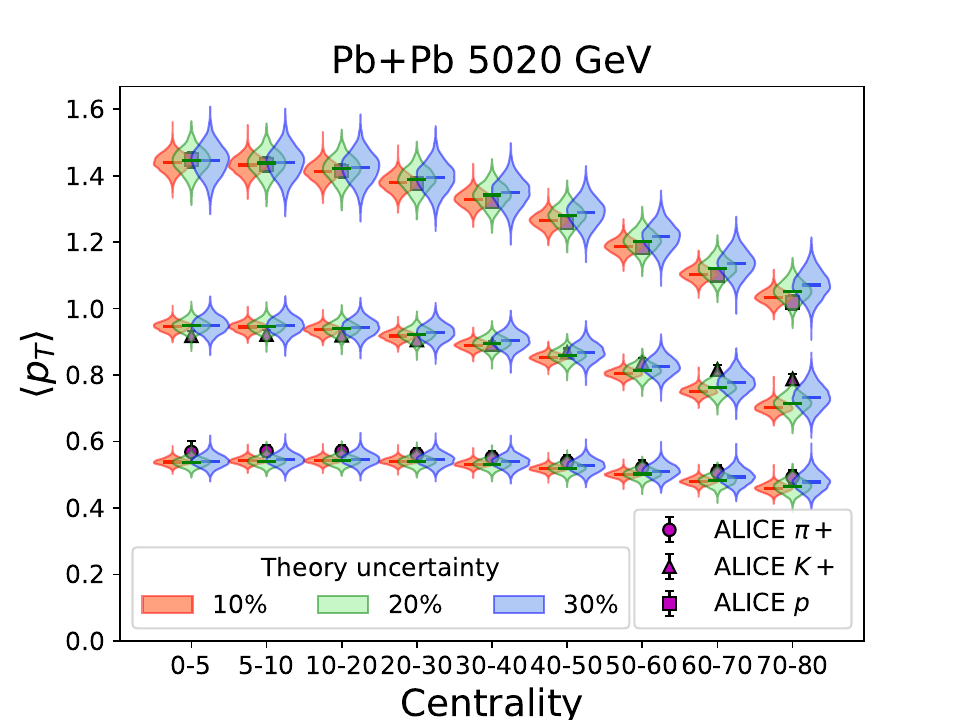}
\caption{Posterior distribution emulator predictions for identified particle transverse momenta in 200 GeV Au+Au (top panel), 2.76 TeV Pb+Pb (middle panel), and 5.02 TeV Pb+Pb (bottom panel) collisions.
PHENIX data from \cite{Adler:2003cb}. 
ALICE data from \cite{Abelev:2013vea} (2.76 TeV) and \cite{ALICE:2019hno} (5.02 TeV).}
\label{F:pTid}
\end{figure}

The data comparison for the transverse momenta of pions, kaons and protons is presented in  Fig.~\ref{F:pTid}. Here the overall agreement with the data is very good for all three particle species, the main problem being the proton $\langle p_T \rangle$ at $200$ GeV Au+Au collisions which is systematically overestimated. Similarly to the charged particle $\langle p_T \rangle$ in Xe+Xe collisions, it seems easier to increase $\langle p_T \rangle$ rather than decrease it within the posterior distribution, so the increase in theoretical uncertainty allows more variation towards higher $\langle p_T \rangle$ values.

\begin{figure}
\includegraphics[width=7.25cm]{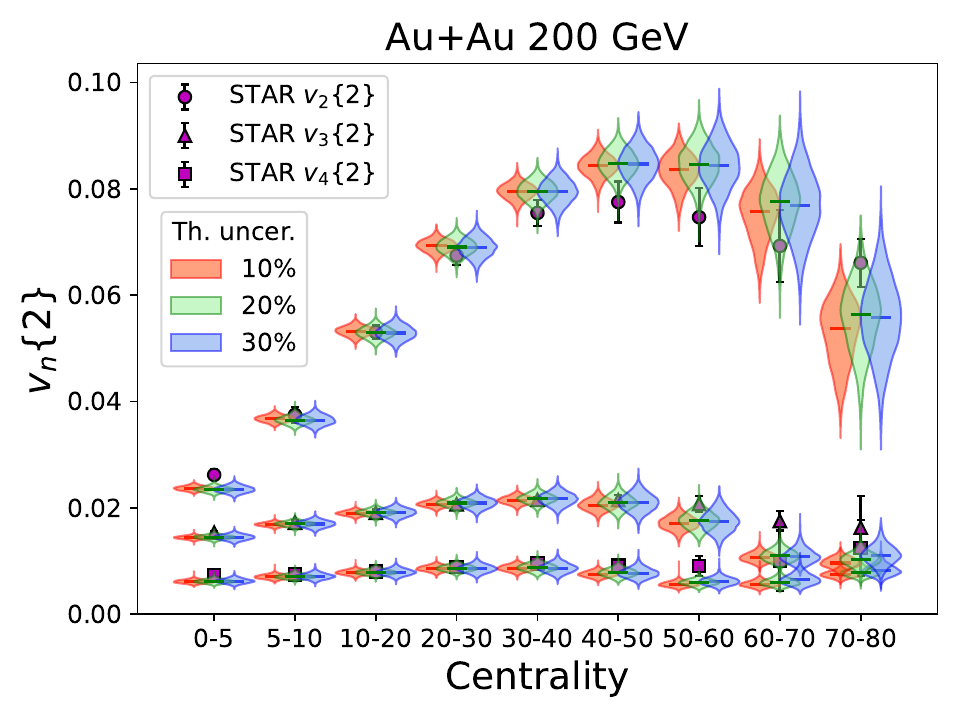}
\includegraphics[width=7.25cm]{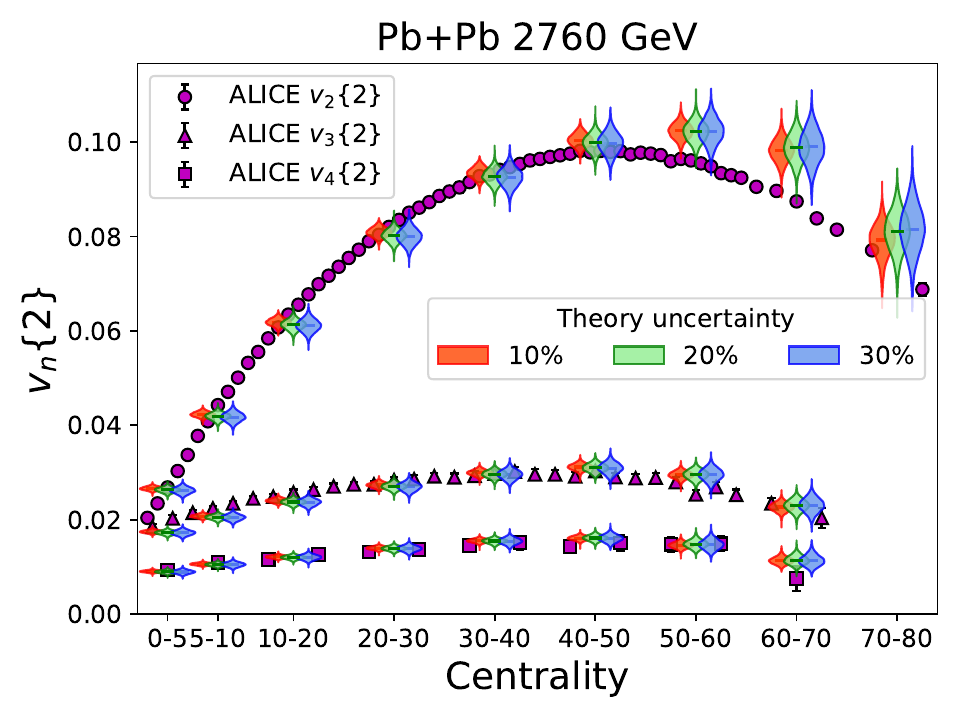}
\includegraphics[width=7.25cm]{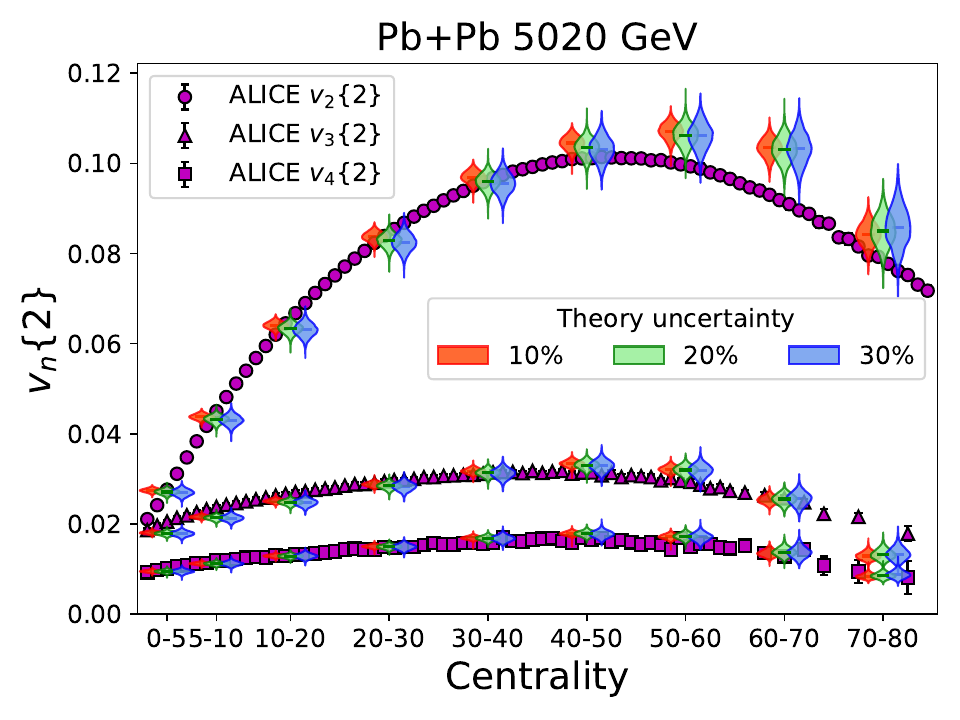}
\includegraphics[width=7.25cm]{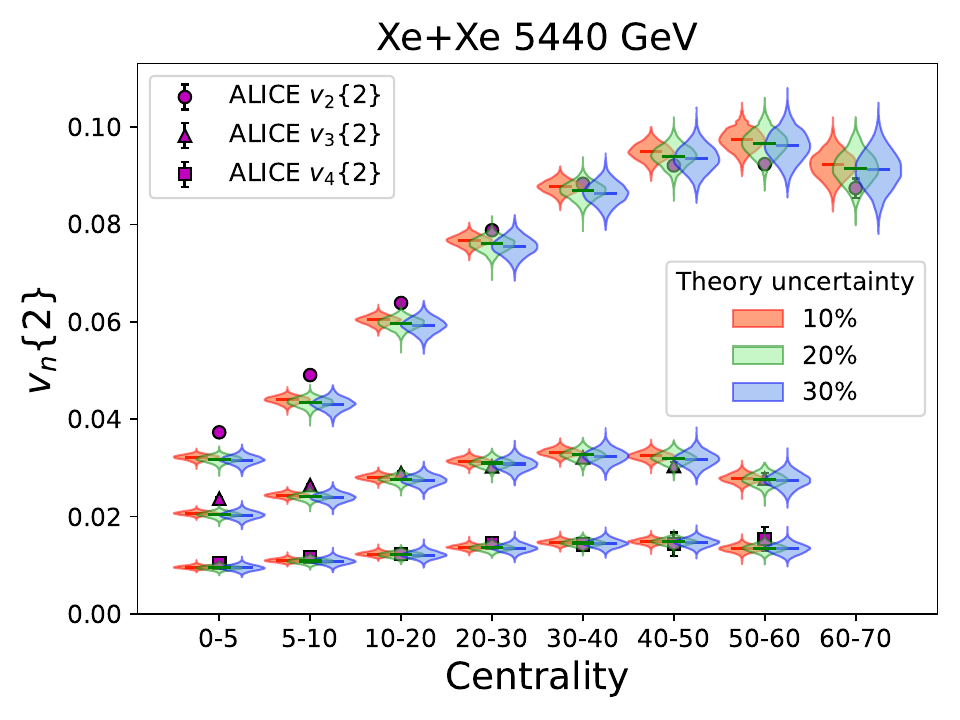}
\caption{Posterior distribution emulator predictions for charged particle flow at RHIC Au+Au (top panel), LHC Pb+Pb (middle panels) and Xe+Xe (bottom panel) collisions.
STAR data from \cite{STAR:2017idk}. ALICE data from \cite{ALICE:2018rtz} (Pb+Pb) and \cite{ALICE:2018lao} (Xe+Xe).}
\label{F:vn}
\end{figure}

The flow harmonics $v_n$ for $n=2,3,4$ are shown in Fig.~\ref{F:vn}. The overall agreement with the data over multiple collision systems and three different harmonics is quite good. However, the centrality dependence is again slightly off; there's too little flow in the most central collisions, and too much $v_2$ produced in more peripheral collisions. The underestimation of flow in the most central collisions may necessitate implementing nuclear deformation effects for Au+Au and Pb+Pb collisions, and further deformations in the case of Xe+Xe. Including nucleon substructure would likely improve the $v_3/v_2$ ratio, as we have demonstrated in a recent study \cite{Hirvonen:2024zne}. We note that in particular $v_3$ and $v_4$ in Au+Au collisions are in a significantly better agreement with the measurements than in our previous non-Bayesian analysis using the same model setup \cite{Hirvonen:2022xfv}.

\begin{figure}
\includegraphics[width=8cm]{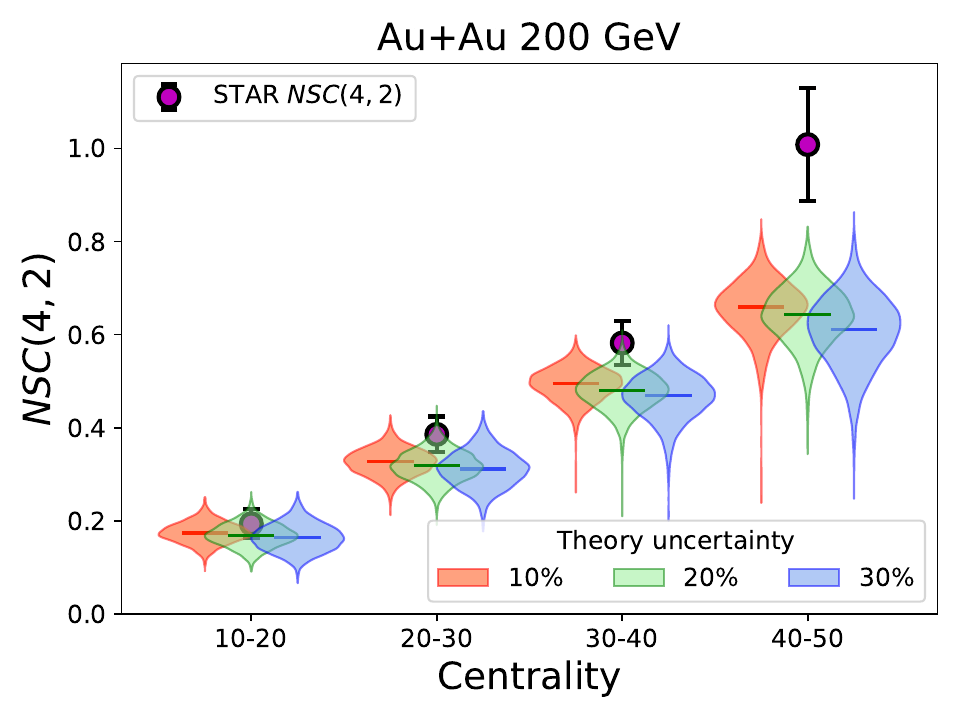}
\includegraphics[width=8cm]{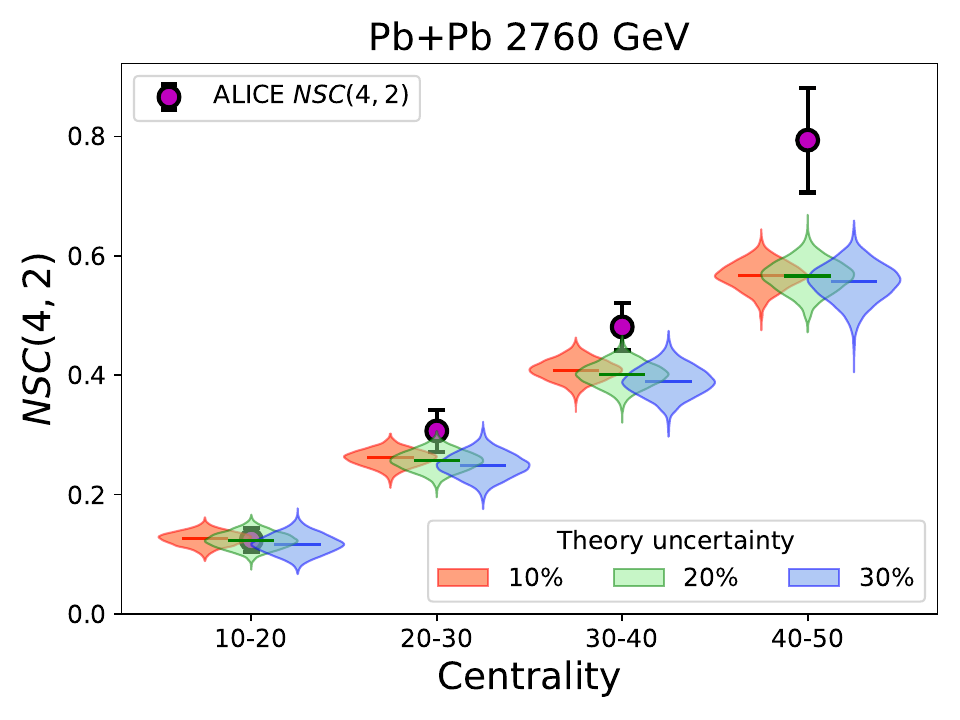}
\includegraphics[width=8cm]{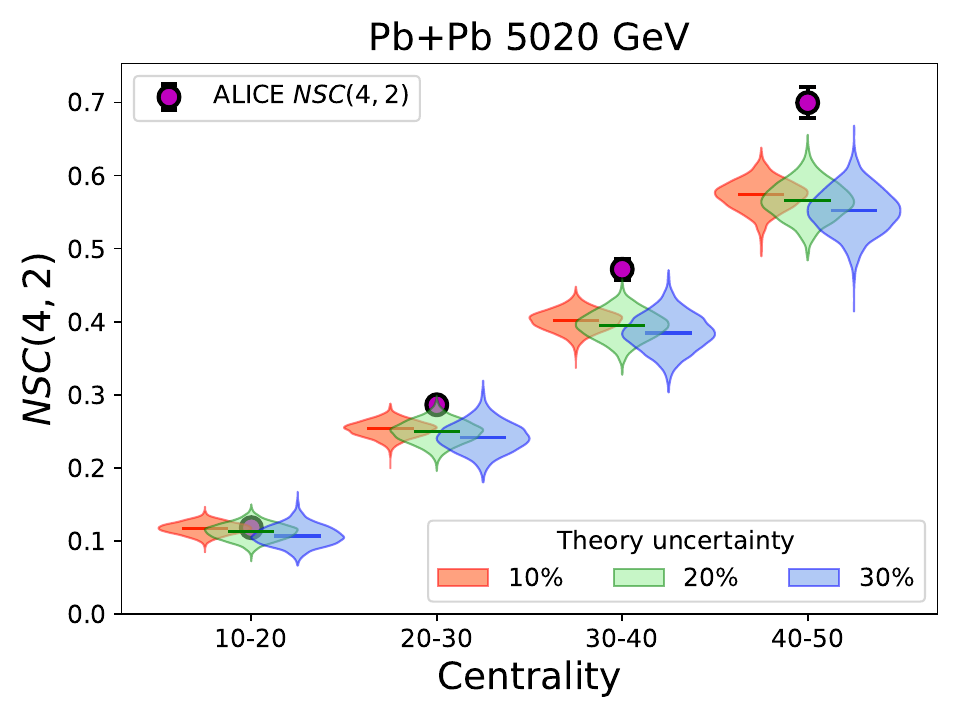}
\caption{Posterior distribution emulator predictions for normalized symmetric cumulant $NSC(4,2)$  in 200 GeV Au+Au (top panel), 2.76 TeV Pb+Pb (middle panel), and 5.02 TeV Pb+Pb (bottom panel) collisions.
STAR data from \cite{STAR:2018fpo}. ALICE data from \cite{ALICE:2016kpq} (2.76 TeV) and \cite{ALICE:2021adw} (5.02 TeV).}
\label{F:nsc}
\end{figure}

To complete the data comparison, we show the emulator predictions for the normalized symmetric cumulant $NSC(4,2)$ in Fig.~\ref{F:nsc}. We get a good description of the data in the 10-20 \% centrality class, but the rise towards more peripheral collisions is too weak compared to measurements.

\subsection{Correlations}

\subsubsection{Correlations between parameters and observables}

Correlations between the parameter-parameter pairs and parameter-observable pairs can serve as a useful guide when investigating the posterior distribution. Strong correlations between parameters can indicate either redundancies in the parametrization (multiple combinations are producing the same effect) or particularly strong constraints from the data, limiting the posterior parameter space and ``forcing'' a correlation. Correlations between parameters and observables provide further guidance where to look for these data constraints.

In Table~\ref{T:correlations} we list the notable Pearson correlations between the parameters within the posterior distribution,
\begin{equation}
r_{ij}=\frac{\sum_k (x_{i,k}-\langle x_i \rangle)(x_{j,k}-\langle x_j \rangle)}{\sqrt{{\sum_k (x_{i,k}-\langle x_i \rangle)^2}{\sum_k (x_{j,k}-\langle x_j \rangle)^2}}}
\end{equation}
where ``notable'' means $|r|>0.3$ for at least two of the three investigated scenarios of varying theoretical uncertainty. In addition, summary tables for all notable correlations between model parameters and final state observables for 20\% theoretical uncertainty can be found in Appendix~\ref{A:correlations_par_obs}.

\begin{table}[h]
\centering
\begin{tabular}{|c|c|c|c|c|}
\hline
\multicolumn{2}{|c|}{Parameter} & \multicolumn{3}{|c|}{Th. uncertainty} \\
 \hline
 A & B & 10\% & 20\% & 30\% \\
\hline
$K_{\mathrm{sat}}$ & $\sigma_n$ & -0.41 & -0.44 & -0.47 \\
\hline
$K_{\mathrm{sat}}$ & $T_H$ & 0.42 & 0.30 & 0.25 \\
\hline
$K_{\mathrm{sat}}$ & $W_{\mathrm{min}}$ & 0.31 & 0.40 & 0.35 \\
\hline
$\sigma_n$ & $T_{\mathrm{chem}}$ & 0.46 & 0.35 & 0.33 \\
\hline
$T_H$ & $P_H$ & 0.60 & 0.45 & 0.33 \\
\hline
$(\eta/s)_{\mathrm{min}}$ & $C_{\mathrm{Kn}}$ & 0.57 & 0.65 & 0.71 \\
\hline
$(\zeta/s)_{\mathrm{max}}$ & $T^{\zeta/s}_{\mathrm{max}}$ & 0.35 & 0.38 & 0.32 \\
\hline
$T^{\zeta/s}_{\mathrm{max}}$ & $(\zeta/s)_{\mathrm{width}}$ & 0.65 & 0.58 & 0.47 \\
\hline
\end{tabular}
\caption{Notable Pearson correlations ($|r|>0.3$ for at least two of the three scenarios of different theoretical uncertainty) between parameters in the posterior distribution.}
\label{T:correlations}
\end{table}

\begin{figure}
\includegraphics[width=7.5cm]{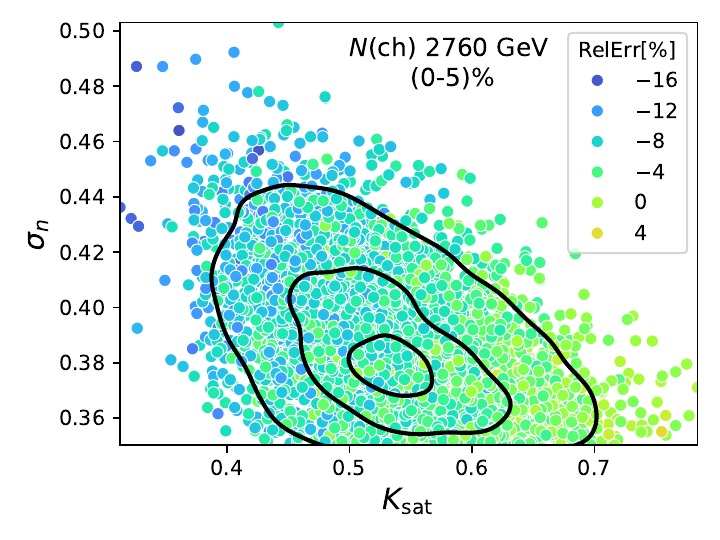}
\includegraphics[width=7.5cm]{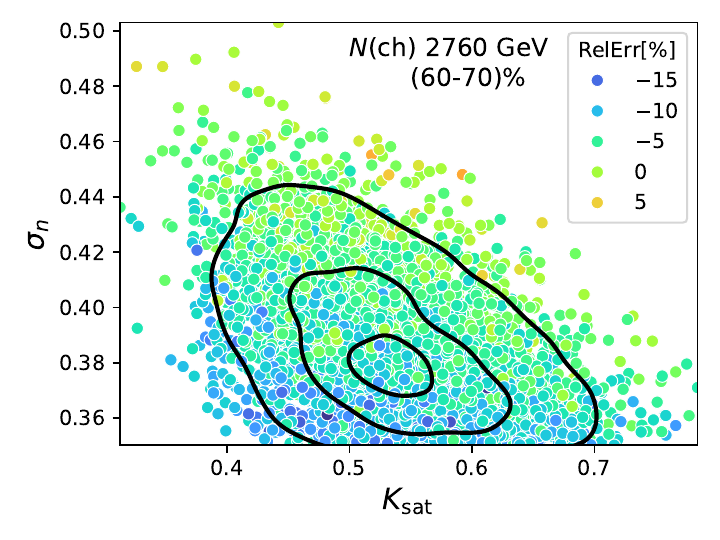}
\caption{Relative deviation from the measured value of charged particle multiplicity $N(\mathrm{ch})$ in Pb+Pb collisions at $\sqrt{s_{NN}}=2.76$ TeV for different combinations of initial state parameters $K_{\mathrm{sat}}$ and $\sigma_n$, sampled from the posterior distribution with 20\% theoretical uncertainty. Black contours indicate the 10\%, 50\% and 90\% levels of point density, such that the highest 10\% of point density
is within the innermost curve. Top panel: 0-5 \% centrality. Bottom panel: 60-70 \% centrality.}
\label{F:Nch_ic_correlation_2760}
\end{figure}

There are three pairs of parameters with particularly strong correlations which appear regardless of the magnitude of the theoretical uncertainty.
The initial state parameters $K_{\mathrm{sat}}$ and $\sigma_n$ are anticorrelated.
This emerges from the tension between multiplicity data in central and peripheral collisions,
as illustrated in Fig.~\ref{F:Nch_ic_correlation_2760} for the Pb+Pb collisions at 2.76 TeV.
Central collisions favor the combination of large $K_{\mathrm{sat}}$ and small $\sigma_n$,
while for the multiplicities in peripheral collisions, smaller values of $K_{\mathrm{sat}}$ combined with larger values of $\sigma_n$ are preferred.

\begin{figure}
\includegraphics[width=7.5cm]{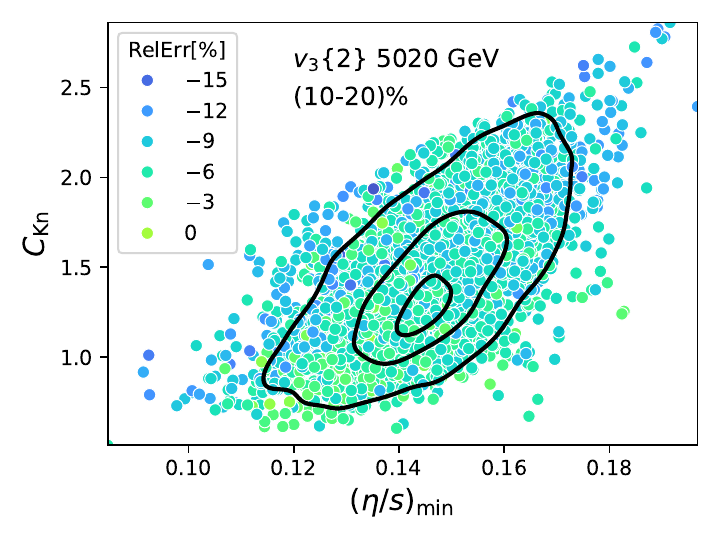}
\includegraphics[width=7.5cm]{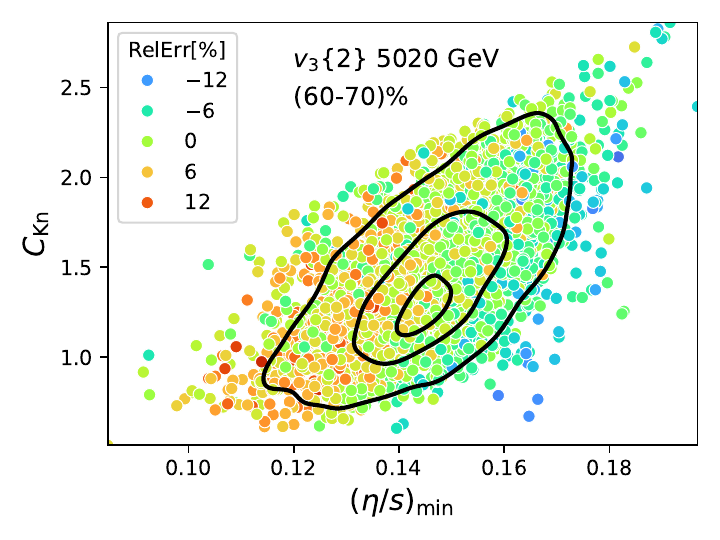}
\includegraphics[width=7.5cm]{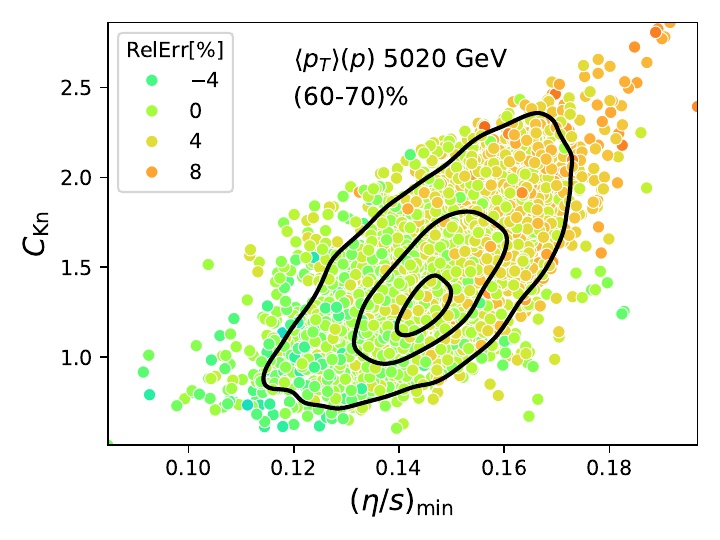}
\caption{Relative deviation from the measured value of charged particle $v_3\{2\}$ (top and middle panel) and proton $\langle p_T \rangle$ (bottom panel) in Pb+Pb collisions at $\sqrt{s_{NN}}=5.02$ TeV for different combinations of $(\eta/s)_{\mathrm{min}}$ and $C_{\mathrm{Kn}}$.}
\label{F:vnpT_etasminKn_correlation_5020}
\end{figure}

Minimum value of the specific shear viscosity $(\eta/s)_{\mathrm{min}}$ correlates with the freeze-out Knudsen number $C_{\mathrm{Kn}}$.
Here the allowed combinations are limited by mean transverse momentum of protons, which imposes constrains to the upper limit of $(\eta/s)_{\mathrm{min}}$ in particular, and the centrality dependence of $v_3$; we demonstrate this in Fig.~\ref{F:vnpT_etasminKn_correlation_5020} for $5.02$~TeV Pb+Pb collisions.

\begin{figure}
\includegraphics[width=9cm]{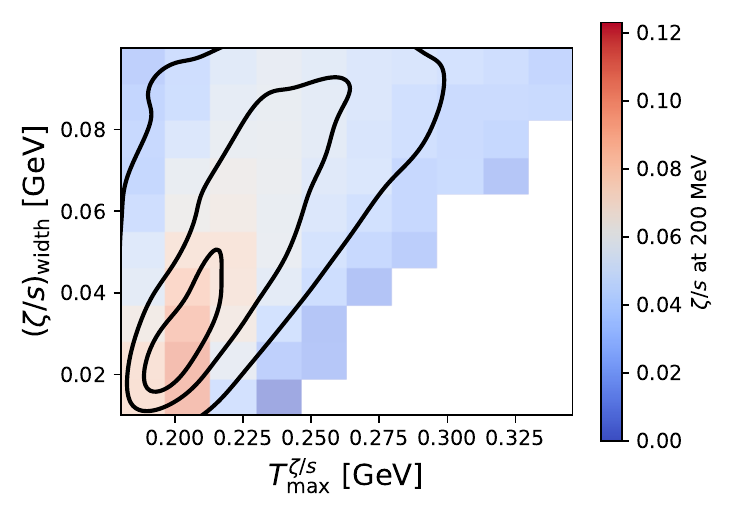}
\caption{Value of bulk viscosity coefficient $\zeta/s$ at $T=200$~MeV for different combinations of the peak temperature $T^{\zeta/s}_{\mathrm{max}}$
and peak width $(\zeta/s)_{\mathrm{width}}$ from the posterior distribution.}
\label{F:zetasT_at_200MeV}
\end{figure}

The temperature where the bulk viscosity reaches its peak $T^{\zeta/s}_{\mathrm{max}}$ is correlated with the width of the peak $(\zeta/s)_{\mathrm{width}}$. This correlation does not stem from data constraints, as these parameters have hardly any notable correlations with any of the observables. Rather, the analysis is driven to maximise
$\zeta/s$ at $T\approx 200$ MeV which makes the combination of narrow peak located at high temperature unpreferable, see Fig.~\ref{F:zetasT_at_200MeV}.

\begin{figure}
\includegraphics[width=7.2cm]{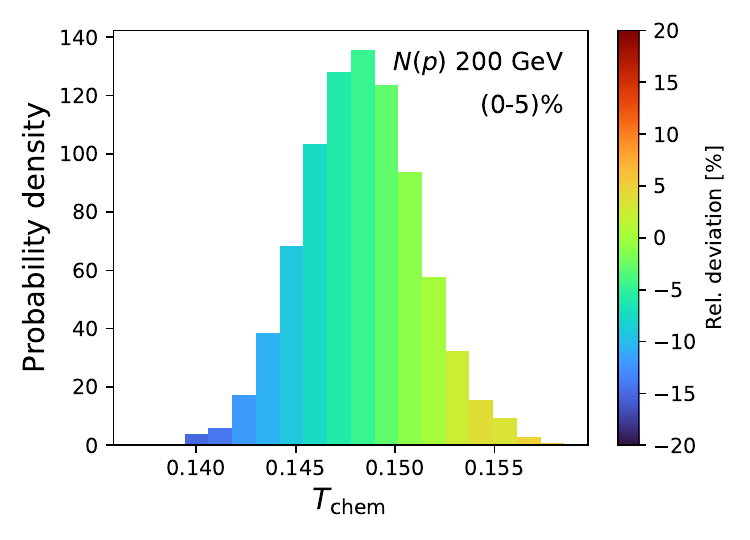} \\
\includegraphics[width=7.2cm]{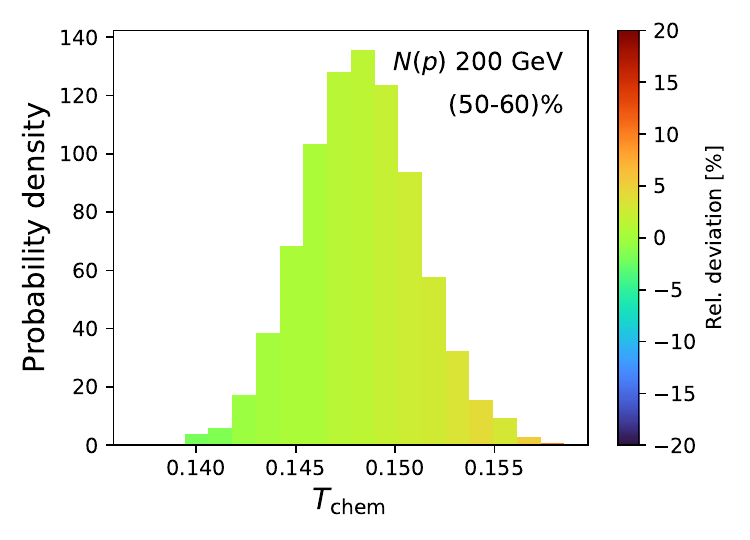} \\
\includegraphics[width=7.2cm]{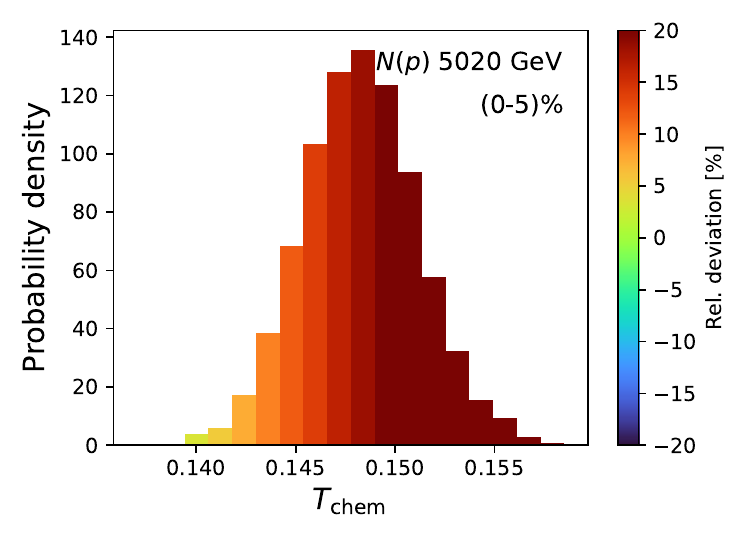} \\
\includegraphics[width=7.2cm]{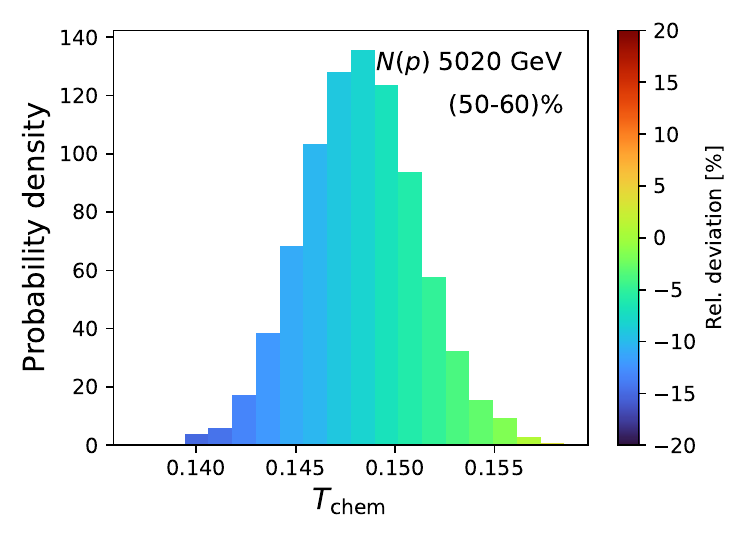} \\
\caption{The posterior distribution of $T_{\rm chem}$ (same in all panels). The color indicates the average relative deviation of proton multiplicity $N(p)$ from the measured value in 200 GeV Au+Au (top panels) and 5.02 TeV Pb+Pb collisions (bottom panels) in 0-5 \% and 50-60 \% centrality classes for different values of $T_{\mathrm{chem}}$. Average is taken over the posterior samples within each bin.}
\label{F:Np_Tchem_correlation}
\end{figure}

Further investigating the correlations between model parameters and observables, there is a particularly strong correlation between chemical freeze-out temperature $T_{\mathrm{chem}}$ and proton multiplicity $N(p)$. We find notable tension in the $N(p)$ data regarding the optimal value of $T_{\mathrm{chem}}$, both in terms of collision energy and centrality, as shown in Fig.~\ref{F:Np_Tchem_correlation}. This tension sets strong constraints on $T_{\mathrm{chem}}$, and also suggests that the chemical freeze-out temperature might depend on the collision energy.

We already noted above that $(\eta/s)_{\mathrm{min}}$ is constrained by $\langle p_T \rangle (p)$ and $v_3$,
and it is not surprising that $\langle p_T \rangle (K)$ and $v_4$ are also correlated with $(\eta/s)_{\mathrm{min}}$. However, it is interesting to note that $\langle p_T \rangle (\pi)$ and $v_2$ do {\em not} have notable correlations with $(\eta/s)_{\mathrm{min}}$. Instead, both show sensitivity to the initial state parameters on one hand, and the conditions at freeze-out on the other; as a correlation to $\Delta (\eta/s)_{H}$ in the case of $\langle p_T \rangle (\pi)$, and as a correlation to $C_{\mathrm{Kn}}$ in the case of $v_2$.

\begin{figure*}
\includegraphics[width=14cm]{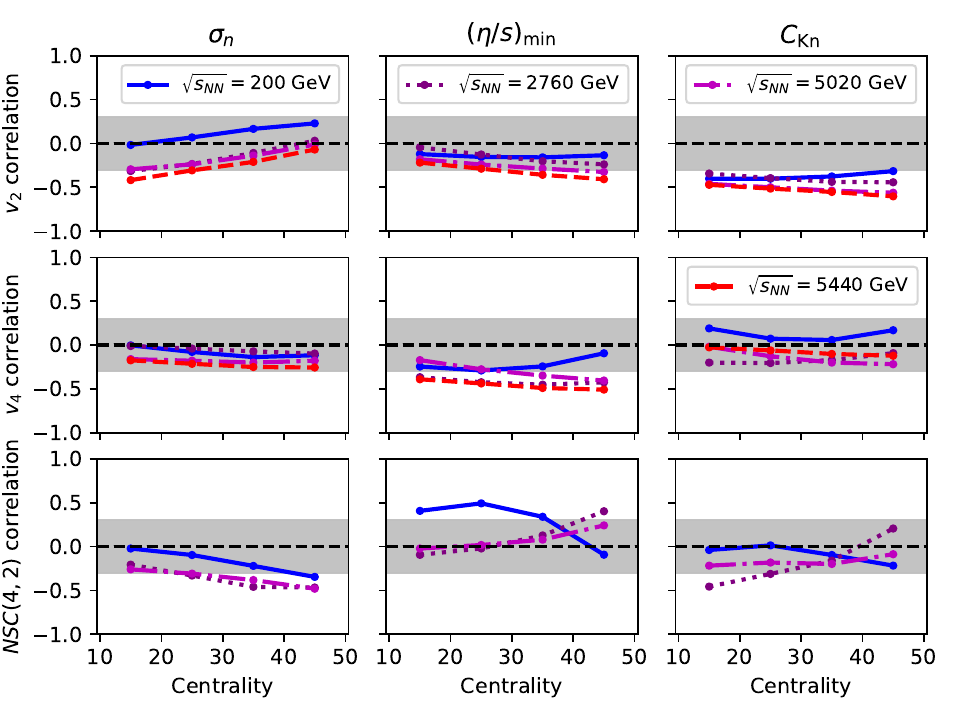}
\caption{Centrality dependence of Pearson correlations for $v_2\{2\}$ (top row), $v_4\{2\}$ (middle row), and $NSC(4,2)$ (bottom row) with parameters $\sigma_n$, $(\eta/s)_{\mathrm{min}}$, and $C_{\mathrm{Kn}}$ (left, middle, and right column, respectively). Gray band indicates negligible correlation strength.}
\label{F:nsc24_correlation}
\end{figure*}

As $v_2$ is correlated with $\sigma_n$ and $C_{\mathrm{Kn}}$, while $v_4$ is correlated with $(\eta/s)_{\mathrm{min}}$, the question arises if these correlations are reflected in $NSC(4,2)$. This turns out to be true only partially, and the dependence on centrality and collision energy is quite different, as shown in Fig.~\ref{F:nsc24_correlation}.
For $v_2$, the $\sigma_n$ correlation is strongest in more central collisions, while for $NSC(4,2)$ it is notable in more peripheral collisions.
For $v_4$, the correlation with $(\eta/s)_{\mathrm{min}}$ is strongest at higher collision energies, for $NSC(4,2)$ 200 GeV has the strongest correlation.
The consistently strong correlation between $v_2$ and $C_{\mathrm{Kn}}$ is greatly diminished for $NSC(4,2)$ and survives only for the 2.76 TeV Pb+Pb collisions at 10-20 \% centrality.

\subsubsection{Correlations between observables and viscous coefficients at different temperatures}

While there are some notable correlations between observables and parameters related to the extrema of the shear viscosity coefficient, we notice that many of the parameters controlling the temperature dependence of $\eta/s$ and $\zeta/s$ have no systematic correlations with observables. In the following, we investigate whether the observables are correlated with $\eta/s$ and $\zeta/s$ in any particular temperature regions. In addition, tabulated correlation data for a selection of temperatures can be found in Appendix~\ref{A:correlations_obs_tcoeffT}.

\begin{figure}
\includegraphics[width=8.5cm]{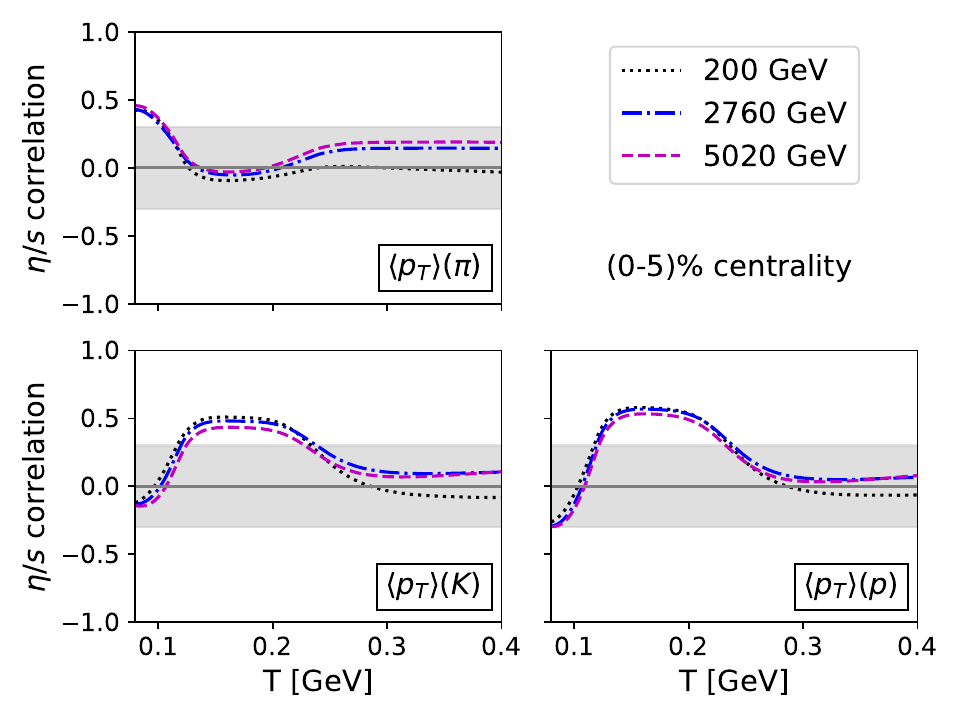}\\
\includegraphics[width=8.5cm]{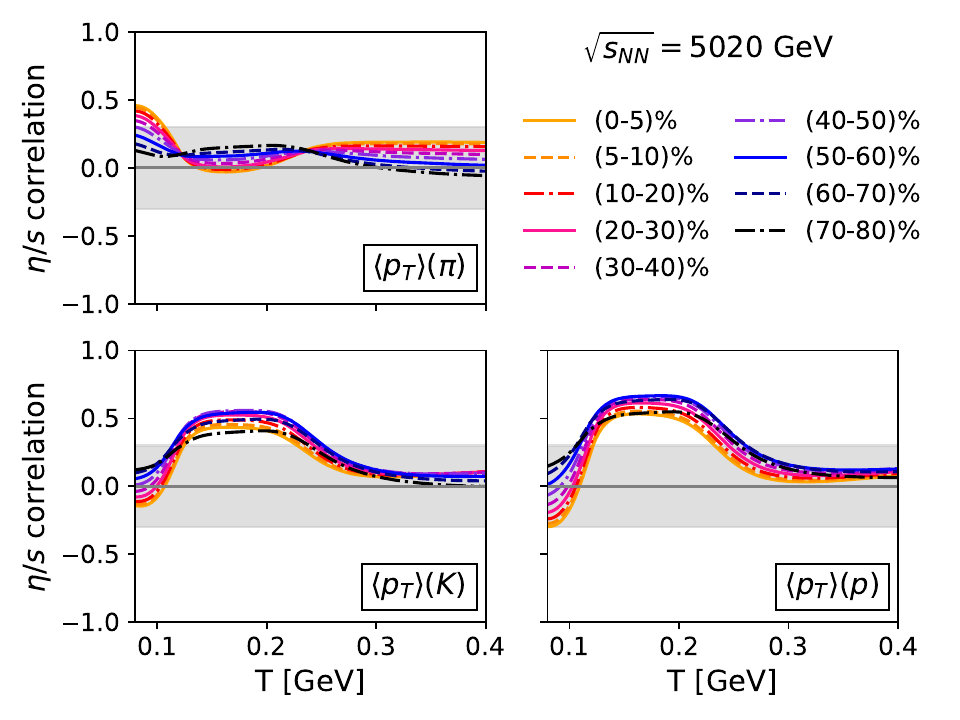}
\caption{Temperature dependence of Pearson correlations between mean transverse momenta and $\eta/s$.
Top three panels: Collision energy dependence of correlations at 0-5 \% centrality.
Bottom three panels: Centrality dependence of correlations at $\sqrt{s_{NN}}=5.02$ TeV collision energy.}
\label{F:corr_meanpt_etas_T}
\end{figure}

For the shear viscosity, the observables with notable correlations are the mean transverse momenta $\langle p_T \rangle$ of pions, kaons and protons and the flow observables $v_n\{2\}$ and $NSC(4,2)$.
We show the correlations between $\langle p_T \rangle$ and $\eta/s$ in the temperature range 80-400 MeV in Fig.~\ref{F:corr_meanpt_etas_T}. Here the findings are consistent with what we already saw in the correlations between model parameters and observables; pion $\langle p_T \rangle$ correlates with $\eta/s$ at low temperatures, as was suggested by the observed correlation between $\langle p_T \rangle(\pi)$ and $\Delta (\eta/s)_{H}$, while $\langle p_T \rangle(K)$ and $\langle p_T \rangle(p)$ correlate with $\eta/s$ in the plateau range $\approx 130-230$ MeV,
as suggested by their correlations with $(\eta/s)_{\mathrm{min}}$. These correlations have a very weak dependence on collision energy,
and also the centrality dependence is notable only at the lowest temperatures, effecting mostly the correlation between $\langle p_T \rangle(\pi)$ and $\eta/s$,
which is strongest at the most central collisions and becomes insignificant around 40-50 \% centrality.

\begin{figure}
\includegraphics[width=8.5cm]{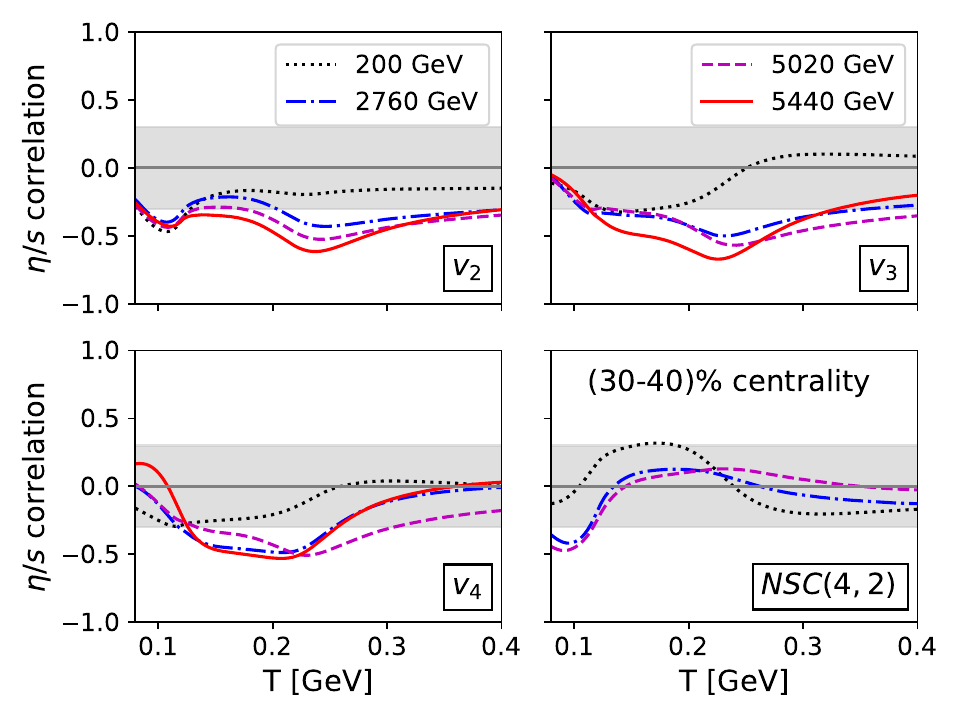}\\
\includegraphics[width=8.5cm]{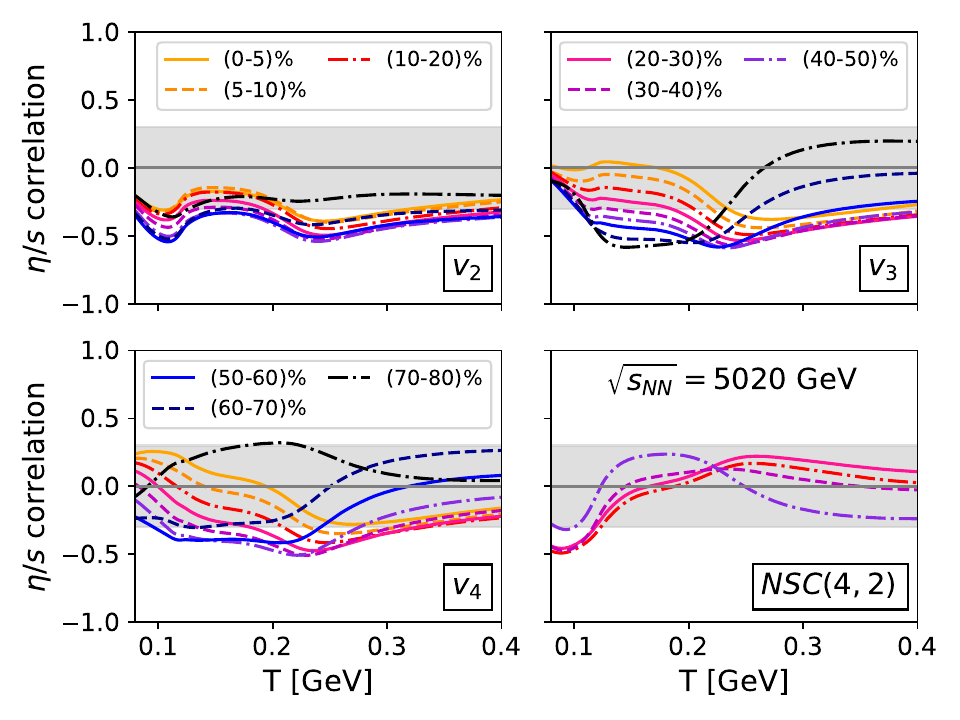}
\caption{Temperature dependence of Pearson correlations between flow observables and $\eta/s$.
Top four panels: Collision energy dependence of correlations at 30-40\% centrality.
Bottom four panels: Centrality dependence of correlations at $\sqrt{s_{NN}}=5.02$ TeV collision energy.}
\label{F:corr_vn_etas_T}
\end{figure}

The flow observables, however, show more variable behavior; see Fig.~\ref{F:corr_vn_etas_T}. While at RHIC the correlations tend to vanish beyond temperatures $T>200$ MeV, at LHC energies they persist up to 300 MeV. At these high temperatures, the correlation strengths for $v_3$ and $v_4$ seem to first increase with collision energy,
but then abruptly decrease again when moving from $5.02$ TeV Pb+Pb collisions to $5.44$ TeV Xe+Xe collisions, which may hint to a system size dependence. We also observe similar weakening of the correlations at high temperatures when moving to more peripheral collisions, supporting the idea that the flow coefficients are not probing these temperatures in smaller systems. It also turns out that the correlator $NSC(4,2)$ is not sensitive to $\eta/s$ at high temperatures either, having the strongest correlations with $\eta/s$ around $T\approx 100$ MeV.

\begin{figure}
\includegraphics[width=8.5cm]{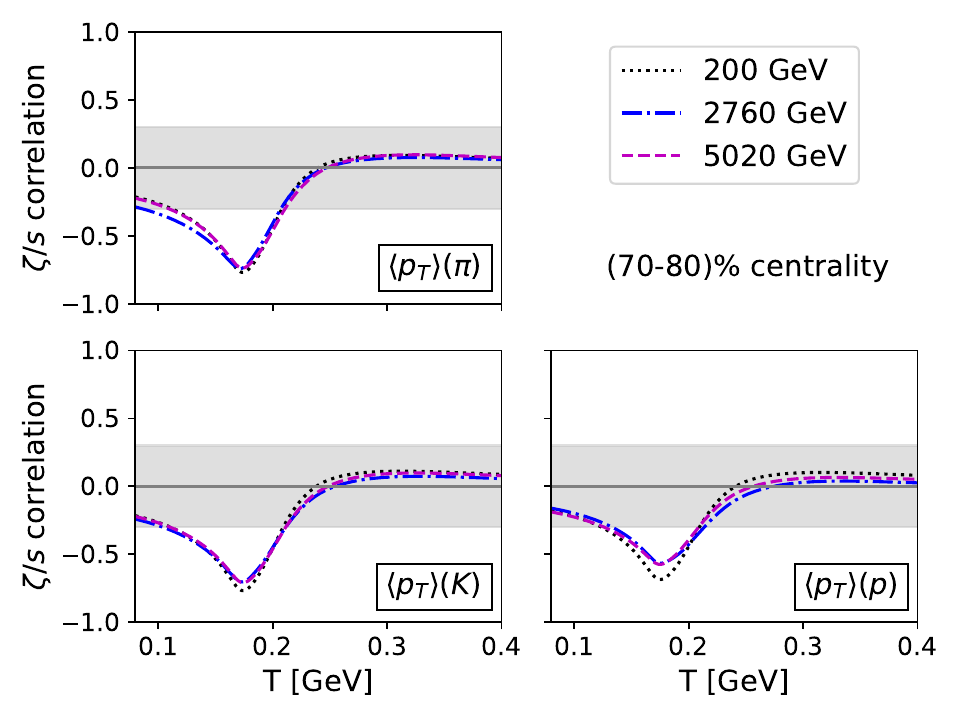}\\
\includegraphics[width=8.5cm]{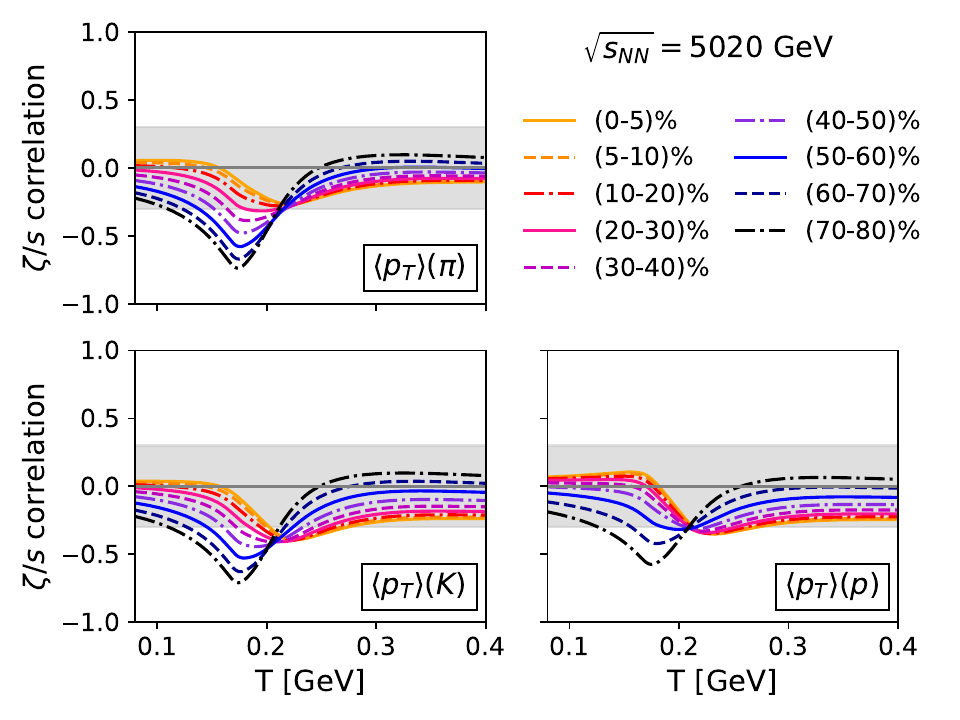}
\caption{Temperature dependence of Pearson correlations between mean transverse momenta and $\zeta/s$.
Top three panels: Collision energy dependence of correlations at 70-80 \% centrality.
Bottom three panels: Centrality dependence of correlations at $\sqrt{s_{NN}}=5.02$ TeV collision energy.}
\label{F:corr_meanpt_zetas_T}
\end{figure}

Finally we move on to the bulk viscosity $\zeta/s$ whose temperature dependence remains poorly constrained.
Indeed, it only shows notable correlations with mean transverse momenta, which we present in Fig.~\ref{F:corr_meanpt_zetas_T}. Here all three investigated particle species tell a remarkably similar story: the correlation is the strongest in the most peripheral collisions and peaks between 150 and 200 MeV, which is the temperature range where the lower limit of our 90\% credible interval for $\zeta/s$ becomes nonzero. The correlation peak does move towards higher temperatures in more central collisions, but the overall strength of the correlation also weakens, suggesting that no observable in the dataset used in this analysis is sensitive to $\zeta/s$ at temperatures beyond $T \gtrsim 250$ MeV.

\section{Summary and conclusions}
\label{S:conclusions}

We have introduced a novel neural network -enhanced framework for performing a Bayesian global analysis to determine the QCD matter properties from the high-energy heavy-ion collision data. The main feature of the framework is the replacement of the computationally demanding hydrodynamical simulations with neural networks that can predict the experimental observables directly from the initial energy density profile event by event. The new feature of the networks compared to our earlier work~\cite{Hirvonen:2023lqy} is that they now take also the matter properties as an input, so that it is no longer necessary to train separate networks for each viscosity parametrization and dynamical freeze-out conditions. This makes them very suitable for global analysis. The NN-generated simulation data are then applied to the training of Gaussian process emulators, which in turn are used to evaluate likelihoods in the Bayesian analysis. Even if the training of the networks still requires a considerable amount of full hydrodynamical simulations, the number of simulated events (here 40 events $\times$ four collision systems $\times$ 1050 parameter sets = $\mathcal{O}(10^5)$ events in total) is orders of magnitude less than what would be needed in computing simulations directly for the training of GP emulators ($10^5 \times 4 \times 1050=\mathcal{O}(10^8)$ events in total).

Here we have demonstrated the feasibility of the framework by performing a Bayesian analysis on the parameters of the EbyE-EKRT + (2+1) D viscous hydrodynamics model with dynamical freeze-out conditions.
We incorporate data from four collision systems, and thanks to the neural networks, are able to include statistically demanding observables such as quadrangular flow $v_4$ and the normalized symmetric cumulant $NSC(4,2)$. We obtain tightest constraints for the initial state saturation strength $K_{\mathrm{sat}}$ and the nucleon width $\sigma_n$, chemical freeze-out temperature $T_{\mathrm{chem}}$ and most importantly, as the QCD matter property, for the temperature dependence of $\eta/s$. In particular, we find the data favoring a relatively narrow nucleon width  $0.35$ $< \sigma_n < 0.47$ fm, chemical freeze-out in the range $143-156$ MeV, and $\eta/s$ with minimum-value plateau extending from $150$ MeV to at least $230$ MeV, with $0.12 < (\eta/s)_{\mathrm{min}} < 0.18$. Bulk viscous coefficient $\zeta/s$ is non-zero at $T = 200-300$ MeV, but otherwise remains unconstrained. Overall, our findings for $(\eta/s)(T)$ and $(\zeta/s)(T)$ are consistent with the other recent Bayesian analyses \cite{JETSCAPE:2020mzn,Heffernan:2023gye,Nijs:2022rme,Virta:2024avu}. 
Our results are also robust against the variations in theoretical uncertainty between 10\% and 30\% of the experimental data values.

The uncertainties are also reflected in the correlations between parameters and observables, as the most constrained parameters also have most notable correlations with the observables included in the analysis. Investigating the correlations of observables with transport coefficients $\eta/s$ and $\zeta/s$ at different temperatures, we find most notable correlations at temperatures $\approx 150-250$ MeV. For $\eta/s$, the strongest constraints in this temperature range appear to come from the mean transverse momenta of kaons and protons, as well as the flow observables $v_n\{2\}$, while the $v_2$-$v_4$ correlator $NSC(4,2)$ and mean transverse momentum of pions are more sensitive to $\eta/s$ at low temperatures. For $\zeta/s$, the constraining observable is $\langle p_T \rangle$, which shows a very similar correlation peak between $150-200$ MeV for all three investigated particle species.

There remains some room for improvement both in the model and the analysis. For a better agreement with the data, implementing further nuclear deformation effects is likely required, as well as updating the initial state model to the MC-EKRT model \cite{Kuha:2024kmq} with hotspots \cite{Hirvonen:2024zne}. To obtain further constraints in the future analyses, we aim to incorporate even more statistically demanding observables such as $NSC(4,3)$ or mixed harmonic cumulants, and consequently we may need to further improve the architecture of the neural networks and the quality of the training data, for example, utilizing active learning approach \cite{JETSCAPE:2024cqe}.

\acknowledgments
J.A. acknowledges support by the program Excellence Initiative--Research University of the University of Wroc\l{}aw of the Ministry of Education and Science.
The project is co-financed by the Polish National Agency for Academic Exchange, number of proposal: BNI/ULM/2024/1/00179/DEC/I.
H.H. is supported by Vanderbilt University and by the U.S. Department of Energy, Office of Science, under Award Numbers DE-SC-0024347 and DE-SC-0024711, and in part by the National Science Foundation under Grant No. DMS-2406870.
We acknowledge the financial support from the Research Council of Finland Project No. 330448 (K.J.E.). This research was funded as a part of the Center of Excellence in Quark Matter of the Research Council of Finland (Projects No. 346325 and 364192). This research is part of the European Research Council Project No. ERC-2018-ADG- 835105 YoctoLHC. We acknowledge the computation resources from the Finnish IT Center for Science (CSC), project jyy2580, and from the Finnish Computing Competence Infrastructure (FCCI), persistent identifier urn:nbn:fi:research-infras-2016072533.

\appendix

\section{Emulator validation}
\label{A:GPtests}
Following Ref.~\cite{Roch:2024xhh}, we define the emulator prediction error as
\begin{equation}
\mathcal{E} \equiv \sqrt{\left\langle \left( \frac{\mathrm{prediction}-\mathrm{truth}}{\mathrm{truth}} \right)^2 \right\rangle},
\label{eq:rms_error}
\end{equation}
and the emulator error estimate accuracy as
\begin{equation}
\mathcal{H} \equiv \ln \left( \sqrt{\left\langle \left( \frac{\mathrm{prediction}-\mathrm{truth}}{\mathrm{GP\ uncertainty}} \right)^2 \right\rangle} \right),
\end{equation}
where the angle brackets indicate the average over test points and the GP uncertainty $\sigma_{\text{GP}}$ is defined in Eq.~\eqref{E:gp_error}.

We summarize the results of emulator validation for the 20 points sampled from the initial posterior distribution in Table~\ref{T:GPvalidation}.
The quality of emulator predictions is good for all the investigated observables.
The RMS errors $\mathcal{E}$ presented here are on average $\approx 40$\% smaller compared to the the initial analysis with 900 training points sampled from the prior.
Notably the emulator error is systematically overestimated for this set; this happens because in the hyperparameter tuning one has to account also for the edges of the prior, where the quality of predictions degrade due to sparsity of training points.

\begin{table*}
\centering
\begin{tabular}{||c|c|c|c|c|c|c|c|c||}
\hline
Observable &  $\langle \mathcal{E} \rangle_{200}$ & $\langle \mathcal{H} \rangle_{200}$ &  $\langle \mathcal{E} \rangle_{2760}$ & $\langle \mathcal{H} \rangle_{2760}$ &  $\langle \mathcal{E} \rangle_{5023}$ & $\langle \mathcal{H} \rangle_{5023}$ &  $\langle \mathcal{E} \rangle_{5440}$ & $\langle \mathcal{H} \rangle_{5440}$\\
\hline
$N(\mathrm{ch})$ & 0.018 & -0.261 & 0.014 & -0.306 & 0.014 & -0.293 & 0.017 & -0.567 \\
\hline
$N(\pi)$ & 0.024 & -0.155 & 0.016 & -0.275 & 0.016 & -0.255 & - & - \\
\hline
$N(K)$ & 0.027 & 0.001 & 0.017 & -0.155 & 0.017 & -0.186 & - & - \\
\hline
$N(p)$ & 0.035 & -0.185 & 0.030 & -0.114 & 0.027 & -0.200 & - & - \\
\hline
$\langle p_T \rangle (\mathrm{ch})$ & - & - & - & - & - & - & 0.008 & -0.327 \\
\hline
$\langle p_T \rangle (\pi)$ & 0.014 & -0.222 & 0.015 & -0.054 & 0.016 & -0.057 & - & - \\
\hline
$\langle p_T \rangle (K)$ & 0.010 & -0.361 & 0.009 & -0.434 & 0.009 & -0.328 & - & - \\
\hline
$\langle p_T \rangle (p)$ & 0.013 & -0.164 & 0.010 & -0.341 & 0.012 & -0.145 & - & - \\
\hline
$v_2$ & 0.029 & -0.206 & 0.017 & -0.096 & 0.017 & 0.040 & 0.017 & -0.094 \\
\hline
$v_3$ & 0.053 & -0.106 & 0.025 & -0.066 & 0.026 & -0.173 & 0.022 & -0.141 \\
\hline
$v_4$ & 0.084 & -0.120 & 0.036 & -0.139 & 0.040 & -0.001 & 0.034 & -0.257  \\
\hline
$NSC(4,2)$ & 0.088 & -0.203 & 0.052 & -0.108 & 0.053 & -0.014 & - & - \\
\hline
\end{tabular}
\caption{Summary of centrality-averaged emulator quality parameters for a test set of 20 points sampled from the initial posterior distribution. 
$\mathcal{E}$ reflects the emulator prediction error relative to the true value,
while $\mathcal{H}$ measures the accuracy of the uncertainty estimate reported by the Gaussian process.
The negative and positive values of $\mathcal{H}$ indicate overestimation and underestimation of error, respectively.}
\label{T:GPvalidation}
\end{table*}

\section{Closure tests}
\label{A:closuretests}

Closure tests serve as a consistency check for the full analysis procedure.
For the target data, we take the model output at the same 20 points which were used for the GP emulator quality tests, and check if the analysis is able to recover the parameter values used to produce the data. Here the relative uncertainties associated with the target data are taken to be the same as in the corresponding experimental data,
while the theoretical uncertainty can be considered negligible when comparing to model data.
We summarize the results in Table~\ref{T:closuretests}.

We obtain the narrowest credible intervals ($<25\%$ of the prior width on average) for $K_{\mathrm{sat}}$, $\sigma_n$, $(\eta/s)_{\mathrm{min}}$, $T_{\mathrm{chem}}$, and $C_{\mathrm{Kn}}/(\eta/s)_{\mathrm{min}}$. These intervals are particularly conservative for $K_{\mathrm{sat}}$ and $\sigma_n$, which have their true values within 20\% credible interval in 40\% of the trials. On the other extreme is the parameter $T_H$ which has a very poor inference accuracy; the marginal posterior distribution is both wide (the 90\% credible interval is $\approx 70$\% of the prior width on average) and imprecise (true value is within 90\% credible interval in only 65\% of the cases).

\begin{table}[h]
\centering
\begin{tabular}{|c|c|c|c|c|}
\hline
 & \multicolumn{3}{c|}{$P(\text{Truth in credible interval})$} & 90\% C.I. \\
Parameter & 20\% C.I. & 68\% C.I. & 90\% C.I. & \parbox{2.0cm}{avg \% of prior}\\
\hline
$K_{\mathrm{sat}}$ & 40\% & 75\% & 100\% & 16.6\% \\
\hline
$\sigma_n$ & 45\% & 85\% & 95\% & 18.0\% \\
\hline
$T_H$ & 10\% & 40\% & 65\% & 69.4\% \\
\hline
$(\eta/s)_{\mathrm{min}}$ & 25\% & 85\% & 100\% & 12.9\% \\
\hline
$P_H$ & 10\% & 40\% & 70\% & 27.7\% \\
\hline
$\Delta (\eta/s)_{H}$ & 5\% & 60\% & 85\% & 36.0\% \\
\hline
$S_Q$ & 15\% & 65\% & 90\% & 77.6\% \\
\hline
$W_{\mathrm{min}}$ & 25\% & 55\% & 75\% & 63.4\% \\
\hline
$(\zeta/s)_{\mathrm{max}}$ & 10\% & 35\% & 90\% & 44.6\% \\
\hline
$T^{\zeta/s}_{\mathrm{max}}$ & 15\% & 45\% & 85\% & 53.4\% \\
\hline
$a_{\zeta/s}$ & 20\% & 60\% & 85\% & 75.7\% \\
\hline
$(\zeta/s)_{\mathrm{width}}$ & 20\% & 65\% & 90\% & 77.1\% \\
\hline
$T_{\mathrm{chem}}$ & 15\% & 75\% & 100\% & 17.0\% \\
\hline
$C_{\mathrm{Kn}}/(\eta/s)_{\mathrm{min}}$ & 5\% & 60\% & 85\% & 23.9\% \\
\hline
$C_R/(\eta/s)_{\mathrm{min}}$ & 15\% & 55\% & 85\% & 44.9\% \\
\hline

\end{tabular}
\caption{Probabilities for finding the closure test truth value in a specified credible interval after 20 trials.
The last column shows the average width of the 90\% credible interval as a fraction of the prior range of parameter values.}
\label{T:closuretests}
\end{table}

\section{Notable Pearson correlations between model parameters and observables}
\label{A:correlations_par_obs}

Tables~\ref{T:par_obs_correlations_200}-\ref{T:par_obs_correlations_5440} summarise notable Pearson correlations ($|r|>0.3$ in at least one centrality class) between the model parameters and observables for the posterior distribution from the analysis with 20\% theoretical uncertainty. Upper and lower values in a cell represent the highest and lowest correlations, respectively, among all the investigated centrality classes for each observable.

\begin{table*}[h]
\centering
\begin{tabular}{|c|c|c|c|c|c|c|c|c|c|c|c|c|c|c|c|}
\hline
\vtop{\hbox{\strut Au+Au}\hbox{\strut $200$ GeV}} & $K_{\mathrm{sat}}$ & $\sigma_n$ & $T_H$ & $(\eta/s)_{\mathrm{min}}$ & $P_H$ & $\Delta (\eta/s)_{H}$ & $S_Q$ & $W_{\mathrm{min}}$ & $(\zeta/s)_{\mathrm{max}}$ & $T^{\zeta/s}_{\mathrm{max}}$ & $a_{\zeta/s}$ & $(\zeta/s)_{\mathrm{width}}$ & $T_{\mathrm{chem}}$ & $C_{\mathrm{Kn}}$ & $C_R$ \\
\hline
\multirow{2}{*}{$N(\mathrm{ch})$} & 0.66 & 0.07 & - & - & - & - & - & - & - & - & - & - & -0.33 & - & - \\ & 0.41 & -0.48 & - & - & - & - & - & - & - & - & - & - & -0.49 & - & - \\
\hline
\multirow{2}{*}{$N(\pi)$} & 0.64 & 0.39 & - & - & - & - & - & - & - & - & - & - & -0.37 & - & - \\ & 0.07 & -0.49 & - & - & - & - & - & - & - & - & - & - & -0.55 & - & - \\
\hline
\multirow{2}{*}{$N(K)$} & 0.67 & 0.29 & - & - & - & - & - & - & - & - & - & - & -0.19 & - & - \\ & 0.06 & -0.47 & - & - & - & - & - & - & - & - & - & - & -0.46 & - & - \\
\hline
\multirow{2}{*}{$N(p)$} & - & - & - & - & - & - & - & - & - & - & - & - & 0.70 & - & - \\ & - & - & - & - & - & - & - & - & - & - & - & - & -0.42 & - & - \\
\hline
\multirow{2}{*}{$\langle p_T \rangle (\pi)$} & 0.52 & 0.43 & - & - & - & 0.43 & - & - & - & 0.30 & - & - & 0.38 & - & - \\ & 0.15 & -0.14 & - & - & - & 0.12 & - & - & - & 0.12 & - & - & -0.34 & - & - \\
\hline
\multirow{2}{*}{$\langle p_T \rangle (K)$} & - & 0.41 & - & 0.60 & - & - & - & - & - & 0.32 & - & - & 0.40 & 0.39 & - \\ & - & -0.09 & - & 0.34 & - & - & - & - & - & -0.01 & - & - & -0.27 & 0.14 & - \\
\hline
\multirow{2}{*}{$\langle p_T \rangle (p)$} & - & - & - & 0.67 & - & - & - & - & - & - & - & - & 0.32 & 0.53 & 0.38 \\ & - & - & - & 0.45 & - & - & - & - & - & - & - & - & -0.25 & 0.22 & -0.18 \\
\hline
\multirow{2}{*}{$v_2$} & 0.37 & - & - & - & - & 0.03 & - & - & - & - & - & - & 0.33 & -0.03 & 0.57 \\ & -0.18 & - & - & - & - & -0.31 & - & - & - & - & - & - & -0.10 & -0.40 & -0.29 \\
\hline
\multirow{2}{*}{$v_3$} & - & - & - & 0.56 & - & - & - & - & - & - & - & - & 0.48 & 0.39 & - \\ & - & - & - & -0.33 & - & - & - & - & - & - & - & - & -0.21 & -0.07 & - \\
\hline
\multirow{2}{*}{$v_4$} & - & - & - & 0.65 & - & - & - & 0.01 & - & - & - & - & 0.55 & 0.48 & 0.31 \\ & - & - & - & -0.29 & - & - & - & -0.31 & - & - & - & - & -0.20 & 0.06 & -0.31 \\
\hline
\multirow{2}{*}{$NSC(4,2)$} & 0.30 & -0.02 & - & 0.49 & - & -0.07 & - & - & - & - & - & - & 0.01 & - & - \\ & -0.02 & -0.35 & - & -0.09 & - & -0.42 & - & - & - & - & - & - & -0.31 & - & - \\
\hline
\end{tabular}
\caption{Notable Pearson correlations ($|r|>0.3$) in at least one centrality class) between the model parameters and observables at $200$ GeV Au+Au collisions for the posterior distribution from the analysis with 20\% theoretical uncertainty. Upper and lower values in a cell represent the highest and lowest correlations, respectively, among all the investigated centrality classes for each observable.}
\label{T:par_obs_correlations_200}
\end{table*}

\begin{table*}[h]
\centering
\begin{tabular}{|c|c|c|c|c|c|c|c|c|c|c|c|c|c|c|c|}
\hline
\vtop{\hbox{\strut Pb+Pb}\hbox{\strut $2.76$ TeV}} & $K_{\mathrm{sat}}$ & $\sigma_n$ & $T_H$ & $(\eta/s)_{\mathrm{min}}$ & $P_H$ & $\Delta (\eta/s)_{H}$ & $S_Q$ & $W_{\mathrm{min}}$ & $(\zeta/s)_{\mathrm{max}}$ & $T^{\zeta/s}_{\mathrm{max}}$ & $a_{\zeta/s}$ & $(\zeta/s)_{\mathrm{width}}$ & $T_{\mathrm{chem}}$ & $C_{\mathrm{Kn}}$ & $C_R$ \\\hline
\multirow{2}{*}{$N(\mathrm{ch})$} & 0.63 & 0.55 & - & - & - & - & - & - & - & - & - & - & -0.20 & - & - \\ & -0.03 & -0.42 & - & - & - & - & - & - & - & - & - & - & -0.32 & - & - \\
\hline
\multirow{2}{*}{$N(\pi)$} & 0.62 & 0.62 & - & - & - & - & - & - & - & - & - & - & -0.21 & - & - \\ & -0.16 & -0.46 & - & - & - & - & - & - & - & - & - & - & -0.42 & - & - \\
\hline
\multirow{2}{*}{$N(K)$} & 0.62 & 0.59 & - & - & - & - & - & - & - & - & - & - & - & - & - \\ & -0.18 & -0.42 & - & - & - & - & - & - & - & - & - & - & - & - & - \\
\hline
\multirow{2}{*}{$N(p)$} & - & 0.32 & - & - & - & - & - & - & 0.09 & - & - & - & 0.82 & - & - \\ & - & 0.07 & - & - & - & - & - & - & -0.36 & - & - & - & -0.03 & - & - \\
\hline
\multirow{2}{*}{$\langle p_T \rangle (\pi)$} & 0.41 & 0.43 & - & - & - & 0.44 & - & - & - & - & - & - & 0.40 & - & - \\ & 0.17 & -0.05 & - & - & - & 0.12 & - & - & - & - & - & - & -0.34 & - & - \\
\hline
\multirow{2}{*}{$\langle p_T \rangle (K)$} & - & 0.38 & - & 0.56 & - & - & - & - & -0.08 & - & - & - & 0.39 & 0.33 & - \\ & - & 0.01 & - & 0.39 & - & - & - & - & -0.31 & - & - & - & -0.26 & 0.17 & - \\
\hline
\multirow{2}{*}{$\langle p_T \rangle (p)$} & - & - & - & 0.66 & - & 0.09 & - & - & -0.14 & - & - & - & - & 0.47 & - \\ & - & - & - & 0.53 & - & -0.32 & - & - & -0.31 & - & - & - & - & 0.29 & - \\
\hline
\multirow{2}{*}{$v_2$} & 0.64 & 0.30 & - & - & - & - & - & - & - & - & - & - & - & -0.06 & - \\ & -0.23 & -0.44 & - & - & - & - & - & - & - & - & - & - & - & -0.44 & - \\
\hline
\multirow{2}{*}{$v_3$} & 0.52 & - & - & -0.04 & - & - & - & - & - & - & - & - & - & -0.08 & - \\ & -0.18 & - & - & -0.47 & - & - & - & - & - & - & - & - & - & -0.39 & - \\
\hline
\multirow{2}{*}{$v_4$} & 0.33 & - & - & 0.06 & - & - & - & - & - & - & - & - & - & - & - \\ & -0.12 & - & - & -0.45 & - & - & - & - & - & - & - & - & - & - & - \\
\hline
\multirow{2}{*}{$NSC(4,2)$} & - & -0.21 & - & 0.40 & - & -0.09 & - & - & - & - & - & - & -0.22 & 0.21 & - \\ & - & -0.47 & - & -0.09 & - & -0.45 & - & - & - & - & - & - & -0.32 & -0.46 & - \\
\hline
\end{tabular}
\caption{Same as table~\ref{T:par_obs_correlations_200} but for $2.76$ TeV Pb+Pb collisions.}
\label{T:par_obs_correlations_2760}
\end{table*}

\begin{table*}[h]
\centering
\begin{tabular}{|c|c|c|c|c|c|c|c|c|c|c|c|c|c|c|c|}
\hline
\vtop{\hbox{\strut Pb+Pb}\hbox{\strut $5.02$ TeV}} & $K_{\mathrm{sat}}$ & $\sigma_n$ & $T_H$ & $(\eta/s)_{\mathrm{min}}$ & $P_H$ & $\Delta (\eta/s)_{H}$ & $S_Q$ & $W_{\mathrm{min}}$ & $(\zeta/s)_{\mathrm{max}}$ & $T^{\zeta/s}_{\mathrm{max}}$ & $a_{\zeta/s}$ & $(\zeta/s)_{\mathrm{width}}$ & $T_{\mathrm{chem}}$ & $C_{\mathrm{Kn}}$ & $C_R$ \\\hline
\multirow{2}{*}{$N(\mathrm{ch})$} & 0.59 & 0.60 & - & - & - & - & - & - & - & - & - & - & - & - & - \\ & -0.05 & -0.41 & - & - & - & - & - & - & - & - & - & - & - & - & - \\
\hline
\multirow{2}{*}{$N(\pi)$} & 0.58 & 0.66 & - & - & - & - & - & - & - & - & - & - & -0.18 & - & - \\ & -0.17 & -0.45 & - & - & - & - & - & - & - & - & - & - & -0.40 & - & - \\
\hline
\multirow{2}{*}{$N(K)$} & 0.58 & 0.64 & - & - & - & - & - & - & - & - & - & - & - & - & - \\ & -0.20 & -0.40 & - & - & - & - & - & - & - & - & - & - & - & - & - \\
\hline
\multirow{2}{*}{$N(p)$} & 0.13 & 0.45 & - & - & - & - & - & - & 0.07 & - & - & - & 0.82 & - & - \\ & -0.34 & 0.08 & - & - & - & - & - & - & -0.33 & - & - & - & 0.10 & - & - \\
\hline
\multirow{2}{*}{$\langle p_T \rangle (\pi)$} & 0.39 & 0.41 & - & - & - & 0.46 & - & - & - & 0.31 & - & - & 0.37 & - & - \\ & 0.22 & -0.04 & - & - & - & 0.12 & - & - & - & -0.01 & - & - & -0.31 & - & - \\
\hline
\multirow{2}{*}{$\langle p_T \rangle (K)$} & - & 0.36 & - & 0.56 & - & - & - & - & -0.07 & 0.30 & - & - & 0.39 & 0.33 & - \\ & - & -0.02 & - & 0.40 & - & - & - & - & -0.37 & -0.10 & - & - & -0.30 & 0.16 & - \\
\hline
\multirow{2}{*}{$\langle p_T \rangle (p)$} & - & - & - & 0.66 & - & 0.12 & - & - & -0.09 & - & - & - & - & 0.56 & - \\ & - & - & - & 0.53 & - & -0.32 & - & - & -0.35 & - & - & - & - & 0.33 & - \\
\hline
\multirow{2}{*}{$v_2$} & 0.63 & 0.45 & - & -0.15 & - & -0.20 & - & 0.30 & - & - & - & - & - & -0.12 & - \\ & -0.12 & -0.35 & - & -0.33 & - & -0.34 & - & 0.19 & - & - & - & - & - & -0.56 & - \\
\hline
\multirow{2}{*}{$v_3$} & 0.48 & 0.27 & - & 0.01 & - & - & - & - & - & - & - & - & - & -0.11 & - \\ & -0.27 & -0.32 & - & -0.56 & - & - & - & - & - & - & - & - & - & -0.38 & - \\
\hline
\multirow{2}{*}{$v_4$} & 0.31 & 0.03 & - & 0.30 & - & - & - & - & - & - & - & - & - & 0.37 & - \\ & -0.19 & -0.34 & - & -0.40 & - & - & - & - & - & - & - & - & - & -0.22 & - \\
\hline
\multirow{2}{*}{$NSC(4,2)$} & - & -0.26 & - & - & - & -0.29 & - & - & - & - & - & - & - & - & - \\ & - & -0.48 & - & - & - & -0.48 & - & - & - & - & - & - & - & - & - \\
\hline
\end{tabular}
\caption{Same as table~\ref{T:par_obs_correlations_200} but for $5.02$ TeV Pb+Pb collisions.}
\label{T:par_obs_correlations_5020}
\end{table*}

\begin{table*}[h]
\centering
\begin{tabular}{|c|c|c|c|c|c|c|c|c|c|c|c|c|c|c|c|}
\hline
\vtop{\hbox{\strut Xe+Xe}\hbox{\strut $5.44$ TeV}} & $K_{\mathrm{sat}}$ & $\sigma_n$ & $T_H$ & $(\eta/s)_{\mathrm{min}}$ & $P_H$ & $\Delta (\eta/s)_{H}$ & $S_Q$ & $W_{\mathrm{min}}$ & $(\zeta/s)_{\mathrm{max}}$ & $T^{\zeta/s}_{\mathrm{max}}$ & $a_{\zeta/s}$ & $(\zeta/s)_{\mathrm{width}}$ & $T_{\mathrm{chem}}$ & $C_{\mathrm{Kn}}$ & $C_R$ \\\hline
\multirow{2}{*}{$N(\mathrm{ch})$} & 0.60 & 0.74 & - & - & - & - & - & - & - & - & - & - & - & - & - \\ & -0.26 & -0.43 & - & - & - & - & - & - & - & - & - & - & - & - & - \\
\hline
\multirow{2}{*}{$\langle p_T \rangle (\mathrm{ch})$} & 0.33 & 0.43 & - & - & - & - & - & - & -0.09 & - & - & - & 0.49 & - & - \\ & 0.09 & 0.10 & - & - & - & - & - & - & -0.31 & - & - & - & 0.01 & - & - \\
\hline
\multirow{2}{*}{$v_2$} & 0.67 & 0.24 & - & -0.11 & 0.30 & -0.17 & - & - & - & - & - & - & - & -0.43 & - \\ & 0.12 & -0.43 & - & -0.47 & 0.03 & -0.36 & - & - & - & - & - & - & - & -0.61 & - \\
\hline
\multirow{2}{*}{$v_3$} & 0.52 & 0.06 & - & -0.29 & - & - & - & - & - & - & - & - & - & -0.27 & - \\ & 0.06 & -0.31 & - & -0.58 & - & - & - & - & - & - & - & - & - & -0.41 & - \\
\hline
\multirow{2}{*}{$v_4$} & - & - & - & -0.34 & - & - & - & - & - & - & - & - & - & - & - \\ & - & - & - & -0.51 & - & - & - & - & - & - & - & - & - & - & - \\
\hline
\end{tabular}
\caption{Same as table~\ref{T:par_obs_correlations_200} but for $5.44$ TeV Xe+Xe collisions.}
\label{T:par_obs_correlations_5440}
\end{table*}

\section{Notable Pearson correlations between observables and transport coefficients at various temperatures}
\label{A:correlations_obs_tcoeffT}

Tables \ref{T:etasT_par_correlations} and \ref{T:zetasT_par_correlations} summarise notable Pearson correlations ($|r|>0.3$ in at least one centrality class) between model parameters and values of specific shear viscosity $\eta/s$ and bulk viscosity $\zeta/s$, respectively, at various temperatures, for the posterior distribution from the analysis with 20\% theoretical uncertainty.

Similarly, tables~\ref{T:etasT_obs_correlations_all} and \ref{T:zetasT_obs_correlations_all} summarise notable Pearson correlations between observables and viscous coefficients at different temperatures. Upper and lower values in a cell represent the highest and lowest correlations, respectively, among all the investigated centrality classes for each observable.

\begin{table}[h]
\centering
\begin{tabular}{|c|c|c|c|c|c|}
\hline
 & \multicolumn{5}{|c|}{$(\eta/s)(T)$ at $T=X$ MeV} \\
\hline
Parameter & $100$ & $150$ & $200$ & $250$ & $300$ \\
 \hline
$K_{\mathrm{sat}}$ & - & - & - & -0.32 & -0.46 \\
\hline
$\sigma_n$ & - & - & - & - & - \\
\hline
$T_H$ & - & - & - & - & - \\
\hline
$(\eta/s)_{\mathrm{min}}$ & 0.31 & 0.97 & 0.93 & - & - \\
\hline
$P_H$ & - & - & - & - & - \\
\hline
$\Delta (\eta/s)_{H}$ & 0.81 & - & - & 0.32 & - \\
\hline
$S_Q$ & - & - & - & - & - \\
\hline
$W_{\mathrm{min}}$ & - & - & - & -0.38 & -0.60 \\
\hline
$(\zeta/s)_{\mathrm{max}}$ & - & - & -0.30 & - & - \\
\hline
$T^{\zeta/s}_{\mathrm{max}}$ & - & - & - & - & - \\
\hline
$a_{\zeta/s}$ & - & - & - & - & - \\
\hline
$(\zeta/s)_{\mathrm{width}}$ & - & - & - & - & - \\
\hline
$T_{\mathrm{chem}}$ & - & - & - & - & - \\
\hline
$C_{\mathrm{Kn}}$ & 0.51 & 0.66 & 0.66 & 0.41 & - \\
\hline
$C_R$ & 0.31 & - & - & - & - \\
\hline
\end{tabular}
\caption{Notable Pearson correlations between $\eta/s$ at various temperatures and parameters for the posterior distribution from the analysis with 20\% theoretical uncertainty.}
\label{T:etasT_par_correlations}
\end{table}

\begin{table}[h]
\centering
\begin{tabular}{|c|c|c|c|c|c|}
\hline
 & \multicolumn{5}{|c|}{$(\zeta/s)(T)$ at $T=X$ MeV} \\
\hline
Parameter & $100$ & $150$ & $200$ & $250$ & $300$ \\
 \hline
$K_{\mathrm{sat}}$ & - & - & - & - & - \\
\hline
$\sigma_n$ & - & -0.37 & -0.32 & - & - \\
\hline
$T_H$ & - & - & - & - & - \\
\hline
$(\eta/s)_{\mathrm{min}}$ & - & - & -0.30 & - & - \\
\hline
$P_H$ & - & - & - & - & - \\
\hline
$\Delta (\eta/s)_{H}$ & - & - & - & - & - \\
\hline
$S_Q$ & - & - & - & - & - \\
\hline
$W_{\mathrm{min}}$ & - & - & - & - & - \\
\hline
$(\zeta/s)_{\mathrm{max}}$ & - & - & 0.53 & 0.69 & 0.56 \\
\hline
$T^{\zeta/s}_{\mathrm{max}}$ & 0.46 & - & -0.35 & 0.66 & 0.84 \\
\hline
$a_{\zeta/s}$ & 0.63 & 0.44 & - & - & - \\
\hline
$(\zeta/s)_{\mathrm{width}}$ & 0.51 & 0.56 & -0.47 & 0.34 & 0.58 \\
\hline
$T_{\mathrm{chem}}$ & -0.37 & -0.50 & - & - & - \\
\hline
$C_{\mathrm{Kn}}$ & - & - & - & - & - \\
\hline
$C_R$ & - & - & - & - & - \\
\hline
\end{tabular}
\caption{Notable Pearson correlations between $\zeta/s$ at various temperatures and parameters for the posterior distribution from the analysis with 20\% theoretical uncertainty.}
\label{T:zetasT_par_correlations}
\end{table}

\begin{table*}[h]
\centering
\begin{tabular}{|c||c|c|c|c|c||c|c|c|c|c||c|c|c|c|c||c|c|c|c|c||}
\hline
 & \multicolumn{5}{c||}{\vtop{\hbox{\strut Au+Au $200$ GeV}\hbox{\strut $(\eta/s)(T)$ at $T=X$ MeV}}}
 & \multicolumn{5}{c||}{\vtop{\hbox{\strut Pb+Pb $2.76$ TeV}\hbox{\strut $(\eta/s)(T)$ at $T=X$ MeV}}}
 & \multicolumn{5}{c||}{\vtop{\hbox{\strut Pb+Pb $5.02$ TeV}\hbox{\strut $(\eta/s)(T)$ at $T=X$ MeV}}}
 & \multicolumn{5}{c||}{\vtop{\hbox{\strut Xe+Xe $5.44$ TeV}\hbox{\strut $(\eta/s)(T)$ at $T=X$ MeV}}} \\
\hline
 & $100$ & $150$ & $200$ & $250$ & $300$
 & $100$ & $150$ & $200$ & $250$ & $300$
 & $100$ & $150$ & $200$ & $250$ & $300$
 & $100$ & $150$ & $200$ & $250$ & $300$ \\
\hline
\multirow{2}{*}{$N(\mathrm{ch})$} & - & - & - & - & - & - & - & - & - & - & - & - & - & - & - & - & - & - & - & - \\ & - & - & - & - & - & - & - & - & - & - & - & - & - & - & - & - & - & - & - & - \\
\hline
\multirow{2}{*}{$N(\pi)$} & - & - & - & - & - & - & - & - & - & - & - & - & - & - & - & - & - & - & - & - \\ & - & - & - & - & - & - & - & - & - & - & - & - & - & - & - & - & - & - & - & - \\
\hline
\multirow{2}{*}{$N(K)$} & - & - & - & - & - & - & - & - & - & - & - & - & - & - & - & - & - & - & - & - \\ & - & - & - & - & - & - & - & - & - & - & - & - & - & - & - & - & - & - & - & - \\
\hline
\multirow{2}{*}{$N(p)$} & - & - & - & - & - & - & - & - & - & - & - & - & - & - & - & - & - & - & - & - \\ & - & - & - & - & - & - & - & - & - & - & - & - & - & - & - & - & - & - & - & - \\
\hline
\multirow{2}{*}{$\langle p_T \rangle (\mathrm{ch})$} & - & - & - & - & - & - & - & - & - & - & - & - & - & - & - & - & - & 0.30 & - & - \\ & - & - & - & - & - & - & - & - & - & - & - & - & - & - & - & - & - & 0.25 & - & - \\
\hline
\multirow{2}{*}{$\langle p_T \rangle (\pi)$} & 0.35 & - & - & - & - & 0.33 & - & - & - & - & 0.37 & - & - & - & - & - & - & - & - & - \\ & 0.15 & - & - & - & - & 0.14 & - & - & - & - & 0.09 & - & - & - & - & - & - & - & - & - \\
\hline
\multirow{2}{*}{$\langle p_T \rangle (K)$} & - & 0.60 & 0.59 & - & - & - & 0.56 & 0.56 & - & - & - & 0.55 & 0.55 & 0.31 & - & - & - & - & - & - \\ & - & 0.33 & 0.35 & - & - & - & 0.38 & 0.39 & - & - & - & 0.38 & 0.41 & 0.18 & - & - & - & - & - & - \\
\hline
\multirow{2}{*}{$\langle p_T \rangle (p)$} & - & 0.68 & 0.65 & - & - & - & 0.65 & 0.64 & 0.31 & - & - & 0.65 & 0.66 & 0.38 & - & - & - & - & - & - \\ & - & 0.45 & 0.46 & - & - & - & 0.52 & 0.53 & 0.22 & - & - & 0.53 & 0.49 & 0.17 & - & - & - & - & - & - \\
\hline
\multirow{2}{*}{$v_2$} & -0.02 & - & - & - & - & -0.16 & - & -0.06 & -0.12 & -0.09 & -0.31 & -0.15 & -0.23 & -0.23 & -0.19 & -0.33 & -0.11 & -0.23 & -0.46 & -0.27 \\ & -0.47 & - & - & - & - & -0.44 & - & -0.32 & -0.43 & -0.38 & -0.52 & -0.34 & -0.42 & -0.53 & -0.44 & -0.50 & -0.47 & -0.54 & -0.60 & -0.48 \\
\hline
\multirow{2}{*}{$v_3$} & - & 0.53 & 0.56 & 0.32 & - & - & -0.05 & -0.13 & -0.19 & 0.01 & - & 0.03 & -0.09 & -0.12 & 0.13 & - & -0.27 & -0.40 & -0.43 & -0.17 \\ & - & -0.33 & -0.29 & -0.04 & - & - & -0.46 & -0.48 & -0.48 & -0.41 & - & -0.58 & -0.55 & -0.57 & -0.46 & - & -0.56 & -0.64 & -0.62 & -0.43 \\
\hline
\multirow{2}{*}{$v_4$} & 0.33 & 0.65 & 0.63 & 0.33 & - & - & 0.05 & 0.07 & 0.07 & - & 0.26 & 0.26 & 0.32 & 0.22 & 0.18 & - & -0.32 & -0.39 & -0.08 & - \\ & -0.25 & -0.31 & -0.24 & -0.03 & - & - & -0.44 & -0.49 & -0.39 & - & -0.32 & -0.40 & -0.47 & -0.47 & -0.35 & - & -0.49 & -0.53 & -0.39 & - \\
\hline
\multirow{2}{*}{$NSC(4,2)$} & -0.04 & 0.44 & 0.44 & - & - & -0.04 & 0.38 & 0.36 & - & - & -0.30 & - & - & - & - & - & - & - & - & - \\ & -0.30 & -0.10 & -0.14 & - & - & -0.52 & -0.14 & -0.08 & - & - & -0.47 & - & - & - & - & - & - & - & - & - \\
\hline
\end{tabular}
\caption{Notable Pearson correlations ($|r|>0.3$ in at least one centrality class) between specific shear viscosity $\eta/s$ at various temperatures and observables, for the posterior distribution from the analysis with 20\% theoretical uncertainty. Upper and lower values in a cell represent the highest and lowest correlations, respectively, among all the investigated centrality classes for each observable.}
\label{T:etasT_obs_correlations_all}
\end{table*}

\begin{table*}[h]
\centering
\begin{tabular}{|c||c|c|c|c|c||c|c|c|c|c||c|c|c|c|c||c|c|c|c|c||}
\hline
 & \multicolumn{5}{c||}{\vtop{\hbox{\strut Au+Au $200$ GeV}\hbox{\strut $(\zeta/s)(T)$ at $T=X$ MeV}}}
 & \multicolumn{5}{c||}{\vtop{\hbox{\strut Pb+Pb $2.76$ TeV}\hbox{\strut $(\zeta/s)(T)$ at $T=X$ MeV}}}
 & \multicolumn{5}{c||}{\vtop{\hbox{\strut Pb+Pb $5.02$ TeV}\hbox{\strut $(\zeta/s)(T)$ at $T=X$ MeV}}}
 & \multicolumn{5}{c||}{\vtop{\hbox{\strut Xe+Xe $5.44$ TeV}\hbox{\strut $(\zeta/s)(T)$ at $T=X$ MeV}}} \\
\hline
 & $100$ & $150$ & $200$ & $250$ & $300$
 & $100$ & $150$ & $200$ & $250$ & $300$
 & $100$ & $150$ & $200$ & $250$ & $300$
 & $100$ & $150$ & $200$ & $250$ & $300$ \\
\hline
\multirow{2}{*}{$N(\mathrm{ch})$} & - & - & - & - & - & - & - & - & - & - & - & - & - & - & - & - & - & - & - & - \\ & - & - & - & - & - & - & - & - & - & - & - & - & - & - & - & - & - & - & - & - \\
\hline
\multirow{2}{*}{$N(\pi)$} & - & - & - & - & - & - & - & - & - & - & - & - & - & - & - & - & - & - & - & - \\ & - & - & - & - & - & - & - & - & - & - & - & - & - & - & - & - & - & - & - & - \\
\hline
\multirow{2}{*}{$N(K)$} & - & - & - & - & - & - & - & - & - & - & - & - & - & - & - & - & - & - & - & - \\ & - & - & - & - & - & - & - & - & - & - & - & - & - & - & - & - & - & - & - & - \\
\hline
\multirow{2}{*}{$N(p)$} & 0.31 & 0.51 & - & - & - & 0.32 & 0.41 & - & 0.09 & - & - & 0.30 & - & - & - & - & - & - & - & - \\ & -0.19 & -0.24 & - & - & - & -0.22 & -0.35 & - & -0.30 & - & - & -0.32 & - & - & - & - & - & - & - & - \\
\hline
\multirow{2}{*}{$\langle p_T \rangle (\mathrm{ch})$} & - & - & - & - & - & - & - & - & - & - & - & - & - & - & - & -0.12 & -0.24 & -0.41 & - & - \\ & - & - & - & - & - & - & - & - & - & - & - & - & - & - & - & -0.32 & -0.58 & -0.51 & - & - \\
\hline
\multirow{2}{*}{$\langle p_T \rangle (\pi)$} & - & -0.00 & -0.16 & - & - & 0.07 & 0.03 & -0.23 & - & - & - & 0.02 & -0.22 & - & - & - & - & - & - & - \\ & - & -0.54 & -0.44 & - & - & -0.34 & -0.58 & -0.41 & - & - & - & -0.53 & -0.46 & - & - & - & - & - & - & - \\
\hline
\multirow{2}{*}{$\langle p_T \rangle (K)$} & - & -0.06 & -0.23 & - & - & - & -0.05 & -0.29 & - & - & - & 0.01 & -0.28 & 0.01 & - & - & - & - & - & - \\ & - & -0.53 & -0.46 & - & - & - & -0.53 & -0.47 & - & - & - & -0.51 & -0.47 & -0.35 & - & - & - & - & - & - \\
\hline
\multirow{2}{*}{$\langle p_T \rangle (p)$} & - & 0.03 & -0.14 & - & - & - & 0.05 & -0.20 & - & - & - & 0.10 & -0.18 & -0.02 & - & - & - & - & - & - \\ & - & -0.46 & -0.44 & - & - & - & -0.39 & -0.43 & - & - & - & -0.41 & -0.39 & -0.33 & - & - & - & - & - & - \\
\hline
\multirow{2}{*}{$v_2$} & - & - & 0.09 & - & - & - & - & - & - & - & - & - & - & - & - & - & - & - & - & - \\ & - & - & -0.31 & - & - & - & - & - & - & - & - & - & - & - & - & - & - & - & - & - \\
\hline
\multirow{2}{*}{$v_3$} & - & -0.01 & - & - & - & - & - & - & - & - & - & - & - & - & - & - & - & - & - & - \\ & - & -0.34 & - & - & - & - & - & - & - & - & - & - & - & - & - & - & - & - & - & - \\
\hline
\multirow{2}{*}{$v_4$} & - & 0.08 & 0.22 & - & - & - & - & - & - & - & - & - & - & - & - & - & - & - & - & - \\ & - & -0.42 & -0.32 & - & - & - & - & - & - & - & - & - & - & - & - & - & - & - & - & - \\
\hline
\multirow{2}{*}{$NSC(4,2)$} & - & - & - & - & - & - & - & - & - & - & - & - & - & - & - & - & - & - & - & - \\ & - & - & - & - & - & - & - & - & - & - & - & - & - & - & - & - & - & - & - & - \\
\hline
\end{tabular}
\caption{Like table \ref{T:etasT_obs_correlations_all} but for $\zeta/s$ instead of $\eta/s$.}
\label{T:zetasT_obs_correlations_all}
\end{table*}

\end{document}